\documentclass[]{aastex701} 

\usepackage[T1]{fontenc}

\let\tablenum\relax
\usepackage{siunitx}
\DeclareSIUnit\angstrom{\text{\AA}}
\DeclareSIUnit\erg{erg}

\usepackage{amsmath}
\usepackage{hyperref}
\usepackage{multirow}
\usepackage{booktabs}

\newcommand{\n}{\\}
\newcommand{\tsub}{\textsubscript}

\newcommand{\A}  [1]{\qty{#1}{\angstrom}} 
\newcommand{\kms}[1]{\qty{#1}{\km\per\s}} 
\newcommand{\um} [1]{\qty{#1}{\um}}       

\newcommand{\td} [1]{
\begin{tabular}{@{\hspace{-16pt}}c}#1\end{tabular}}
\newcommand{\tdd}[2]{
\begin{tabular}{@{\hspace{-20pt}}r}#1\n\textpm#2\n\end{tabular}}
\newcommand{\uFluxLam}[1]{\qty[separate-uncertainty-units=bracket]
{#1e-17}{\erg\per\s\per\cm\squared\per\angstrom}} 

\begin{document}

\title{A New Member of the Fast and Furious Family: \\ A Relativistic and Time-Variable UV Outflow in a Luminous Quasar}

\author[0009-0006-5718-7023]{Lucas M. Seaton}
\affiliation{Department of Physics and Astronomy, York University, 4700 Keele St., Toronto, ON M3J 1P3, Canada}
\email{lucas.seaton@hotmail.com}

\author[0000-0002-1763-5825]{Patrick B.\ Hall}
\affiliation{Department of Physics and Astronomy, York University, 4700 Keele St., Toronto, ON M3J 1P3, Canada}
\email{phall@yorku.ca}

\author{Liliana Flores}
\affiliation{Physical Sciences Division, School of STEM, University of Washington Bothell, WA, 98011, USA}
\email{floresl2@uw.edu}

\author[0000-0003-0677-785X]{Paola Rodr\'iguez Hidalgo}
\affiliation{Physical Sciences Division, School of STEM, University of Washington Bothell, WA, 98011, USA}
\email{paola@uw.edu}

\author[0009-0009-4362-8782]{Marianna Veltri}
\affiliation{Department of Physics and Astronomy, York University, 4700 Keele St., Toronto, ON M3J 1P3, Canada}
\email{mariannaveltri@hotmail.com}

\author[0009-0007-7109-4293]{Zezhou Zhu}
\affiliation{Department of Physics and Astronomy, York University, 4700 Keele St., Toronto, ON M3J 1P3, Canada}
\email{zzz@my.yorku.ca}

\author[0000-0001-7351-6540]{Javier Serna}
\affiliation{Homer L. Dodge Department of Physics and Astronomy, University of Oklahoma, Norman, OK 73019, USA}
\email{jserna@ou.edu}

\author{W. Niel Brandt}
\affiliation{Department of Astronomy \&\ Astrophysics, 525 Davey Lab, The Pennsylvania State University, University Park, PA 16802, USA}
\affiliation{Institute for Gravitation and the Cosmos, The Pennsylvania State University, University Park, PA 16802, USA}
\affiliation{Department of Physics, 104 Davey Lab, The Pennsylvania State University, University Park, PA 16802, USA}
\email{wnbrandt@gmail.com}

\author{Scott Anderson}
\affiliation{Department of Astronomy, University of Washington, Box 351580, Seattle, WA 98195, USA}
\email{sfander@uw.edu}

\author{Roberto J. Assef}
\affiliation{Instituto de Estudios Astrof\'isicos, Facultad de Ingenier\'ia y Ciencias, Universidad Diego Portales, Av. Ej\'ercito Libertador 441, Santiago, Chile}
\email{roberto.assef@mail.udp.cl}

\author[0000-0002-2931-7824]{Eduardo Ba{\~n}ados} 
\affiliation{Max-Planck-Institut für Astronomie, Königstuhl 17, D-69117, Heidelberg, Germany}
\email{banados@mpia.de}

\author[0000-0001-9920-6057]{Catherine~J.~Grier}
\affiliation{Department of Astronomy, University of Wisconsin-Madison, Madison, WI 53706, USA}
\email{kate.grier@astro.wisc.edu}

\author{Yasaman Homayouni}
\affiliation{Department of Astronomy \&\ Astrophysics, 525 Davey Lab, The Pennsylvania State University, University Park, PA 16802, USA}
\email{ybh5251@psu.edu}

\author[0000-0002-6770-2627]{Sean Morrison}
\affiliation{Department of Astronomy, University of Illinois at Urbana-Champaign, Urbana, IL 61801, USA}
\email{smorris0@illinois.edu}

\author[0000-0002-1656-827X]{C.~Alenka Negrete}
\affiliation{Instituto de Astronomía, Universidad Nacional Autónoma de México, A.P. 70-264, 04510, Mexico, D.F., México}
\email{alenka@astro.unam.mx}

\author[0000-0002-2091-1966]{Amy L. Rankine}
\affiliation{Institute for Astronomy, University of Edinburgh, Royal Observatory, Edinburgh EH9 3HJ, UK}
\email{amy.rankine@ed.ac.uk}

\author{Jessie Runnoe}
\affiliation{Department of Physics and Astronomy, Vanderbilt University, VU Station 1807, Nashville, TN 37235, USA}
\email{jessie.c.runnoe@vanderbilt.edu}

\author{Donald P. Schneider}
\affiliation{Department of Astronomy \&\ Astrophysics, 525 Davey Lab, The Pennsylvania State University, University Park, PA 16802, USA}
\affiliation{Institute for Gravitation and the Cosmos, The Pennsylvania State University, University Park, PA 16802, USA}
\email{dps7@psu.edu}

\author{Yue Shen}
\affiliation{Department of Astronomy, University of Illinois at Urbana-Champaign, Urbana, IL 61801, USA}
\email{shenyue@illinois.edu}

\author{Matthew Temple}
\affiliation{Department of Astronomy \&\ Astrophysics, 525 Davey Lab, The Pennsylvania State University, University Park, PA 16802, USA}
\email{matthew.temple@mail.udp.cl}

\author{Benny Trakhtenbrot}
\affiliation{School of Physics and Astronomy, Tel Aviv University, Tel Aviv 69978, Israel}
\email{benny.trakht@gmail.com}

\author[0000-0002-1410-0470]{Jonathan R. Trump}
\affiliation{Department of Physics, 196A Auditorium Road, Unit 3046, University of Connecticut, Storrs, CT 06269, USA}
\email{jonathan.trump@uconn.edu}

\author{Erik Weiss}
\affiliation{Department of Physics and Astronomy, York University, 4700 Keele St., Toronto, ON M3J 1P3, Canada}
\email{9erik1@gmail.com}

\begin{abstract}
	We report the fastest quasar outflow first detected in the ultraviolet, via variable \ion{C}{4} and \ion{Si}{4} absorption at outflow velocities $-77,000$ \kms{} to at least $-90,000$ \kms{}, in the radio-quiet quasar SDSS J231854.31+243954.2 (J2318).
	J2318 is a weak-lined quasar in the rest-frame ultraviolet, but Gemini GNIRS spectroscopy reveals an H$\alpha$ redshift of $z=2.6781 \pm 0.0004$.
    A twenty-year photometric time series shows peak-to-peak variability of 0.5 mag in the $g$ band.
	The \ion{C}{4} outflow strengthened monotonically over three epochs spanning $\sim$2.2 rest-frame years.
    The existence of such a high-velocity outflow implies that models of quasar outflows must be able to either accelerate gas to $0.3c$ while still preserving \ion{C}{4} and \ion{Si}{4} ions, or enable the formation of \ion{C}{4} and \ion{Si}{4} ions in gas which has been accelerated to $0.3c$.
    Virial estimates reveal a black-hole mass of $1.65\times10^9M_\odot$, which leads to an Eddington luminosity and Eddington ratio of $2.4\times10^{47}$ erg s$^{-1}$ and $0.45$, respectively.
    Using very conservative assumptions, the UV-absorbing outflow alone has an estimated mass loss of $>0.82~M_\odot~{\rm yr}^{-1}$ and a kinetic luminosity ratio $L_{kin}/L_{bol}\geq0.75$\%. 
    The lower limit is just above the threshold usually cited for significant feedback on the host galaxy. 
    Comparison to PDS 456, the only other known quasar with a UV-absorbing outflow at $0.3c$, suggests that the true $\dot{M}$ and $L_{kin}/L_{bol}$ could be up to two orders of magnitude larger.
\end{abstract}

\section{Introduction} \label{sec:intro}
A quasar is a supermassive black hole (SMBH) located at the centre of a massive galaxy and surrounded by a rapidly rotating accretion disk. 
Quasars are extremely luminous, ultimately powered by the conversion of gravitational potential energy into thermal and non-thermal radiation as matter spirals through the disk into the SMBH (e.g., \citealt{ss73}). 
Quasar’s luminosities allow us to observe active galactic nuclei (AGNs) at large redshifts, enabling us to study the inner regions of galaxies in the earlier universe.
Quasars usually have broad emission lines (BELs; e.g., \citealt{2011AJ....141..167R}) and often broad absorption lines (BALs; e.g., \citealt{2020MNRAS.492.4553R}). 
BAL troughs arise from material from the accretion disk or the nuclear environment that is accelerated by radiation pressure \citep[e.g.,][]{proga2003}, causing the absorbing material to travel towards us and appear blueshifted in the quasar rest frame. 
Connection between the nuclear region and its surrounding host galaxy is necessary to understand the coevolution of SMBHs and their host galaxies, as evidenced by the correlation between the SMBH mass and the stellar spheroids of the host galaxy 
\citep[e.g.,][]{2000ApJ...539L..13G,2002ApJ...574..740T}.
Furthermore, simulations of galactic evolution require nuclear feedback from AGN outflows and jets as a regulating mechanism \citep[e.g.,][]{2005Natur.433..604D,hop06}.
Thus, studying outflows can provide observational tests to quasar-host-galaxy models, since the outflows will impact their host galaxy and are thought to provide a key component of the feedback required for galaxy formation models to match observations \citep[e.g.,][]{Morganti2017}.

Quasar outflows can show remarkable variations in periods of time as short as days \citep[e.g.,][]{2013MNRAS.429.1872C,2015ApJ...806..111G,2019ApJ...872...21H,Yi+2022}. 
These changes are possibly due to absorbing clouds moving across our line of sight \citep[e.g.,][]{Vivek+2012,2013MNRAS.429.1872C,Yi+2022}, or changes in the ionization of the gas clouds \citep[e.g.,][]{2012ApJ...757..114F,2013ApJ...777..168F,2015ApJ...806..111G,wywf,DeCicco+2018,2022SciA....8.3291H}, or both \citep[e.g.,][]{J0230,2017arXiv170503019M,2019ApJ...872...21H}.
This variability can explicitly be seen through obvious changes in quasar spectra \citep[e.g.,][]{Vivek+2012,2013MNRAS.429.1872C,2013ApJ...777..168F,2015ApJ...806..111G,2016ApJ...824..130G,J0230,jarthesis,2017arXiv170503019M,DeCicco+2018,2019ApJ...872...21H,2019MNRAS.482.1121S,Yi+2022}
or through the appearance or disappearance of BAL troughs \citep[e.g.,][]{Vivek+2012,2012ApJ...757..114F,J0230,jarthesis,2017arXiv170503019M,DeCicco+2018,2019MNRAS.482.1121S}.

Over the past few decades, attempts have been made to quantify the observed(intrinsic) fraction of \ion{C}{4} BAL quasars within the total AGN population.
The observed fraction is determined by the amount of BAL quasars identified in the respective quasar sample, whereas the intrinsic fraction, which usually results in a higher fraction, takes into account sample selection effects.
The reported fractions vary in the literature and we provide two examples for illustrative purposes. \citet{allenbal} report an observed fraction of $8.0\pm1.0\%$ and an intrinsic fraction of $40.7\pm5.4\%$ in their sample of 68733 Sloan Digital Sky Survey (SDSS) quasars (QSOs), while \citet{2023ApJ...952...44B} report an observed and intrinsic fraction of $\simeq20\%$ and $\sim47\%$, respectively, for their sample of 1935 luminous quasars, and also note that there could be an increasing BAL fraction at higher redshift.

BAL quasars were originally defined as systems exhibiting absorption blueshifted by \kms{-3000} to $-25,000$ \kms{} \citep{wea91}; where the convention is that a more negative velocity indicates a faster outflow that is moving away from the quasar and towards the observer along our sightline.
This traditional limit of $-25,000$ \kms{} was artificial, established due to possible confusion of high-velocity \ion{C}{4} absorption with lower-velocity \ion{Si}{4} absorption.
Faster-moving absorption systems observed in both the ultraviolet \citep[UV; e.g.,][]{2011MNRAS.411..247R} and the X-ray \citep[e.g.,][]{Matzeu+23} should of course be considered in studies of quasar feedback, as there is no fundamental physical difference between BALs and these higher velocity outflows (see Appendix A of \citealt{sdss123}).  
In the X-ray, such systems are referred to as ultra-fast outflows (UFOs), which have been detected at low redshifts $z<0.4$ (e.g., \citealt{Matzeu+23})
and in a few mostly gravitationally lensed quasars out to $z=4$ \citep{2021ApJ...920...24C}.
In the UV, extremely high-velocity outflows (EHVOs) are defined as broad absorption line troughs at $v<-30,000$ \kms{} seen in \ion{C}{4} and sometimes \ion{Si}{4} \citep{EHVO}, and they have been detected to the highest redshifts $>6$ \citep[e.g.,][]{Wang+2021,Belladitta+2025} with velocities reaching $0.15<v<0.2c$.
Like traditional BALs, EHVOs are seen to be variable, but EHVO quasars have properties which are in some ways distinct from BAL quasars \citep{EHVORHR}, suggesting that EHVO quasars and BAL quasars have different parent populations despite both exhibiting broad ultraviolet absorption troughs in their spectra.
We refer to the absorption troughs in J2318 as BAL troughs and EHVO troughs interchangeably in this paper, but it should be kept in mind that its outflow is not drawn from the traditional BAL population.

In both BAL and EHVO quasars, how gas accelerates without overionizing so as to be seen in relatively low-ionization stages at high outflow velocities remains a challenge for theoretical models to explain (\citealt{dydadavisproga23,2025MNRAS.541.2393M}, R. Dannen et~al., in preparation). 
\citet{mcgv} put forward a model in which shielding gas absorbs X-rays and enables UV line driving to accelerate gas to $v_{max}\simeq 0.2c$ (see their \S~3) with detached line profiles seen in cases where gas is accelerated before entering our line of sight to the continuum source (end of their \S~2). 
However, \citet{2013MNRAS.435..133H} show that the amount of shielding is too weak to control ionization and accelerations in quasars with weak EHVO absorption. 
\citet{rpc4} also argue against a radiative shield and instead propose radiation pressure compression as an explanation for the observed ionization levels.
Slabs of photoionized gas are compressed and confined along the line of sight by the incident ionizing radiation, such that the gas becomes dense enough to prevent overionization, but ionized enough to produce observed UV absorption lines. 
The acceleration of absorbing gas in BAL and EHVO quasars appears to be governed by the ionizing UV spectrum, with softer spectra producing more effective radiative acceleration \citep{2019MNRAS.483.1808Hamann}. 
Quasars with higher UV luminosities, and lower X-ray luminosities, have faster \ion{C}{4} BAL outflow velocities, larger equivalent widths, and higher modified balnicity indices \citep{gibsonbal}.

The highest-velocity UV-wavelength EHVO quasar identified to date is PDS 456, in which absorption was first identified in the X-ray \citep{2015Sci...347..860N} via Fe K$\alpha$ at $v=(-0.35\pm0.05)c$ but is also present in the ultraviolet \citep{2018MNRAS.476..943H} with \ion{C}{4} at $v=(-0.30\pm0.03)c$.
The highest-velocity EHVO quasar first previously identified from its UV absorption is SDSS J023011.28+005913.6 \citep{J0230}, with \ion{C}{4} in outflow at $v=(-0.186\pm0.015)c$.

In this work, we present the identification of the quasar SDSS J231854.31+243954.2 (hereafter J2318) at $z=2.6781$ as an extremely high-velocity outflow quasar with \ion{C}{4} and \ion{Si}{4} absorption at speeds reaching $v\sim-0.3c$.
J2318 was identified, by visual inspection, as an EHVO quasar using data from the SDSS five (SDSS-V; \citealt{sdssVover}), an all-sky, multi-epoch spectroscopic survey which provides near-infrared and optical multi-object fiber spectroscopy of over six million objects. 
The SDSS-V's Black Hole Mapper (BHM, S. Anderson et al., in prep.) provides a large sample of spectroscopic data on quasars which enables analysis of their spectral variability \citep{sdssv}. 

In this paper, we outline our observations of J2318 in \S~2, present our analysis of the data in \S~3, and discuss our results in \S~4.
We summarize our work and make concluding remarks in \S~5.
Where necessary we assume a flat cosmology with $h=0.676$, $\Omega_M=0.31$, and $\Omega_\Lambda=0.69$, as used in the SDSS DR16 quasar catalog \citep{dr16q}.

\section{Observations} \label{sec:obs}
SDSS-V operates multiple telescopes and spectrographs at two different observatories located in North and South America.
The Apache Point Observatory (APO) in New Mexico, USA, is home to the 2.5m Sloan Foundation Telescope \citep{gun06}.
The Carnegie Observatories' 2.5m du Pont telescope \citep{bv73} is located at the Las Campanas Observatory (LCO) in Chile.
Both observatories have a set of survey instruments for SDSS-V which include a near-IR APOGEE spectrograph \citep{apogeespec} for observations of stars and a pair of optical multi-object fiber spectrographs \citep{bosssmee} used to collect spectra with spectral resolution  $R=\lambda/\Delta\lambda \simeq 2000$ in SDSS-III, SDSS-IV, and now in SDSS-V for BHM, among other projects.

Through the end of 2023, J2318 had 2 imaging and 3 spectroscopic observations in SDSS.
We report spectroscopic observations of J2318 in Table \ref{tab:observ} and photometric observations in Table \ref{tab:phot}.
For Table 1, note that the SDSS-V CatalogID changes with the version of the targeting crossmatch database used \citep{sdssVover}.
Each object considered as a potential SDSS-V target does have a unique identifier (its {\sc sdss\_id}); J2318 has {\sc sdss\_id} = 71909218. 
For completeness, we checked and there is no spectrum of J2318 in DESI DR1~\citep{DESIDR1}.

\begin{table*}
	\centering
	\begin{tabular}{ccrrccllr}
		\hline
		MJD\tsub{Obs} & UT Date\tsub{Obs} & Rest Day & Rest $\Delta t$ & Origin & Plate/Field & FiberID/CatalogID & Pipeline  & SN$_{\um1}$ \\
		\hline
		57328.08      & 2015--11--02   & 0.00            & 0.00    & SDSS-IV & 7670        & 918                & v5\_13\_2 & 20.1        \\
		59188.09      & 2020--12--05   & 505.70          & 505.70   & SDSS-V & 015036      & 4384421494 (v0)        & v6\_2\_0 & 21.7        \\
        60251.11      & 2023--11--03   & 794.71          & 289.01   & SDSS-V & 104623     & 63050395075696130 (v1)  & v6\_1\_1 & 9.9         \\
		60291.25      & 2023--12--13   & 805.63           & 10.91  & GNIRS & -         & -                & - & 10.5        \\
		\hline
	\end{tabular}
	\caption{Spectroscopic observations of J2318 (SDSS J231854.31+243954.2; ${\rm R.A.} = 349.72629$, ${\rm Decl.} = +24.66506$).
        Rest Day is the cumulative rest-frame days elapsed since the first observation. 		
        Rest $\Delta t$ is the rest-frame time (at $z$=2.6781) in days elapsed since the previous spectroscopic observation.
        SDSS-IV spectra are tracked using Plate, MJD, and FiberID. SDSS-V spectra are tracked using Field, MJD, and CatalogID, but CatalogID changes with each targeting crossmatch version; see text. 
        For SDSS spectra we give the version of the spectroscopic pipeline used to reduce each spectrum (\citealt{bosspipe}, S. Morrison et al.\ in prep.). 
        SN$_{\um1}$ is the signal-to-noise ratio per \kms{69} pixel, calculated as the average of the flux divided by the error in the flux over all pixels at wavelengths common to all three spectra: \A{8239.483} to \A{10325.231}.}
	\label{tab:observ}
\end{table*}

\subsection{Spectroscopic observations} \label{sec:spec}
\subsubsection{SDSS observations} \label{sec:sdss} 
J2318 was first observed spectroscopically by SDSS on MJD 57328 (2015 November 2).
During this SDSS-IV phase of the survey \citep{sdssIVoverview}, as in previous phases, the spectrum was obtained using a fiber-optic cable plugged into a specially machined metal  plate \citep{2016AJ....151...44D}.
Its targeting flags indicate it was a Time Domain Spectroscopic Survey (TDSS) Single Epoch Spectroscopy (SES) target \citep{tdssSES}.
The TDSS SES sample was designed to be a probe of general optical variability, using combined imaging from SDSS and Pan-STARRS1 \citep{PS1database} to select a highly pure sample of variable objects unbiased with regard to color or variability pattern.

The MJD 57328 spectrum of J2318 was first released in SDSS Data Release 14 (DR14; \citealt{2018ApJS..235...42A}), although we use the final legacy reduction of the spectrum released in DR17.
J2318 was mistakenly not included in the DR14 quasar catalog \citep{2018A&A...613A..51P}, possibly due to its incorrect DR14 pipeline redshift of $z=2.469$.
J2318 was included in the DR16 quasar catalog \citep{dr16q}, with a visual inspection redshift of $z=2.67$ and an absolute magnitude of $M_i(z=2)=-28.4$.
J2318 is also not currently classified as a quasar within VizieR \citep{vizier}.

J2318 was reobserved spectroscopically by SDSS-V on MJD 59188 (5 Dec., 2020) as part of the All-Quasar Multi-Epoch Spectroscopy program AQMES-wide \citep{dr19}.
In that initial phase of SDSS-V, observations were still obtained using plug plates.
J2318 was observed again by SDSS-V on MJD 60251 (2023 November 3).
By that time in SDSS-V, spectra were obtained using fiber-optic cables held in place by fiber positioner robots 
\citep{2020SPIE11447E..81P}.
The v6\_1\_3 spectroscopic reduction of the MJD 59188 spectrum was publicly released in DR19 \citep{dr19} and the v6\_2\_1 reductions of both SDSS-V spectra will be released in DR20 in 2026.
The J2318 spectra used in our analysis are earlier reductions with slightly higher noise but no systematic offsets.

The Galactic extinction maps of \cite{sfd98} indicate a color excess of $E(B-V)=0.086$ along the line of sight to J2318.
We corrected for this Galactic extinction using the Milky Way extinction curve of \cite{mw89} with $R_V=3.1$.

\begin{figure}
    \includegraphics[width=1.00\textwidth]{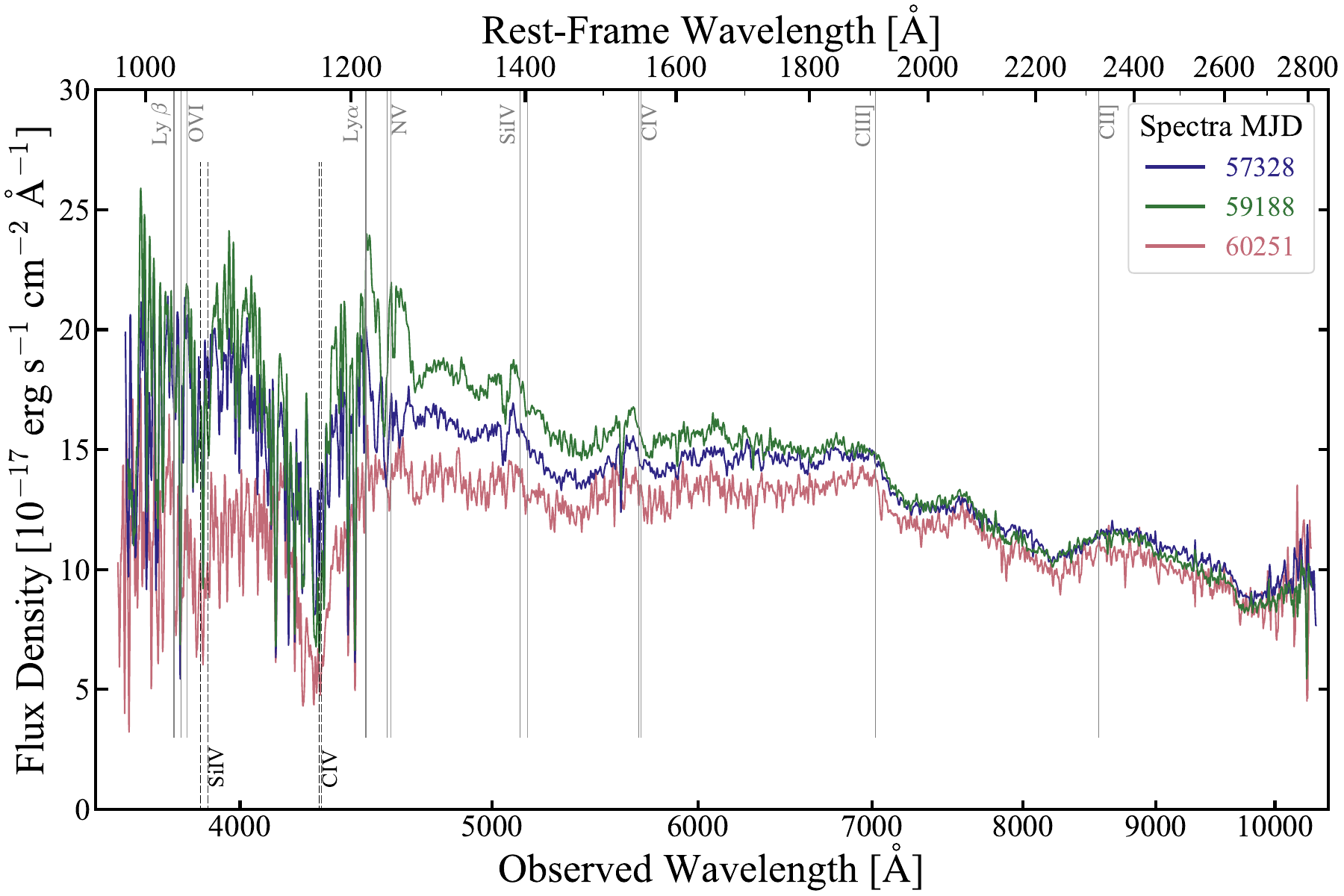} 
	\centering
	\caption{SDSS Spectra.
		The SDSS-IV spectrum, MJD 57328, is shown in blue, and the two SDSS-V spectra, MJD 59188 and 60251, in green and red, respectively.
        The continuum is highest in 59188 and lowest in 60251.
        Emission features along the top are marked at a redshift of $z=2.6781$, and along the bottom \ion{Si}{4} and \ion{C}{4} are marked where seen in absorption.
        }
	\label{fig:J2318_rawSDSS}
\end{figure}

Figure \ref{fig:J2318_rawSDSS}
displays the Galactic-extinction-corrected SDSS spectra of J2318, smoothed by a five-pixel weighted average.
The identification of the absorption trough in the second SDSS-V spectrum (obtained MJD 60251) as an EHVO in \ion{C}{4} is secure due to the presence of \ion{Si}{4} at the same outflow velocity (see discussion in \S \ref{sec:troughs}).
J2318 was brightest and bluest in the MJD 59188 spectrum, and faintest and reddest in the MJD 60251 spectrum.
The $g-i$ color synthesized from the spectra is redder for the MJD 60251 spectrum than for the MJD 57328 and MJD 59188 spectra at $3.4\sigma$ and $4.8\sigma$ significance, respectively (see Table~\ref{tab:phot}).

\begin{figure}
	\includegraphics[width=1.00\textwidth]{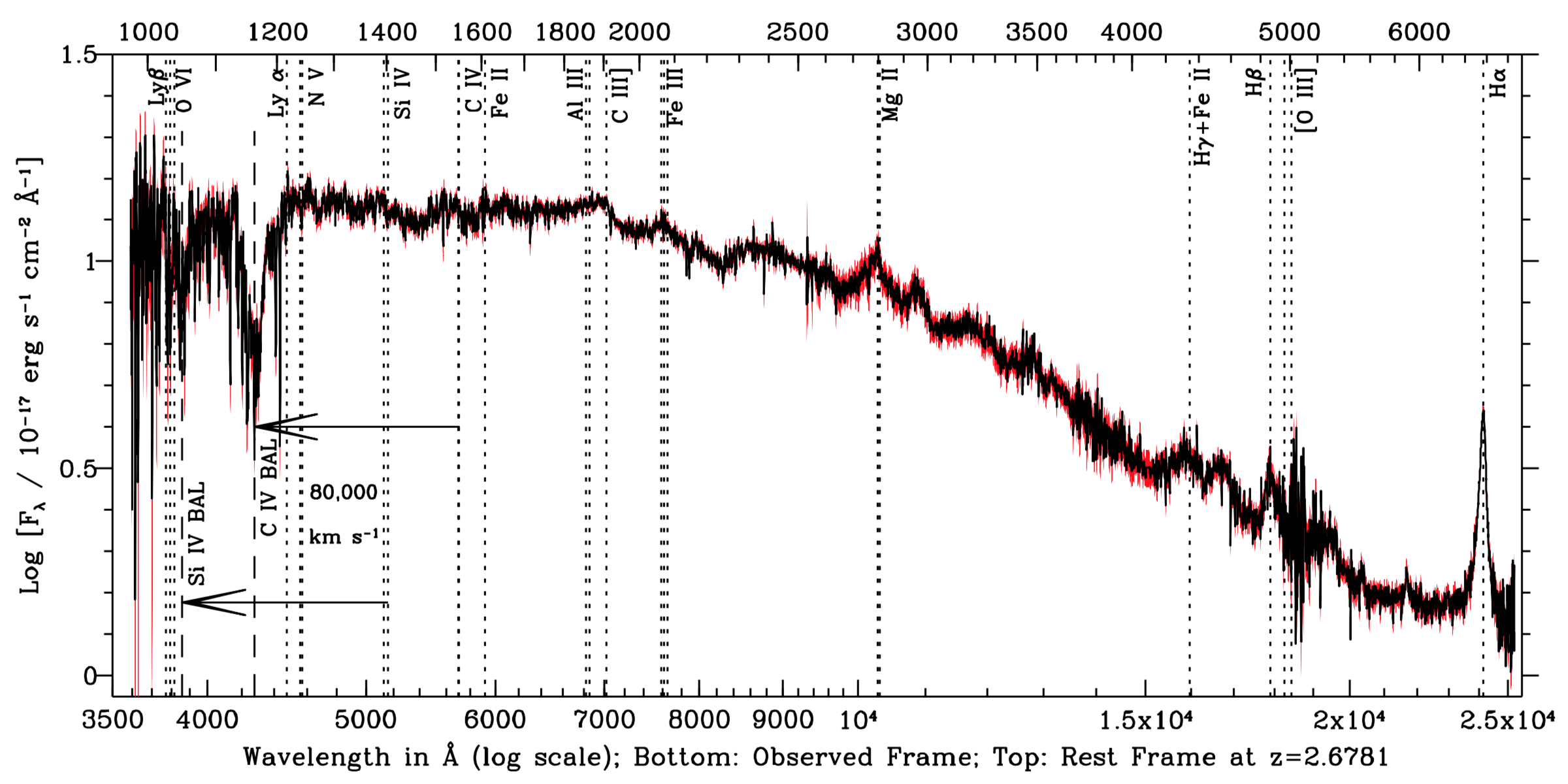}
	\centering
	\caption{Optical-NIR Spectrum. The combined MJD 60251 SDSS-V and MJD 60291 Gemini spectrum created as described in \S \ref{sec:combining}, displayed on a log observed flux and log wavelength scale. 
    The black line is the spectrum and the red envelope is the $\pm 1\sigma$ uncertainty range. 
    The bottom axis shows the observed-frame wavelengths while the top axis shows the rest-frame wavelengths at $z=2.6781$. 
    Emission lines common in quasar spectra are marked with dotted vertical lines; not all are obvious in J2318. 
    The extremely high-velocity C IV and Si IV troughs are marked with dashed vertical lines and with arrows showing their shifts from their rest wavelengths.}
	\label{fig:J2318_SDSS+Gemini}
\end{figure}

\subsubsection{Gemini observations} \label{sec:gemini} 
Due to the uncertain systemic redshift from the weak, broad UV lines, we requested Gemini Director's Discretionary Time near-infrared spectroscopy to secure a redshift from the H$\alpha$ emission line.

J2318 was observed on MJD 60291 (2023 December 13), using the GNIRS instrument \citep{GNIRSdesign,GNIRSperf} on the Gemini North telescope.
A total of eight three-minute exposures were obtained with the short camera in cross-dispersed mode, with spectral coverage from \um{0.824} to \um{2.523} and a 0\farcs45 wide slit yielding resolution $R=\lambda/\Delta\lambda\simeq 1130$ (instrumental FWHM $\simeq$ \kms{265}).

Data were reduced using standard methods in \textsc{iraf} \citep{IRAFV218}.\footnote{IRAF is distributed by the Community Science and Data Center at NSF's NOIRLab, which is managed by the Association of Universities for Research in Astronomy (AURA) under a cooperative agreement with the National Science Foundation.}
Due to an issue with the GNIRS detector, every eigth column was replaced with the average of adjacent columns in one detector quadrant onto which parts of some shorter-wavelength orders of the spectrum were dispersed. 
This interpolation should not affect the extracted spectrum significantly, as the dispersed spectra are oriented $\sim3\arcdeg$ away from lying along columns and the spatial FWHM of the trace of the target is 4 to 5 pixels on the detector.
Telluric lines were removed using the ratio of a spectrum of the star HIP 116478 (obtained immediately after the target spectrum) to a blackbody with the star's temperature of \qty{5620}{K}.

The target spectra from the six cross-dispersed orders were scaled to match the next-longest-wavelength order using a weighted average in each overlap region.
These scaled spectra were shifted to a vacuum wavelength scale, giving us a spectral coverage between \A{2245} to \A{6874}, and combined using a weighted average and interpolation onto a continuation of the logarithmically-spaced SDSS vacuum wavelength scale with fiducial $\lambda_0 = \A{e4}$ and $\Delta\log\lambda = 10^{-4}$. The combined spectrum has a mean SNR of 7 per \kms{69} pixel.

\paragraph{Flux calibration adjustment}
The initial combined GNIRS spectrum was found to be extremely blue,
bluer than any SDSS quasar in the sample of \citet{2007ApJ...668..682D}.
For comparison, as measured by the UKIRT Hemisphere Survey (\S~\ref{sec:phot_ir}), J2318 had an observed color $(J-K)_{AB}=-0.126\pm 0.003$, corresponding to $(f_{\nu,J}/f_{\nu,K})_{UHS}=1.123\pm 0.003$.
However, using those same bandpasses \citep{2006MNRAS.367..454H}, the initial combined GNIRS spectrum yielded $(J-K)_{AB}=-0.400\pm 0.005$, corresponding to $(f_{\nu,J}/f_{\nu,K})_{GNIRS}=1.446 \pm 0.007$.
We are unsure of the origin of this discrepancy. 
Differential atmospheric refraction between 1\micron\ and 2.5\micron\ is unlikely to have been responsible; using the formulas of \citet{fil82}, we calculate that at the airmass range of observation (1.08$-$1.13) the differential atmospheric refraction would have been only $0\farcs12$, considerably smaller than the slit width of $0\farcs45$. 
To be conservative, we assume that the flux calibration of the initial combined GNIRS spectrum is in error and we recalibrate that spectrum to match the UHS observed $J-K$ color.\footnote{The UHS measurements were made 1.7-3 rest-frame years prior to the Gemini spectroscopy, so variability in $J-K$ is a possibility. 
$J$-band variability for this object is poorly constrained; it was found to be $-0.12\pm 0.22$ magnitudes in the $\sim 4$ rest-frame years prior to the UHS epoch (\S~\ref{sec:phot_ir}). 
However, the $\Delta(J-K)=-0.275\pm0.006$ required for the Gemini measurement to be a real color change since the UHS epoch would have happened at approximately the same time as an observed change of $\Delta(g-i)=+0.217\pm0.069$ between ZTF imaging and SDSS-V spectroscopy. 
A quasar getting unprecedentedly blue in the rest optical while getting redder in the rest UV is unlikely.}
We recalibrate the GNIRS spectrum before matching it to optical spectra by multiplying it by the power law $\lambda^X$ required to create a spectrum with that synthesized color. The predicted value of $X$ is found by solving $(\lambda_K/\lambda_J)^X(f_{\nu,J}/f_{\nu,K})_{GNIRS}=(f_{\nu,J}/f_{\nu,K})_{UHS}.$
The predicted value is $X=0.46$, which is close to the empirical value we adopt of $X=0.44$. The difference is likely due to the interplay between features in the quasar spectrum and the UHS filter response curves.

\subsubsection{Matching the SDSS spectra continuum shapes} \label{sec:specmatch}
The rest-frame UV spectrum changed in flux level and spectral shape between observations. 
This spectral variability is likely to be real; quasars are known to be variable and SDSS spectra are well calibrated.  
Using calibration star observations, SDSS spectra obtained with fibers plugged into plates have been shown to have a spectrophotometric accuracy of $\pm$5\% with only 0.56\% of observations having flux $\ge$50\% below that expected for the object's magnitude \citep{sdssrm0}. 
These anomalies are suspected to be due to dropping of the fiber during science exposures. 
In SDSS-V spectra taken with fiber positioner robots, that failure mode will be absent and - while new failure modes may be present - anecdotally, no dramatic increase in cases of spurious spectral variability has been noticed in SDSS-V.

As one option to investigate how absorption-trough measurements are affected by the flux and spectral shape changes, we undertook to `morph' the MJD 57328 and 59188 SDSS spectra to match the MJD 60251 SDSS spectrum.
For this process we used wavelength regions outside the obvious \ion{Si}{4} and \ion{C}{4} absorption troughs at 3755--\A{3930} and 4160--\A{4350}, respectively.
The MJD 57328 spectrum is only moderately well matched to the MJD 60251 spectrum by correcting the MJD 57328 spectrum for a change in power-law normalization and slope between the two epochs ($\chi^2_\nu = 1.145$, dof=4137).
A better match is found by multiplying the SDSS-IV spectrum by a parabolic function in $\log\lambda$ ($\chi^2_\nu = 1.06$, dof=4136).
The same trends as above are seen in matching the MJD 59188 spectrum to the MJD 60251 spectrum.
A change in power-law normalization and slope yields $\chi^2_\nu = 1.47$, dof=4129.
Multiplying by a parabolic function in $\log\lambda$ yields $\chi^2_\nu = 1.35$, dof=4128.
(We note that larger average $\chi^2$ values are seen at the shortest and longest wavelengths even when higher-order fits are tested, possibly due to  larger systematic spectrophotometric calibration uncertainties at those wavelengths \citep{sdssrm0,tpcorr} not being reflected in the uncertainties in the spectra.) 
These `morphed' versions of the MJD 57328 and MJD 59188 spectra multiplied by their best-fit parabolic functions were then normalized and used to study systematic uncertainties in broad absorption trough measurements; see \S \ref{sec:norm}, \S \ref{sec:direct} and \S \ref{sec:abs}.

\subsubsection{Combining optical and near-infrared spectra} \label{sec:combining}
The combined GNIRS spectrum was corrected for Galactic extinction using the same extinction curve used for the SDSS spectra.
The GNIRS spectrum was then scaled to match the MJD 60251 SDSS spectrum using a simple multiplicative scaling by the ratio of the weighted-average fluxes in the wavelength overlap region (\A{8239.483} to \A{10325.231}).
For purposes requiring the full wavelength coverage of our spectra, we use a combined spectrum consisting of the MJD 60251 SDSS spectrum (smoothed by a 5-pixel boxcar) at wavelengths $<\A{9700}$ and the scaled but unsmoothed GNIRS spectrum at longer wavelengths.

We also produce combined SDSS+GNIRS spectra for MJD 57328 and MJD 59188 by scaling the GNIRS spectrum to match the relevant SDSS spectrum using the same procedure as above.
We do this for both the Galactic-extinction-corrected SDSS spectra and the spectra which have had their continuum shapes matched to that of MJD 60251. 
For the latter the same scaling as for MJD 60251 is used for the GNIRS spectrum at $>\A{9700}$.
We caution that only the MJD 60251+60291 SDSS+GNIRS spectrum is quasi-simultaneous (see Figure \ref{fig:J2318_SDSS+Gemini}).

\subsection{Observations in other surveys and at other wavelengths} \label{sec:other}
Photometric data on J2318 (Table \ref{tab:phot}) were collected from various archives to study the variability of the quasar.

\subsubsection{Optical photometry} \label{sec:phot_opt}
\begin{figure}
	\centering
	\includegraphics[width=1\linewidth]{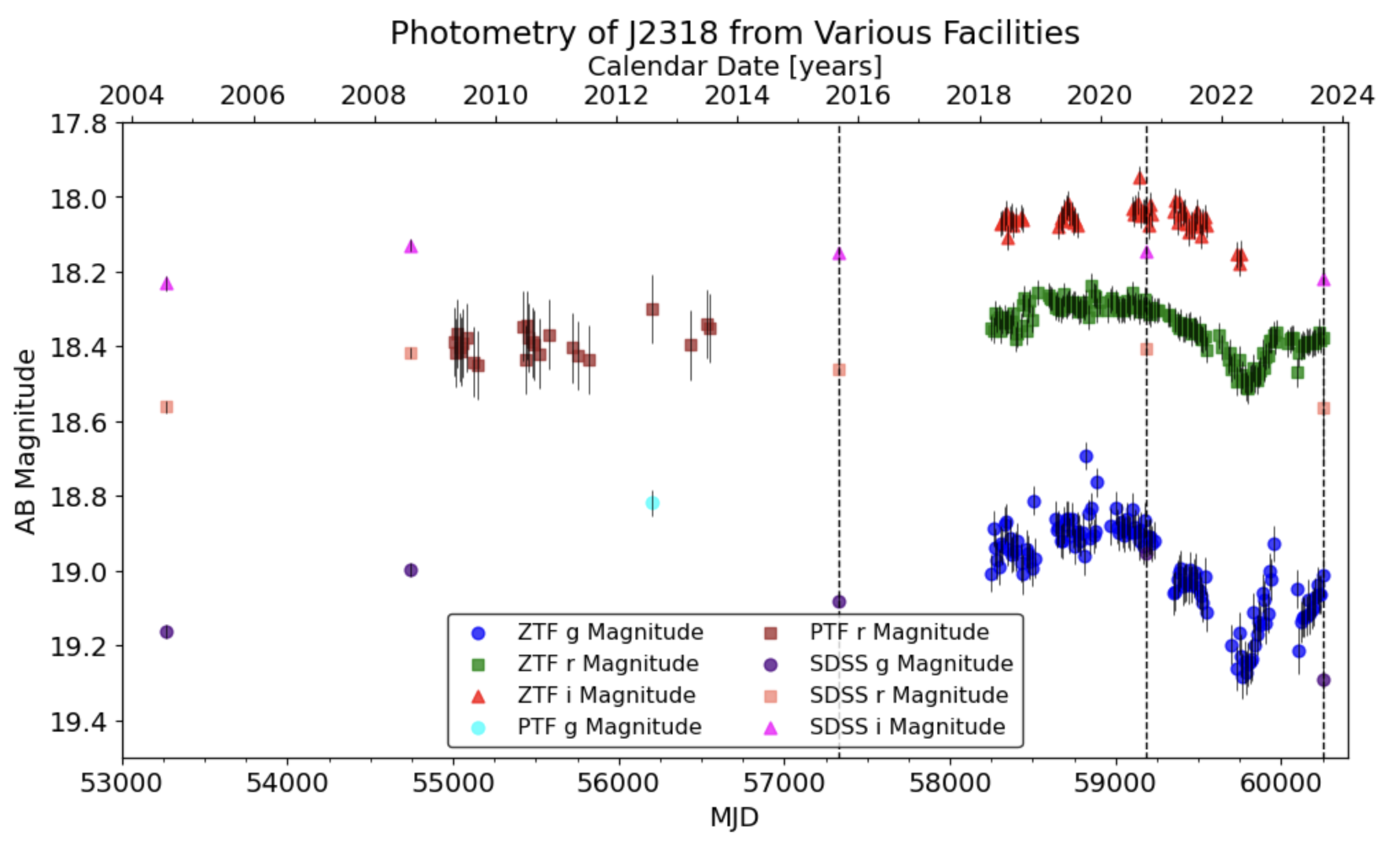}
	\caption{Photometric time series. 
    The earliest observations are the primary and secondary SDSS imaging epochs. Dates of spectroscopic observations obtained by SDSS-IV and SDSS-V are represented by the dashed vertical lines. The photometry at those epochs is synthesized from the spectroscopy.
    The PTF and ZTF observations are averaged over 11-day periods to reduce scatter. 
    In this figure, magnitudes have not been corrected for Galactic extinction.
    }
	\label{fig:J2318_Mag}
\end{figure}

Using data from several different surveys, observed-frame optical variability on timescales of months to decades can be observed in Figure \ref{fig:J2318_Mag}.
Although Pan-STARRS \citep{PS1database} data were used to select this quasar as a TDSS target, we found those data too sparse and noisy for inclusion in Figure \ref{fig:J2318_Mag}.
The earliest data in Figure \ref{fig:J2318_Mag} are from the two SDSS imaging epochs that exist for J2318.

From 2008-2014, photometric data of J2318 were obtained by the Palomar Transit Factory (PTF), a multiepochal robotic survey that acquired photometric data of transient and variable astronomical phenomena in the northern sky \citep{PTF,PTF_Lc}. 
The PTF photometric data were taken in the $g$ and Mould $R$ bands. 
A conversion from Mould $R$ to $r$ magnitude was done using equation 4 in \cite{PTFphotcal} using their average $\alpha_{c,R}$ and an average value for $r-i$ for J2318 calculated using the photometric data from the Zwicky Transient Facility (ZTF).

ZTF is a wide-field optical time-domain survey that uses the Palomar 48 inch Schmidt telescope \citep{ZTF}. Photometry in the $g$, $r$ and $i$ bands was acquired from ZTF DR20 \citep{ZTF_Lc}.

J2318 does not have time-series photometry in Gaia DR3 \citep{gaiaDR3}.

J2318 is observed to have photometric variability of a few tenths of a magnitude, as is typical for quasars over a time-span of e.g., over 15 years.
J2318 increased in brightness from the SDSS imaging epochs into the PTF and ZTF epochs, followed by a dimming and reddening from approximately MJD 59000 to 59750, followed by a slight recovery in brightness up to the second SDSS-V spectroscopic epoch.  
There appears to be a small offset causing the SDSS synthesized spectroscopic magnitudes to be $\sim$$0.1 \pm 0.1$ mag fainter than the ZTF imaging magnitudes; this could represent a difference in photometric systems.

J2318 was detected in Palomar Observatory Sky Survey photographic plates from POSS-I (in 1953) and POSS-II (in 1991, 1992, and 1995), as digitized and compiled into the USNO B-1 photometric catalog \citep{usnob1}.
We calculated its expected photographic magnitudes in the absence of variability from its 2004 SDSS magnitudes using the conversions of \cite{sesar06}.
The rms scatter between observed and predicted photographic magnitudes is 0.42~mag, about a factor of two larger than the scatter found by \cite{sesar06}.
This scatter could arise from intrinsic variability in J2318 of a few tenths of a magnitude, or from one anomalous measurement: the scatter is dominated by a POSS-I red magnitude fainter than expected by 0.86 mag, despite the POSS-I blue magnitude being 0.07 mag brighter than expected. 
In any event, optical photometry of J2318 on several-decade timescales finds variability $<1$~mag.

\begin{table*}
	\centering
	\setlength{\tabcolsep}{2pt}
	\begin{tabular}{ccccccccccc}
		\hline
		MJD\tsub{Obs}
		 & UT Date\tsub{Obs}      & Type
		 & $g-i$               & $u$                 & $g$                 & $r$                 & $i$                 & $z$
		 & W1                  & W2                                                                                                          \\
		\hline
		53265.22
		 & 2004--09--17        & \td{SDSS Imaging\n(primary)}
		 & \tdd{ 0.793}{0.028} & \tdd{19.717}{0.049} & \tdd{18.878}{0.019} & \tdd{18.365}{0.016} & \tdd{18.085}{0.021} & \tdd{17.956}{0.027} 
		 & n/a                 & n/a                                                                                                         \\
		54740.38
		 & 2008--10--01        & \td{SDSS Imaging\n(secondary)}
  	 & \tdd{ 0.730}{0.024} & \tdd{19.639}{0.057} & \tdd{18.713}{0.018} & \tdd{18.220}{0.016} & \tdd{17.983}{0.016} & \tdd{17.791}{0.024} 
		 & n/a                 & n/a                                                                                                         \\
		\td{55203 to\n55414}
		 & \td{2010--01--07 to\n2010--08--06} & \td{WISE Imaging\n(forced photometry)}
		 & n/a                 & n/a                 & n/a                 & n/a                 & n/a                 & n/a
		 & \tdd{18.172}{0.025} & \tdd{17.953}{0.041}                                                                                         \\
		57328.08
		 & 2015--11--02        & SDSS-IV Spectrum
		 & \tdd{ 0.794}{0.016} & n/a                 & \tdd{18.798}{0.014} & \tdd{18.264}{0.010} & \tdd{18.004}{0.008} & n/a                 
		 & n/a                 & n/a                                                                                                         \\
		\td{58301 to\n58311}
		 & \td{2018--07--02 to\n2018--07--12} & ZTF Imaging
		 & \tdd{ 0.719}{0.057} & n/a                 & \tdd{18.644}{0.047} & \tdd{18.162}{0.033} & \tdd{17.925}{0.033} & n/a                 
		 & n/a                 & n/a                                                                                                        \\
         59188.09
		 & 2020--12--05        & SDSS-V Spectrum \#1
		 & \tdd{ 0.670}{0.040} & n/a                 & \tdd{18.670}{0.037} & \tdd{18.210}{0.020} & \tdd{18.000}{0.015} & n/a                 
		 & n/a                 & n/a                                                                                                         \\
		\td{59750 to\n59760}
		 & \td{2022--06--20 to\n2022--06--30} & ZTF Imaging
		 & \tdd{ 0.938}{0.067} & n/a                 & \tdd{18.944}{0.058} & \tdd{18.295}{0.038} & \tdd{18.006}{0.035} & n/a
		 & n/a                 & n/a                                                                                                         \\
		60251.11
		 & 2023--11--03        & SDSS-V Spectrum \#2
		 & \tdd{ 0.936}{0.039} & n/a                 & \tdd{19.009}{0.033} & \tdd{18.367}{0.141} & \tdd{18.073}{0.020} & n/a                 
		 & n/a                 & n/a                                                                                                         \\
		\hline
	\end{tabular}
	\caption{Photometry of J2318.
		All magnitudes are on the AB magnitude scale and have been corrected for Galactic extinction.
		SDSS imaging magnitudes are PSFMAG values.
		For SDSS spectra, magnitudes are calculated from the SPECTROFLUX values (total fluxes for point sources).
		The $g-i$ color can be compared directly across all epochs.
        Only two of the numerous ZTF Imaging epochs are listed, as examples.
        The WISE magnitudes are from unWISE forced photometry; see \S \ref{sec:other}.}
	\label{tab:phot}
\end{table*}

\subsubsection{Infrared photometry} \label{sec:phot_ir}
J2318 does not have a photometric detection in 2MASS \citep{2miss}, but is nonetheless faintly visible in 2MASS Atlas images \citep{2MASS_PSC} taken on MJD 51087 (1998 October 1).
We downloaded these images and performed aperture photometry on J2318 and on an unresolved object 8\farcs5 to the east-southeast of J2318 (SDSS J231854.87+243950.4, likely a star).
We used the star to set the zeropoint of this photometry, which used the same 4$''$ radius aperture as the 2MASS magnitudes, and converted to AB magnitudes. 
For the J2318 magnitude uncertainties, in each band we adopted the star's magnitude uncertainty times the flux ratio of the star to J2318. 
Including the correction for Galactic extinction, the resulting 2MASS photometry for J2318 is $J_{AB} = 18.00 \pm 0.22$, $H_{AB} = 17.87 \pm 0.26$, and $K_{s,AB} = 18.07\pm 0.42$.

J2318 was observed twice in both $J$ and $K$ bands in the UKIRT Hemisphere Survey \citep[UHS;][]{2018MNRAS.473.5113D,UHSDR2}. 
The measurements were taken within 13 rest-frame days of each other and are consistent within the uncertainties, so we combined them. After converting to AB magnitude following \cite{2006MNRAS.367..454H} and accounting for Galactic extinction, we find $J_{AB}=17.884\pm 0.014$ (MJD 56211+56260) and $K_{AB}=18.055\pm 0.029$ (MJD 57992+58013). 
The $J$ photometry from 2MASS and UHS differs by only $0.12\pm 0.22$ magnitudes over $\sim 4$ rest-frame years.

Both J2318 and the nearby star were detected by the Wide-field Infrared Survey Explorer (WISE) space telescope \citep{wise,WISE_SC}.
The WISE photometry of J2318 in Table \ref{tab:phot} in the W1 and W2 bands (3.4 and 4.6 microns) is unWISE \citep{unWISE_TDC,unWISE_c} forced photometry (DR13 version) from \cite{unWISE.SDSS}.

\begin{figure}
	\centering
	\includegraphics[width=1\linewidth]{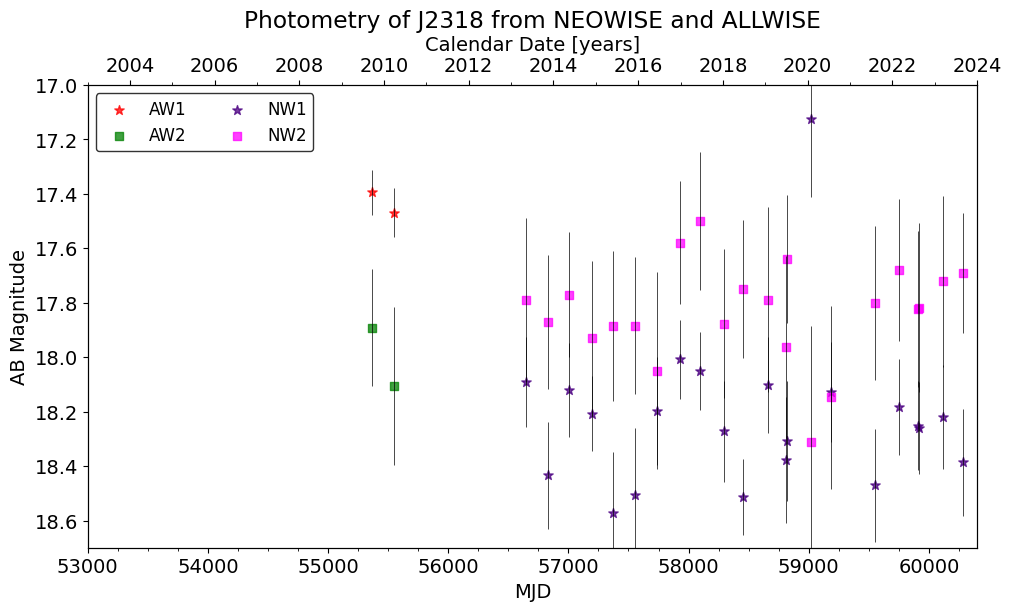}
	\caption{NEOWISE and ALLWISE mid-infrared photometric data on J2318. W1-band magnitudes are shown as stars, and W2-band magnitudes are shown as squares. All values were converted from Vega magnitude to AB magnitude.}
	\label{fig:J2318_NEO_ALL_WISE_final}
\end{figure}

To search for time variability in these bands, data from ALLWISE\footnote{https://wise2.ipac.caltech.edu/docs/release/allwise/expsup/index.html} \citep{ALLWISE_MP} and NEOWISE \citep{NEOWISE,NEOWISEreact,NEOWISE_SEST} are presented in Figure \ref{fig:J2318_NEO_ALL_WISE_final},
cleaned of occasional epochs of photometry of the nearby star being reported as photometry of J2318. 
The apparent dimming in the W2 band between ALLWISE and NEOWISE is not statistically significant, given the reported uncertainties (which are consistent with the rms scatter in that band's observations). 
There is a possible marginal decrease in flux in the W1 band between the ALLWISE and NEOWISE epochs is significant at the 2.6$\sigma$ level accounting for the scatter in the NEOWISE photometry. 

ALLWISE photometry in the W3 and W4 bands was reported for MJDs 55368 to 55370. 
The resulting average AB magnitudes were $W3=16.22\pm0.37$ and $W4=14.33\pm0.46$.
These values are in agreement within the uncertainties with the unWISE values based on the same data of $W3=16.37\pm0.13$ and $W4=16.5\pm1.2$.

\subsubsection{Observations at other wavelengths}
At radio wavelengths, J2318 was not in the sky area covered by FIRST \citep{1997ApJ...475..479W} and was not detected by the NVSS \citep{con98}.
J2318 was observed by the Very Large Array Sky Survey \citep[VLASS;][]{VLASS} but was not detected in data release 1.2 or 3.2.\footnote{\url{https://science.nrao.edu/science/surveys/vlass}}

Assuming a single-pass $5\sigma$ upper limit of 0.6 mJy \citep{VLASS}, that corresponds to $\nu L_\nu \leq 6 \times 10^{41}$ erg s$^{-1}$ at 3 GHz and a radio-to-$i$-band logarithmic flux ratio $R_i\leq 0.5$ \citep{sdss1st}, below the traditional boundary for radio-quiet quasars of $R_i=1$. 

At $X$-ray wavelengths, from the Chandra Source Catalog version 2.0 \citep{CSC} we find that J2318 was not detected in shallow observations with the ACIS instrument on Chandra, with a corresponding luminosity limit of $L(0.5-\qty{7}{\keV}) < \qty{1.73e45}{\erg\per\s}$ to rule out a marginal detection (a confidence limit of 97.7\%, yielding on average one false detection per pointing).
This limit makes J2318 either X-ray normal or X-ray weak; see \S~\ref{sec:sed}.
There are no previous X-ray observations of J2318 from XMM-Newton \citep{XMM}.
J2318 is not located in the sky coverage of eROSITA DR1 \citep{eROSITADR1}.

Low-resolution near- to mid-infrared spectrophotometry of J2318 was obtained throughout MJD 60839 - 60875 (midpoint MJD 60857), by the Spectro-Photometer for the History of the Universe, Epoch of Reionization, and Ices Explorer (SPHEREx) mission \citep{2025arXiv251102985B,SPHEREx_QR2}, covering observed-frame wavelengths of approximately $0.75-5.0$ \micron~ (rest-frame $\sim2000-14,000$ \A{} at $z = 2.678$). 
Although the spectral resolution is modest ($R\sim41-130$) with moderate $\sim6''$ angular resolution, prominent rest-frame optical emission features such as H$\alpha$ and \ion{Mg}{2} are identifiable, and the overall continuum shape provides additional constraints on the spectral energy distribution beyond SDSS-V and WISE photometry.
Given the limited spectral resolution and S/N at the longest wavelengths, we primarily view the SPHEREx spectrum as a qualitative extension of the SED rather than a tool for detailed line measurements. We therefore defer discussion of it to \S~\ref{sec:sed}.

No data at other wavelengths are listed for J2318 in NED, the NASA Extragalactic Database (e.g., J2318 was not in the sky-area that was observed by GALEX).

\section{Analysis} \label{sec:analyze}
\subsection{Redshift determination} \label{sec:redshift}
Due to the weakness of the [\ion{O}{3}] emission and the noise in the region, this line was not used to determine the redshift of the quasar and instead the H$\alpha$ emission line was used.
The Gemini spectrum was fit at $>2.05~\rm{\mu m}$ with a power-law continuum, a single Gaussian for \ion{Na}{1}$\,\lambda\lambda$5891,5897, and several options for H$\alpha$: either two Gaussians, three Gaussians, or two Gaussians plus relatively narrow emission in H$\alpha$ and either the [\ion{N}{2}] $\lambda\lambda 6549,6585$ doublet or the [\ion{S}{2}] $\lambda\lambda 6718,6732$ doublet.
To model the narrow emission we use a set of three Gaussians constrained to have the wavelength ratios of H$\alpha$ and the chosen doublet, identical linewidths, and a line ratio of 3.05 for the [\ion{N}{2}] doublet \citep{2023AdSpR..71.1219D} or 0.44 for the [\ion{S}{2}] doublet (the high-density limit; \citealt{2018ApJ...856...46R}).

The best-fit amplitudes for both [\ion{N}{2}] and [\ion{S}{2}] are vanishingly small, so we exclude models including them from further consideration.
A significant improvement in $\chi^2$ is achieved when fitting three Gaussians rather than two to the H$\alpha$ emission. 
For our large number of degrees of freedom $\nu=887$, the $F$-test \citep{br92} requires that a model with $\Delta p=3$ additional parameters resulting in an improvement of $\Delta\chi^2=\chi^2_{old}-\chi^2_{new}$ must have $F_\chi=(\Delta\chi^2/\Delta p) / \chi^2_{\nu,new} >5.5$ to have a $<0.1\%$ chance of showing that improvement by chance.
As compared to a two-Gaussian model, 
$F_\chi=16$ for the three-Gaussian model ($\chi^2_{new}/DOF=646/887$; i.e., $\chi^2_{\nu,new}=0.729$).
Therefore, we adopt the three-Gaussian model.

\begin{figure}
	\includegraphics[width=1.00\textwidth]{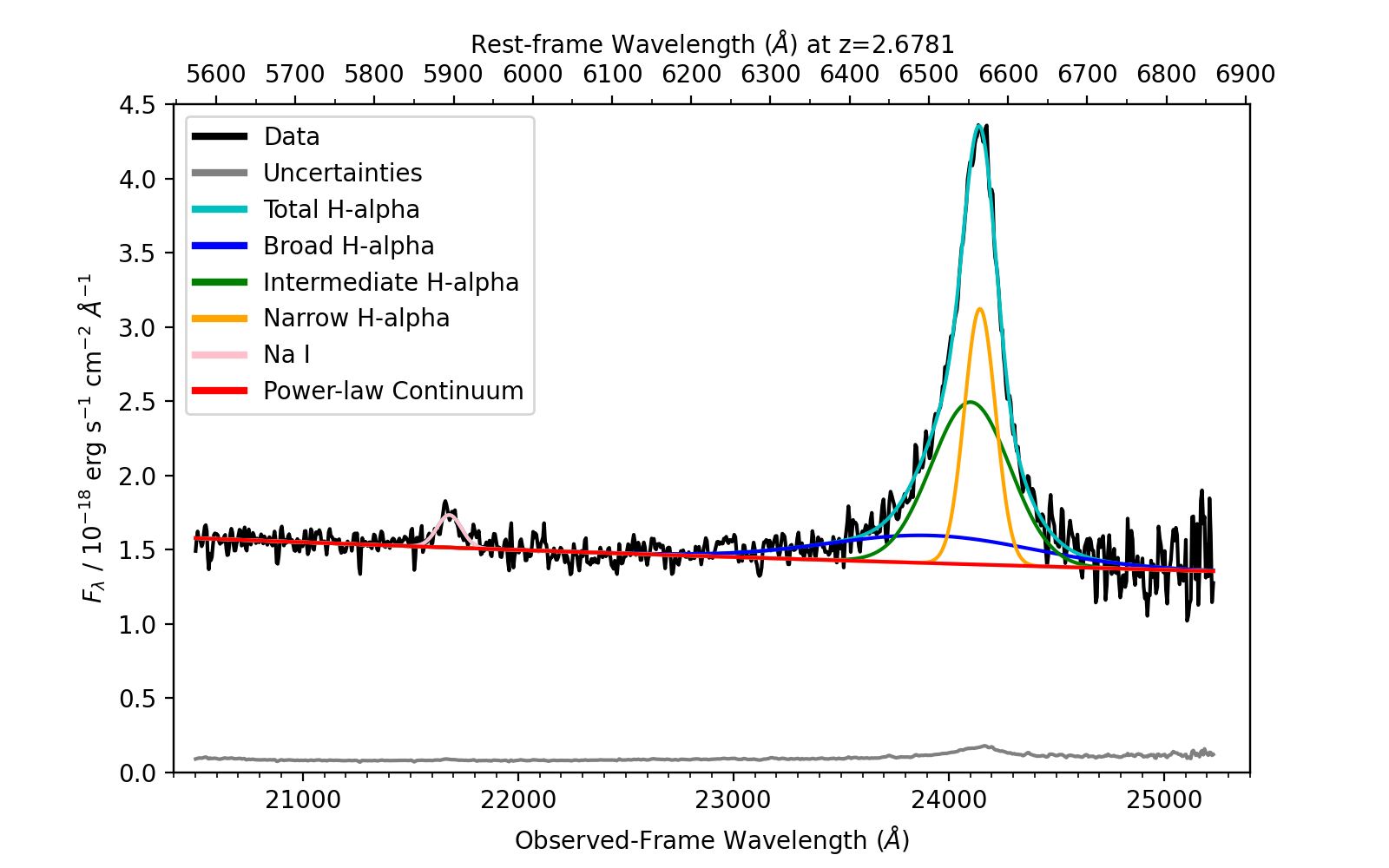}
	\centering
	\caption{H$\alpha$ region. Fit to the H$\alpha$ region of J2318 including the best-fit three-Gaussian model for the H$\alpha$ line. The data are from the GNIRS spectrograph on the Gemini North Telescope.}
	\label{fig:J2318_Halpha}
\end{figure}

The fit to the H$\alpha$ region is shown in Figure \ref{fig:J2318_Halpha} and the results are listed in Table \ref{halphafit}.
The overall fit to the H$\alpha$ line has rest-frame equivalent width of \A{201\pm16} (consistent with the average value of $194.5 \pm 0.6$ \AA\ found for low-redshift SDSS quasars by \citealt{sdss73}) and FWHM = \kms{3000\pm140}.

We take the redshift of the narrowest H$\alpha$ component as the redshift of the quasar: $z = 2.6781 \pm 0.0004$. 
As a check of the H$\alpha$ redshift, we perform Gaussian fitting of the narrowest component of the H$\beta$ ($\lambda=4862.68$\AA) and \ion{Mg}{2} ($\lambda=2798.75$\AA) emission features in the GNIRS spectrum with PyQSOfit \citep{PyQSOFit,2019ApJS..241...34S}, which yield redshift measurements of $z=2.5990\pm0.0017$ and $z=2.6647\pm0.0010$, respectively. 
We then visually inspected the SDSS rest-frame wavelengths using these redshifts and found that they cause the emission features of Ly$\alpha$, \ion{N}{5}, \ion{Si}{4}, \ion{C}{4}, and \ion{Mg}{2} to appear blueshifted with respect to their vacuum wavelengths.
The \ion{C}{4} and \ion{Si}{4} rest-frame emission features are slightly blueshifted with respect to their vacuum wavelengths when using $z=2.6781$, but that is to be expected as these ions are being absorbed at extreme velocities and \ion{C}{4} emission is found to be blueshifted in BALQSOs \citep{2020MNRAS.492.4553R}. 
Furthermore, the Ly$\alpha$, \ion{Mg}{2}, and \ion{N}{5} rest-frame emission features lie at their respective vacuum wavelengths when using $z=2.6781$.

\begin{table*}
	\centering
	\begin{tabular}{cccccc}
		\hline
		Component            & Redshift            & $\Delta v$  & FWHM   & $A$  & EW  \\
                    &            &  (\kms{}) & (\kms{})  &  (\uFluxLam{}) &  (\A{}) \\
		\hline
		Very broad H$\alpha$ & $2.6435\pm0.0093$ & $-2850\pm760$       & $14010\pm1980$ & $0.186\pm0.043$   & $43\pm12$  \\
		Broad      H$\alpha$ & $2.6715\pm0.0012$ & $ -570\pm100$       & $ 5360\pm 340$ & $1.092\pm0.096$   & $97\pm10$  \\
		Narrow     H$\alpha$ & $2.6781\pm0.0004$ & $    0\pm 30$       & $ 2120\pm 110$ & $0.172\pm0.106$   & $61\pm 5$  \\
		\ion{Na}{1}          & $2.6781\pm0.0012$ & $    0\pm 90$       & $ 1840\pm 230$ & $0.219\pm0.024$   & $ 6\pm 1$  \\
		\hline
	\end{tabular}
	\caption{Emission-line fitting parameters of the J2318 H$\alpha$ region.
	The emission lines are in addition to a power-law continuum $f(\lambda)=A{(\lambda/6500)}^\alpha$ with $A=\uFluxLam{1.410\pm0.012}$, $\lambda$ in rest-frame \A{}, and $\alpha=-0.73\pm 0.04$.
	The values of $\Delta v$ are relative to the adopted rest frame of $z=2.6781$, with negative velocities indicating a blueshift. The EW values are in rest-frame \AA.}
	\label{halphafit}
\end{table*}

\subsection{Intervening absorption} \label{sec:intabs}
We visually inspected the spectrum of J2318 for narrow absorption lines. 
The spectrum exhibits two strong narrow absorption lines at wavelengths \A{4403.9} and \A{4430.9}, visible in Figures \ref{fig:J2318_rawSDSS} and \ref{fig:J2318_SDSS+Gemini} between the broad \ion{C}{4} absorption trough and the Ly$\alpha$ emission wavelength. 
We identify these lines as Ly$\alpha$ absorption at $z = 2.6219\pm0.0004$ and $z = 2.6448\pm0.0002$.
The $z = 2.6219$ system also features weak, unsaturated \ion{C}{4} doublet absorption (rest-frame ${\rm EW}=0.55\pm 0.05$\,\AA).
Intervening \ion{C}{4} doublet absorption is also seen at $z=1.959\pm 0.001$ and possibly at $z=2.263\pm 0.001$.

\subsection{Weak-lined quasar} \label{sec:wlq}
The spectrum of J2318 identifies it as a candidate weak-lined quasar (WLQ), and its underlying continuum will be discussed in \S\ref{sec:contfit} and Figures \ref{fig:J2318_DV2_SpecFit}, \ref{fig:PQF_Ha}, and \ref{fig:J2318_DV2_RedZoom_Fit}.
In \cite{2015ApJ...805..122L}, a WLQ is defined as a quasar whose \ion{C}{4} emission line has rest-frame equivalent width of $<\A{10}$.
A weak \ion{C}{4} emission line is apparent in the MJD 57328 spectrum, with its peak blueshifted by about \kms{-2700}.
We measure \ion{C}{4} emission line EWs in our SDSS spectra by establishing a linear continuum between windows at $1465$-\A{1475} \A{} and $1574$-\A{1584}, integrating the flux between \A{1475} and \A{1574}, and accounting for the flux lost to the narrow \ion{C}{4} absorption in that wavelength range.  
The resulting \ion{C}{4} EW was \A{5.0\pm 0.4} on MJD 57328, \A{5.3\pm 0.6} on MJD 59188, and \A{4.2\pm 1.8} on MJD 60251, confirming that J2318 is a WLQ.
The \ion{C}{4} emission-line blueshift is \kms{3760\pm 150} using the definition of \cite{2020MNRAS.492.4553R}; i.e., the blueshift of the wavelength at which half the total cumulative line flux is reached.
(For comparison, the H$\alpha$ blueshift is \kms{420\pm 50}.)
This \ion{C}{4} blueshift is larger than average for WLQs but within the range of \kms{0}$-$\kms{8000} seen in them; see Figure 8 of \citet{2015ApJ...805..123P} and Figure 3 of \citet{2023ApJ...950...97H}.

As for rest-frame optical lines, we measure the H$\beta$ EW for J2318 by establishing a linear continuum between windows at $4750$-\A{4790} and $4960$-\A{5000} and integrating the flux between \A{4790} and \A{4960}.
The H$\beta$ EW is \A{32.4\pm 8.7}, with the uncertainty dominated by the uncertain continuum level. 
In their studies of WLQs,
\citet{2015ApJ...805..123P} found H$\beta$ EWs in the range $15$-\A{40} and \citet{2024ApJ...972..191C} reported an average H$\beta$ EW of \A{52\pm 2}, as compared to \A{64\pm 8} for non-WLQ SDSS quasars (their Figure 3).
J2318 fits the pattern that H$\beta$ EWs in WLQs are biased toward the low end of the distribution seen in normal quasars. 

At the rest wavelengths of [\ion{O}{3}], the GNIRS spectrum is very noisy due to poor atmospheric transmission. 
Integrating over the [\ion{O}{3}] wavelength regions used by \cite{sdss73}, we do not formally detect [\ion{O}{3}] emission above the noise. 
We find EWs of (\A{1.4\pm2.6}, \A{2.3\pm3.4}) for $\lambda 4960,5008$. 
This is consistent with the finding that WLQs have on average weaker [\ion{O}{3}] than normal quasars of similar redshifts and luminosities (Figure 3 of \citealt{2024ApJ...972..191C}).

The H$\alpha$ EW of J2318 was measured to be \A{201\pm 16} (\S~\ref{sec:redshift}). This is consistent with the average found for low-redshift SDSS quasars by \citet{sdss73} and falls within the range of H$\alpha$ EWs of $150$-\A{250} found for WLQs in the study of \citet{2015ApJ...805..123P}. Thus, the H$\alpha$ EW of J2318 is unremarkable.

\subsection{Continuum Fitting} \label{sec:contfit}
The fitting and normalization of quasar continua is not an exact science and can be performed in a multitude of ways \citep[e.g.,][]{trump06,allenbal,2014ApJ...788..123L,2019ApJ...872...21H,WuShen22}.
This process introduces some systematic uncertainty associated with each fitting routine and subsequent normalization
\citep[e.g.,][]{2023ApJ...952...44B}.
Having multiple continuum models allows approximation of those uncertainties in the fitting process, normalization, and spectral analysis (e.g., black-hole mass and broad absorption measurements).
We emphasize that the continuum models explored here are not intended to represent unique or fully physical descriptions of the intrinsic quasar spectrum or dust geometry. 
Instead, they are used as empirical tools to establish a stable local continuum for accurate normalization of absorption features. 
While some best-fit parameters may not be physically meaningful in isolation, the derived absorption measurements and outflow properties are robust to the choice of continuum model.

For the proper measurement and comparison of the changing absorption features, we model the underlying continuum of the quasar for each SDSS spectrum combined with the GNIRS spectrum (\S~\ref{sec:combining}).
We converted the observed wavelengths into the rest frame of $z=2.6781$ (\S~\ref{sec:redshift}).
We adopt the rest-frame wavelength relatively line free (RLF) regions outlined by \cite{gibsonbal}, and adjust them accordingly to avoid any strong emission or absorption features.
The RLF regions we used were $1070$-\A{1100}, $1260$-\A{1300}, $1425$-\A{1475}, $2227$-\A{2252}$^\dagger$, $3010$-\A{3060}$^\dagger$, $3300$-\A{3425}$^\dagger$, $3820$-\A{4020}, $4100$-\A{4125}, $4660$-\A{4760}, $5553$-\A{5628}, $6018$-\A{6083}, and $6705$-\A{6805}, where the regions marked with a dagger are \ion{Fe}{2} RLFs that lie within \ion{Fe}{2} emission complexes.
We found that fitting the continuum using the \ion{Fe}{2} RLFs produced poorer power-law continuum fits for most of the models that we tested.

To account statistically for Ly$\alpha$ forest absorption in the fitting, the fluxes for all pixels in the RLF region shortward of \A{1216} rest-frame were multiplied by $e^{\tau_{\rm eff}}$, where $\tau_{\rm eff}$ is the effective optical depth of the Ly$\alpha$ forest at the redshift $z=2.39$ of our Ly$\alpha$ forest RLF region ($e^{\tau_{\rm eff}} = 1/\langle F\rangle$ in Eq.~2 of \citealt{2021MNRAS.501.5811K}, in which $\langle F\rangle$ is the mean transmitted flux in the Ly$\alpha$ forest).
Our adopted correction factor of 1.112 arises from adopting a value of $\langle F\rangle$ which is $4\sigma_{\langle F\rangle}$ above the average value of the three fits presented in \cite{2021MNRAS.501.5811K} as the red, orange, and black curves in their Figure 10; our correction factor for the error values was increased by 1.9\% to 1.134 to reflect the scatter among these fits.
Our adopted value of $\langle F\rangle$ yields a smaller correction factor than the mean for this redshift. Simply put, visual inspection shows that the mean correction factor is not appropriate for the RLF region in this quasar. \cite{2021MNRAS.501.5811K} note in their \S~4.3 that due to the considerable cosmic variance in the Ly$\alpha$ forest, the quantity $\sigma_{\langle F\rangle}$ is not a Gaussian sigma. We estimate that 5/37 sightlines at $z>1.85$ in their sample deviate from the mean $\langle F\rangle$ by $\geq 4\sigma_{\langle F\rangle}$. 

\begin{figure}\centering\includegraphics[width=18cm]{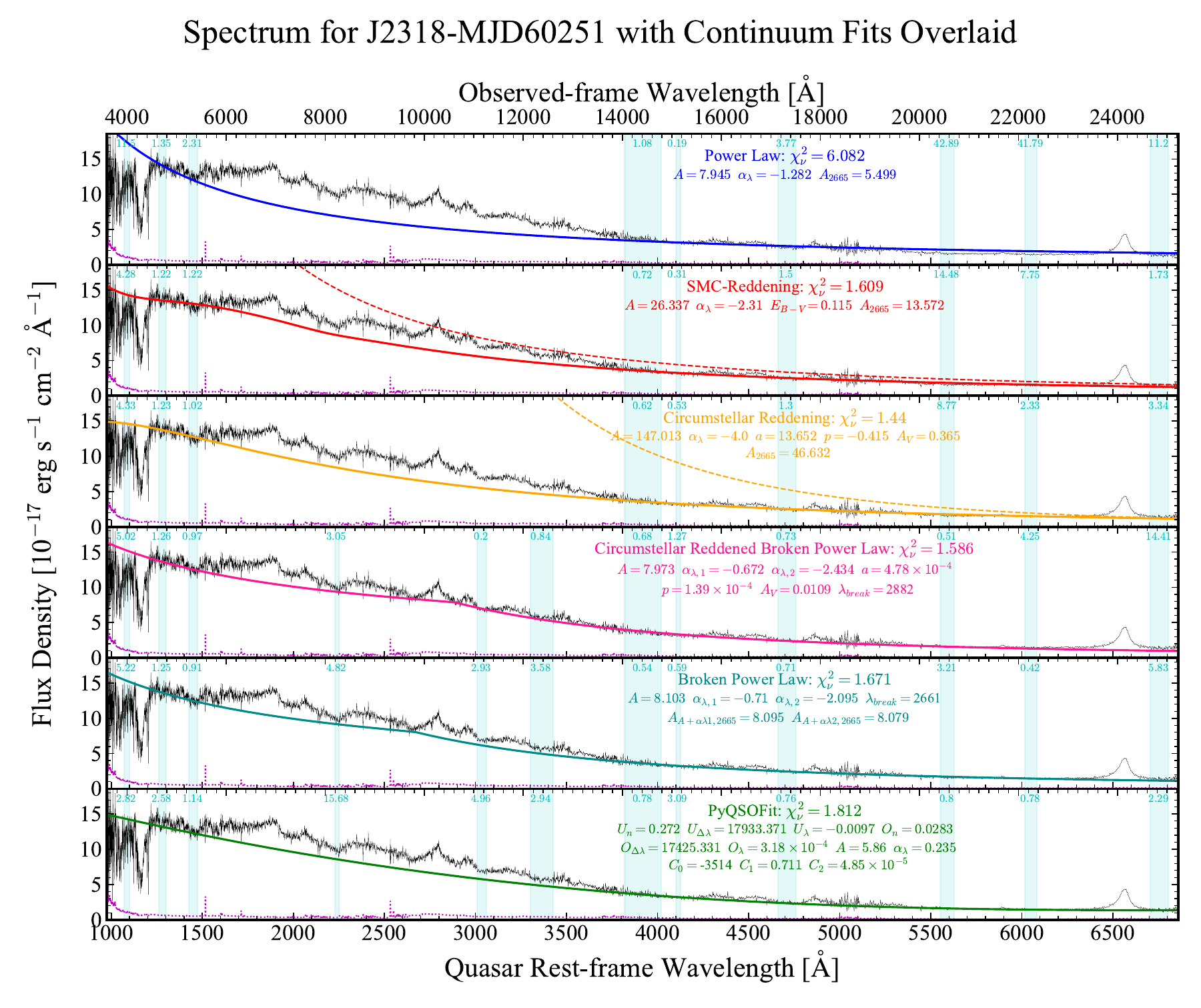}\caption{MJD 60251+60291 Continuum Fitting. The raw continuum (black) has been fit to various models (colored). 
The flux errors are plotted in magenta, the RLF regions used for the fitting are shown in vertical cyan bands (with their $\chi_\nu^2$ values listed), and the $\chi_\nu^2$ and best-fit parameter values for each model are displayed in their respective colors. 
Dashed curves represent the power law before reddening, and comparison to the CSb $A$ parameter in other models should be done using $A_{2665}=A(2665/2000)^{\alpha_\lambda}$.
        Top panel: a simple power law.
		Second panel: a power law reddened with the SMC extinction curve of \cite{pei92}.
		Third panel: a power law reddened with the three-parameter power-law extinction curve of \cite{2008ApJ...686L.103G}.
		Fourth panel: a broken power law reddened with the three-parameter power-law extinction curve of \cite{2008ApJ...686L.103G}.
        Fifth panel: a broken power law with no reddening.
		Sixth panel: the continuum estimate generated by PyQSOFit.
        Due to the lower $\chi_\nu^2$ values, the SMC and CS reddening models are used for analysis.
	}\label{fig:J2318_DV2_SpecFit}\end{figure}

\begin{figure}\centering\includegraphics[width=18cm]{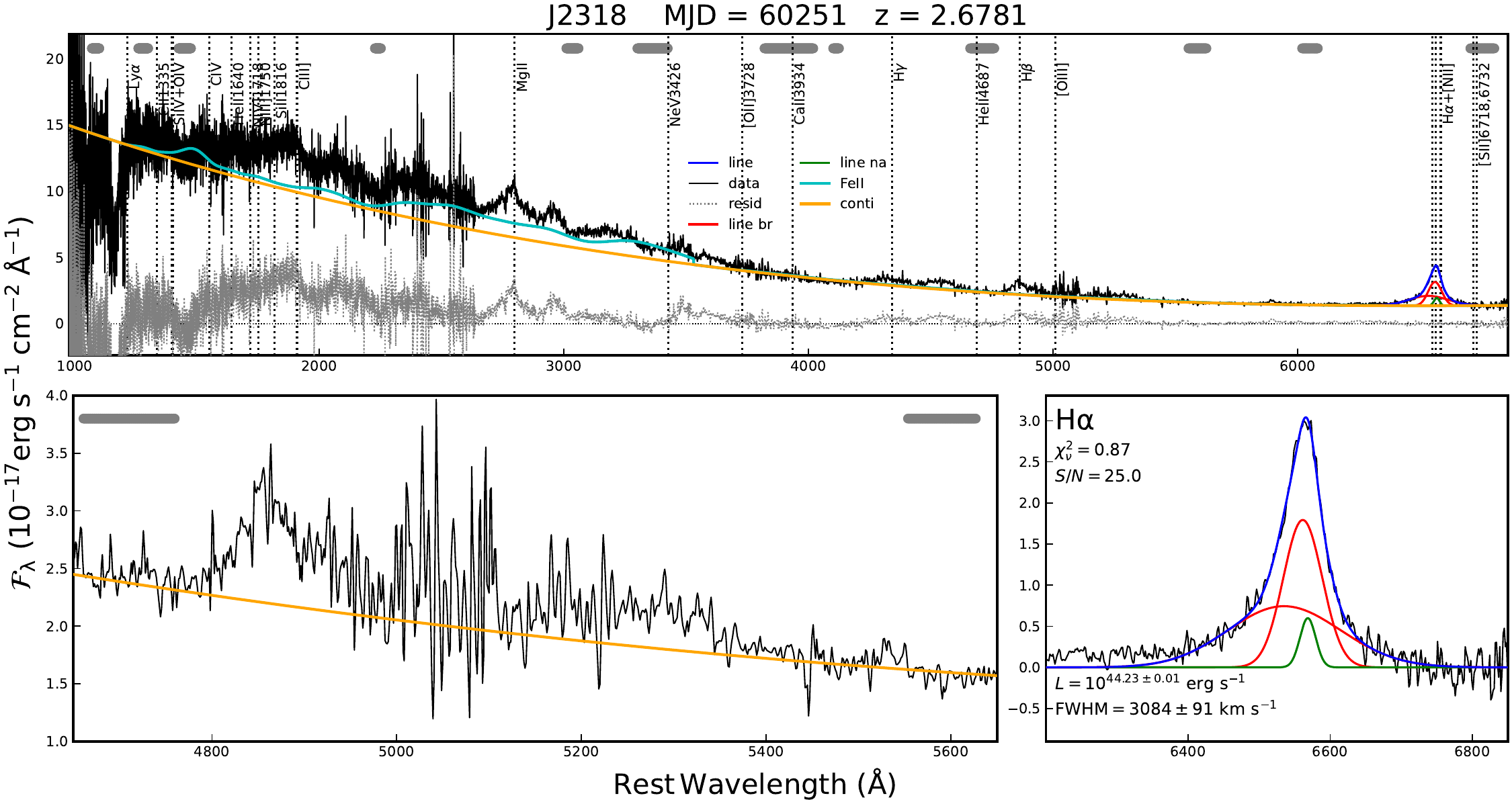}\caption{MJD 60251 PyQSOFit fitting. The unsmoothed spectrum (black) has been fit using PyQSOFit (top panel) to determine the flux at \A{5100} (left panel) and the FWHM of H$\alpha$ (blue, right panel). 
RLFs are indicated by gray bars along the top, the \ion{Fe}{2} template fitting is in cyan, the continuum fit is in orange, and the fitting residuals are plotted in light gray.
}\label{fig:PQF_Ha}\end{figure}

\begin{figure}\centering\includegraphics[width=18cm]{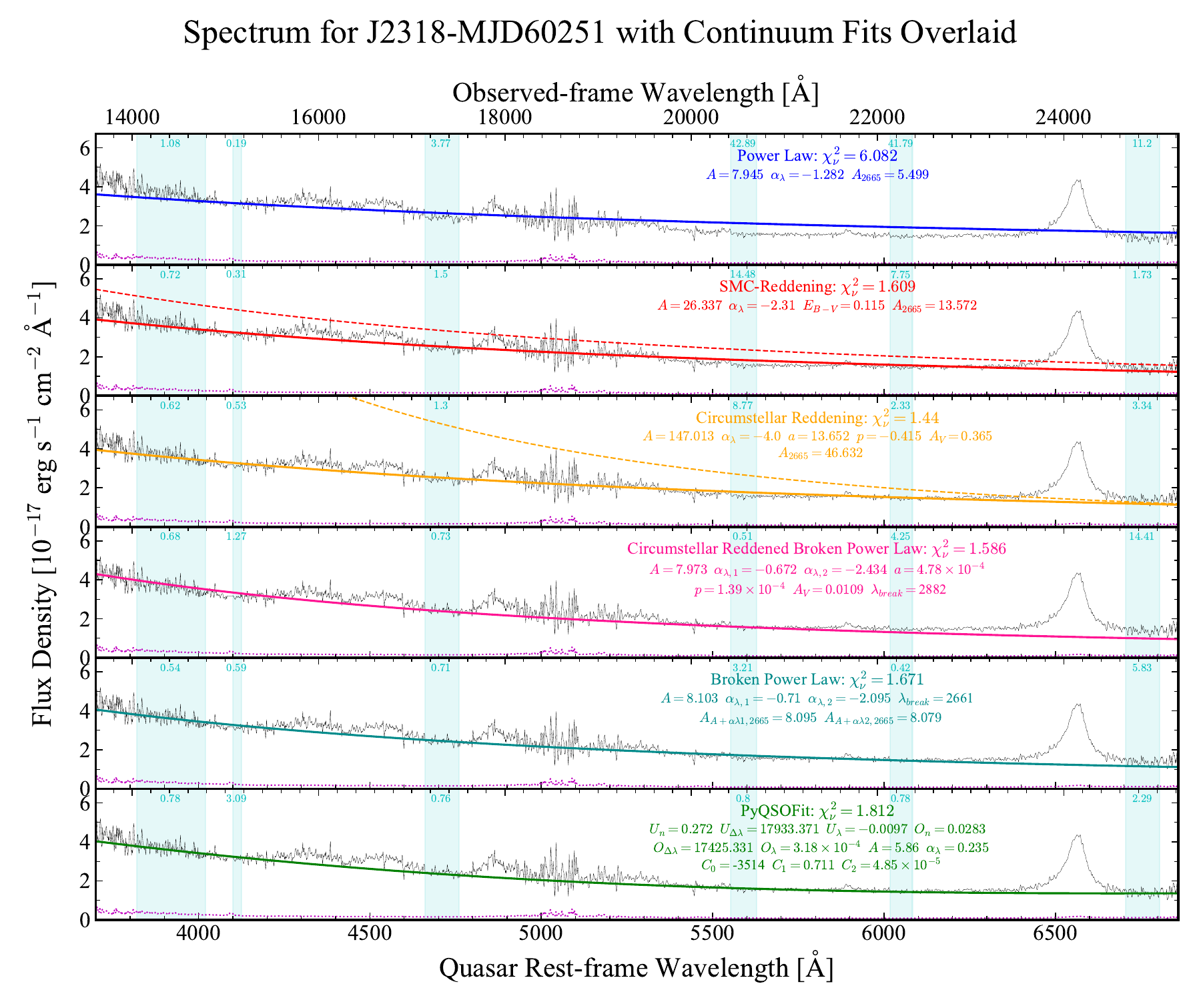}\caption{J2318 MJD 60251+60291 where $\lambda>3700$ \AA. The dereddened quasar flux density (black) of J2318 where the continuum is estimated by the RLF regions (vertical cyan bands) such that it is fit to various models (colored). 
The flux errors are indicated in magenta, and the $\chi_\nu^2$ and best-fit parameter values for each model are displayed in their respective colors. 
Dashed curves represent the power law before reddening, and comparison to the CSb $A$ parameter in other models should be done using $A_{2665}=A(2665/2000)^{\alpha_\lambda}$. Only wavelengths greater than $3700$ \AA\, are displayed.
        Top panel: a simple power law.
		Second panel: a power law reddened with the SMC extinction curve of \cite{pei92}.
		Third panel: a power law reddened with the three-parameter power-law extinction curve of \cite{2008ApJ...686L.103G}.
		Fourth panel: a broken power law reddened with the three-parameter power-law extinction curve of \cite{2008ApJ...686L.103G}.
        Fifth panel: a broken power law with no reddening.
		Sixth panel: the continuum estimate generated by PyQSOFit.
        Due to the lower $\chi_\nu^2$ values, the SMC and CS reddening models are used for analysis.
	}\label{fig:J2318_DV2_RedZoom_Fit}\end{figure}

As described below, we tested six functional continuum fitting models to these RLF regions (Figures \ref{fig:J2318_DV2_SpecFit}, \ref{fig:PQF_Ha}, and \ref{fig:J2318_DV2_RedZoom_Fit}). 
Two of the models use all RLFs and four of them exclude the \ion{Fe}{2} RLFs.
Although the fitting was done for each SDSS MJD combined with the GNIRS spectrum, Figures \ref{fig:J2318_DV2_SpecFit}, \ref{fig:PQF_Ha}, and \ref{fig:J2318_DV2_RedZoom_Fit} only show the results for the quasi-simultaneous MJD 60251+60291 spectrum.
We are interested in each model's accuracy at longer wavelengths (Figure \ref{fig:J2318_DV2_RedZoom_Fit}) since we require luminosity values and the FWHM of H$\alpha$ to estimate the black hole mass, and Eddington luminosity and ratio (\S\ref{sec:mbh}).

Quasar continua are commonly fit to a power law (PL) \citep[e.g.,][]{2015ApJ...806..111G,J0230}. 
\cite{2016A&A...585A..87S} found that a composite spectrum of bright ($r<17$) quasars from $0.1-1~{\rm \mu m}$ was well fit by a power law with $\alpha_\lambda=-1.71\pm 0.01$ and negligible reddening.
The PL model used in this analysis excludes the \ion{Fe}{2} RLFs and has the form,
\begin{equation}
    F_{\rm PL}(\lambda) = A\left(\frac{\lambda}{\A{2000}}\right)^{\alpha_\lambda}~,
\end{equation}
where $A$ and $\alpha_\lambda$ are the amplitude and slope parameters.
The PL (first panel in Figure \ref{fig:J2318_DV2_SpecFit}) does not fit the underlying continuum of J2318 and has $\chi_\nu^2=6.082,~{\rm dof}=915$, which is the highest $\chi_\nu^2$ of our continuum models.
It underestimates the continuum at shorter wavelengths, while overestimating it at longer wavelengths (first panel in Figure \ref{fig:J2318_DV2_RedZoom_Fit}).

Dust reddening can cause quasar continua to deviate from a pure power law.
We adopt two separate dust-reddened power laws that also exclude the \ion{Fe}{2} RLFs in their continuum fitting; the Small Magellanic Cloud (SMC) and the circumstellar (CS) reddened power laws.
Table \ref{tab:Fitting_Parameters} lists the fitting parameter and the $\chi_\nu^2$ results of both reddened power-law models for every spectrum that was fit.

The SMC reddened power law has been used to fit reddened BALQSO spectra \citep[e.g.,][]{2012ApJ...757..114F,2013ApJ...777..168F,2016ApJ...824..130G,2019ApJ...872...21H} and has the form
\begin{equation}
    F_{\rm SMC}(\lambda) = A\left(\frac{\lambda}{\A{2000}}\right)^{\alpha_\lambda} 10^{-0.4\xi(\lambda) (1 + R_V) E_{B-V}}~,
\end{equation}
where $E_{B-V}$ is the reddening, $R_V=2.93$ is the ratio of total-to-selective extinction in the $V$ band, and $\xi(\lambda)$ is the SMC extinction curve of \citet{pei92}.
When fitting we forced the reddening to be positive, $E_{B-V}\geq0$, while both amplitude $A$ and power-law index $\alpha_\lambda$ were left as free parameters.
The SMC is the third-best fit with $\chi_\nu^2=1.609,~{\rm dof=915}$.
Although the SMC extinction curve (solid curve in the second panel of Figure \ref{fig:J2318_DV2_SpecFit}) is a decent fit at all wavelengths, with $A=26.3$ and $\alpha_\lambda=-2.31$, the intrinsic spectrum before reddening (dashed curve) is very luminous and is as blue as the \citet{ss73} prediction for a thin disk (and quasars that blue are rare).

To account for the extra reddening at shorter wavelengths we adopt a CS reddening model used by \citet{2014ApJ...788..123L} to model the unusual near-UV to optical continuum of the BAL quasar Mrk 231, which has low reddening in the optical spectrum and much higher reddening in the near UV.
The authors used a dust extinction model developed by \citet{2008ApJ...686L.103G} to explain the wavelength-dependent reddening observed in Type Ia supernovae (SNe Ia).
A thorough explanation of the physics and motivation for CS reddening can be found in \citet{2008ApJ...686L.103G} or \citet{2005ApJ...635L..33W}.

The CS reddened power-law model used in fitting J2318's spectra excludes the \ion{Fe}{2} RLFs and has the form 
\begin{equation}
    F_{\rm CS}(\lambda) = A \left(\frac{\lambda}{\A{2000}} \right)^{\alpha_\lambda} 10^{-0.4\left[1 - a + a\left(\lambda/\lambda_V\right)^p\right]A_V}~,
\end{equation}
where the form of this three-parameter power-law extinction curve is taken from \citet{2008ApJ...686L.103G}, and $\lambda_V=0.55~{\rm \mu m}$ is the V-band's central wavelength.
Table \ref{tab:Fitting_Parameters} lists the CS best fit parameters and $\chi_\nu^2$ results for the fitting of each spectrum.
We recognize that the best-fit value of $\alpha_\lambda=-4.0$ is unphysical, but it becomes more negative, and the $\chi_\nu^2$ becomes higher, when left as a free parameter.
Thus, $-4.0$ was set as the limiting possible value since it is the long-wavelength slope of a single-temperature blackbody. 
However, the model allows for good fitting of the continuum to aid in measurements of the absorption troughs (third panel in Figure \ref{fig:J2318_DV2_SpecFit}).
Furthermore, the CS model's fitting of the red-end of the spectrum (third panel in Figure \ref{fig:J2318_DV2_RedZoom_Fit}) is comparable to that of the SMC model and with $\chi_\nu^2=1.44,~{\rm dof=915}$ it is the best model of those tested.
When all parameters were left free, \citet{2014ApJ...788..123L} found the reddening parameters to be $A_V=1.54$, $a=0.78$, and $p=-1.72$ for their single BAL quasar Mrk 231.
Our analysis produced smaller values of $A_V$, larger values of $a$, and less negative values of $p$.

\cite{2016MNRAS.460..212C} fit thin accretion disk models to optical-near-infrared quasar spectra and find some curvature relative to pure power-law fits, such that the flux at short wavelengths is smaller than expected from an extrapolation from long wavelengths.
Furthermore, \citet{2007AJ....134..294S} reported that different power-law exponents were often needed shortward and longward of $\sim\A{5600}$, which they referred to as $\alpha_{UVO}$ and $\alpha_{Ored}$, respectively. 
They found that the flux at long wavelengths (RLFs $5600$-\A{5648}, $6198$-\A{6215}, and $6820$-\A{6920}) was larger than expected from an extrapolation from short wavelengths (RLFs $1348$-\A{1358} and $5600$-\A{5648}).
As a simple approximation to these results, we fit a broken power-law model.
The circumstellar reddened broken (CSb) power law uses the entire RLF set and has the following form,
\begin{equation}
	F_{\rm CSb} = \left\{\begin{array}{ll}
		A{\left(\frac{\lambda}{\lambda_{\rm break}}\right)}^{\alpha_{\lambda,1}} 10^{-0.4\left[1-a+a{(\lambda/\lambda_V)}^p\right]A_V} & {\rm when}~\lambda<\lambda_{\rm break}    \\ \\
		A{\left(\frac{\lambda}{\lambda_{\rm break}}\right)}^{\alpha_{\lambda,2}} 10^{-0.4\left[1-a+a{(\lambda/\lambda_V)}^p\right]A_V} & {\rm when}~\lambda\geq\lambda_{\rm break}
	\end{array}\right.~.
\end{equation}
Here, the reddening parameters $a$, $p$, and $A_V$ and the power-law amplitude, $A$, are the same for all wavelengths; thus, it is only the power-law slope, $\alpha_{\lambda,i}$, that is different between the short $\lambda<\lambda_{\rm break}$ and long $\lambda\geq\lambda_{\rm break}$ wavelengths.
A broken power-law continuum could in principle arise from emission from an accretion disk whose temperature profile is a broken power law.
Such a model helps fit at rest-frame ultraviolet and optical wavelengths simultaneously. 
The output of the accretion disk is greater than that of the host galaxy’s contribution in the optical for a quasar as luminous as J2318.
The fourth panel of Figure \ref{fig:J2318_DV2_SpecFit} shows that the broken power-law spectrum with the CS reddening curve (power-law reddening) applied to it is the second-best fit of the models considered, with $\chi_\nu^2=1.586,~{\rm dof=1196}$, and has reasonable parameters for the amplitude and power-law slopes of the unreddened spectrum.
Unfortunately, this model underestimates the continuum at the extreme red-end of the spectrum (fourth panel in Figure \ref{fig:J2318_DV2_RedZoom_Fit}).
Due to $\lambda_{\rm break}$, the CSb model's $A$ parameter is not easily comparable to other models.
Thus, we introduce an additional parameter, $A_{2665}=(2665/2000)^{\alpha_\lambda}$, to the other models to aid in comparison.

We also construct a broken power law (BPL) without any reddening to test how well such a model can fit without any reddening.
Like the CSb model, it is only the power slopes $\alpha_{\lambda,i}$, that are different between the short and longer wavelengths separated by $\lambda_{\rm break}$.
\begin{equation}
    F_{\rm BPL} = A\left[\left(\frac{\lambda}{\lambda_{\rm break}}\right)^{\alpha_{\lambda,1}} + \left(\frac{\lambda}{\lambda_{\rm break}}\right)^{\alpha_{\lambda,2}}\right]
\end{equation}
The BPL model (fifth panel in Figure \ref{fig:J2318_DV2_SpecFit}) has $\chi_\nu^2=1.671,~{\rm dof=1196}$, which is the third worst.

Finally, we employ PyQSOFit (PQF;  \citealt{PyQSOFit,2019ApJS..241...34S}) to fit quasar spectra.
The underlying continuum is fit using a power law and a polynomial, while three separate templates are used to model spectral emission features.
Two of these templates model Fe emission in the UV and optical regimes, while the third models the Balmer series emission. 

Fe emission in the UV region, $1200-\A{3500}$, is modeled using the template of \citet{vw01} modified using three free parameters: a normalization factor, $U_n$, a FWHM, $U_{\Delta\lambda}$, and a wavelength shift, $U_\lambda$.
Fe emission in the optical region, $3686-\A{7484}$, is modeled using the template of \citet{bg92} and three similar parameters: a normalization factor, $O_n$, a FWHM, $O_{\Delta\lambda}$, and a wavelength shift, $O_\lambda$.
A Balmer continuum template failed to accurately model the combined SDSS+GNIRS spectra of J2318, presumably due to the weak emission of the Balmer series, and thus we left it out of the PQF model. 
\citet{WuShen22} also found the Balmer template to be unfavorable for the fitting of their SDSS DR16 \citep{dr16q} quasar sample, however, this was due to limited SN and spectral coverage.

Our PQF model uses all RLFs and has the following form,
\begin{equation}
	F_{\rm PQF} = A\left(\frac{\lambda}{\A{3000}}\right)^{\alpha_\lambda} + \left(C_0 + C_1\lambda + C_2\lambda^2\right) + U_n {\rm Fe}_{\rm UV}\left( \lambda, U_{\Delta\lambda}, U_\lambda \right) + O_n {\rm Fe}_{\rm opt} \left( \lambda,  O_{\Delta\lambda}, O_\lambda \right)
\end{equation}
where $A$ and $\alpha_\lambda$ are the amplitude and slope of the power law, and $C_0$, $C_1$, and $C_2$ are the polynomial coefficients.
The underlying continuum is modeled using only the first two terms, while the emission complexes are modelled using the last two terms.
Monte Carlo resampling of the spectrum based on the initial flux density uncertainties is performed to estimate the errors in the parameters.
The PQF model (sixth panel in Figure \ref{fig:J2318_DV2_SpecFit}) has $\chi_\nu^2=1.812,~{\rm dof=1196}$, which is the second-highest of all models tested.
The model overestimates the continuum at low wavelengths, similar to that of the PL but shifted bluewards, and does reasonably well at longer wavelengths (sixth panel in Figure \ref{fig:J2318_DV2_RedZoom_Fit}).
Figure \ref{fig:PQF_Ha} displays the PQF continuum fit for the unsmoothed spectrum of MJD 60251 along with a Gaussian fit of the H$\alpha$ emission.

\begin{table*}\centering\begin{tabular}{ccccccccccc}
    \hline
     & \multicolumn{4}{c}{SMC Reddening Model} & \multicolumn{6}{c}{CS Reddening Model}\\
     \cmidrule(lr){2-5}\cmidrule(lr){6-11}
    Spectra & $A$ & $\alpha_\lambda$ & $E_{B-V}$ & $\chi_\nu^2$ & $A$ & $\alpha_\lambda$ & $a$ & $p$ & $A_V$ & $\chi_\nu^2$ \\
    \hline
    57328 & $16.031$ & $-1.847$ & $0.056$ & $3.15$ & $183.09$ & $-4.0$ & $18.594$ & $-0.26$ & $0.479$ & $2.386$ \\
    59188 & $17.901$ & $-2.003$ & $0.062$ & $2.543$ & $174.489$ & $-4.0$ & $15.261$ & $-0.287$ & $0.494$ & $2.075$ \\
    60251 & $26.337$ & $-2.31$ & $0.115$ & $1.609$ & $147.013$ & $-4.0$ & $13.652$ & $-0.415$ & $0.365$ & $1.44$ \\
    \hline
    57328 & $23.672$ & $-2.786$ & $0.123$ & $2.67$ & $77.313$ & $-4.0$ & $7.639$ & $-0.581$ & $0.342$ & $2.06$ \\
    59188 & $26.898$ & $-2.891$ & $0.136$ & $2.489$ & $74.868$ & $-4.0$ & $7.229$ & $-0.63$ & $0.32$ & $2.059$ \\
    \hline
\end{tabular}\caption{ SMC and CS Continuum Fitting Parameter and $\chi_\nu^2$ Results. The best-fit parameters for the unsmoothed spectra of all epochs (first three rows) and the spectra matched to MJD 60251+60291 (last two rows).
}
\label{tab:Fitting_Parameters}\end{table*}

\subsection{Continuum Normalization} \label{sec:norm}
\begin{figure}\centering\includegraphics[width=18cm]{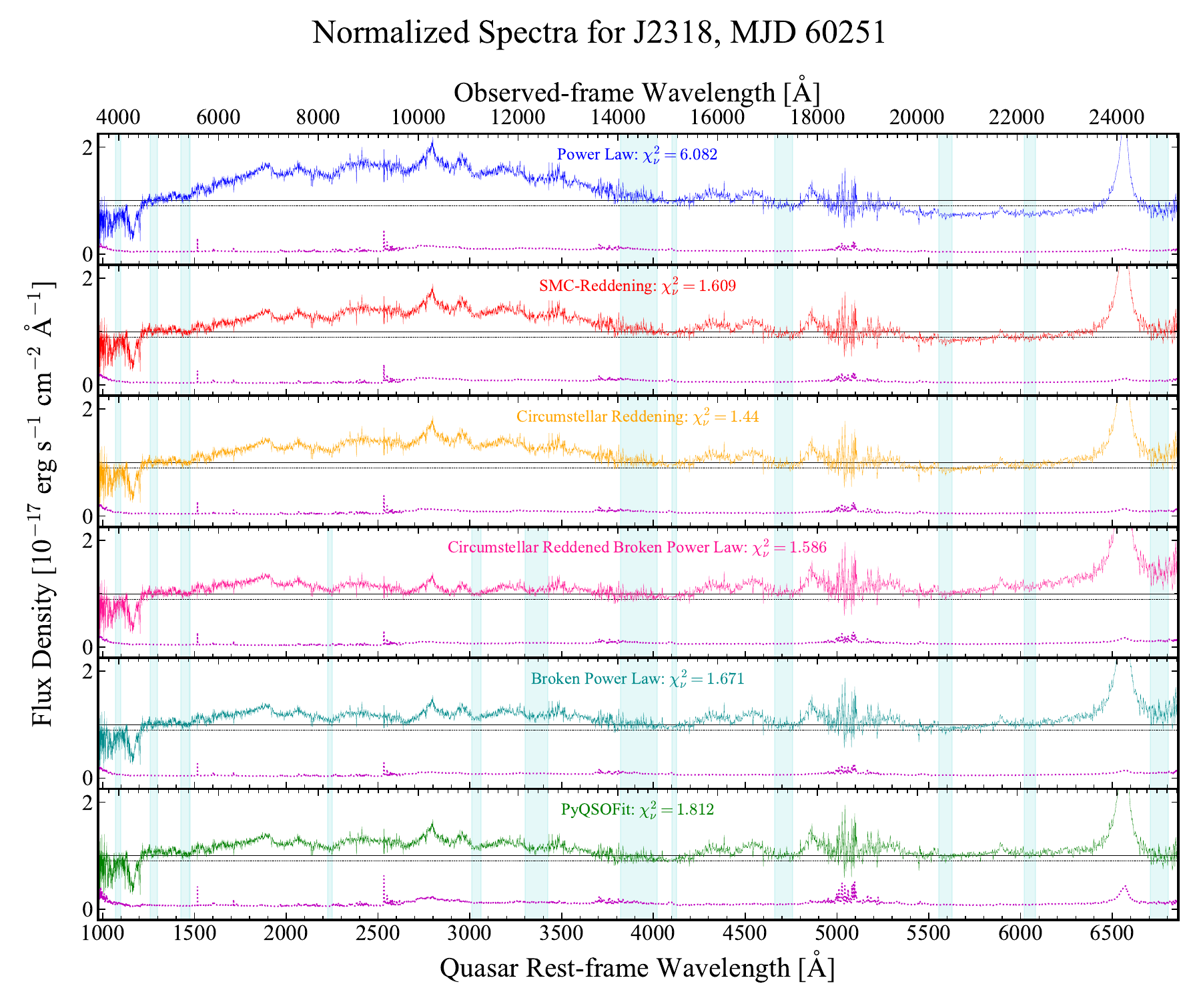}\caption{J2318-MJD 60251+60291 Normalized Continuum. The dereddened quasar flux density of J2318 where the continuum is estimated by the RLF regions (vertical cyan bands) such that it is fit and normalized by various models (colored). The flux errors are indicated in magenta, and the $\chi_\nu^2$ for each model are displayed in their respective colors.
        Top panel: a simple power law.
		Second panel: a power law reddened with the SMC extinction curve of \cite{pei92}.
		Third panel: a power law reddened with the three-parameter power-law extinction curve of \cite{2008ApJ...686L.103G}.
		Fourth panel: a broken power law reddened with the three-parameter power-law extinction curve of \cite{2008ApJ...686L.103G}.
        Fifth panel: a broken power law with no reddening.
		Sixth panel: the continuum estimate generated by PyQSOFit.
        Due to the lower $\chi_\nu^2$ values, the SMC and CS reddening models are used for analysis.
	}\label{fig:J2318_DV2_norm}\end{figure}

To extract information and perform quantitative measurements pertaining to J2318, we must normalize each spectrum by dividing their flux values by their respective continuum fitting flux values to get a continuum value equal to $\sim1$.
We normalize the SDSS+GNIRS spectrum for all three epochs, and Figure \ref{fig:J2318_DV2_norm} displays the normalization using each fit model tested. 
We expect the underlying continuum level to be at a normalized flux density of unity and traditional BAL troughs to be any broad absorption that consistently dips below 0.9 for a velocity width of $\geq$\kms{2000} \citep{wea91}.

For the PL model (first panel in Figure \ref{fig:J2318_DV2_norm}), the Ly$\alpha$ forest and wavelengths longer than \A{4600} dip below the expected continuum level, excluding emission features (e.g., $\rm H\alpha$ and $\rm H\beta$).
The \ion{C}{4} and \ion{Si}{4} BALs lie within the Ly$\alpha$ forest, but these are the only broad absorption troughs present in J2318.
Rest-frame $5350$-\A{6450} is a continuum region and shows no evidence of an outflow.
In the CSb model (fourth panel in Figure \ref{fig:J2318_DV2_norm}), the complex emission of \ion{Fe}{2} is suppressed and wavelengths longer than \A{5100} rise above the expected continuum level.
The former is due to the \ion{Fe}{2} RLFs being used in the fitting.
The BPL model (fifth panel in Figure \ref{fig:J2318_DV2_norm}) exhibits a similar suppression for the complex emission of \ion{Fe}{2} and wavelengths longer than \A{6100} are above the expected continuum level.
Due to these anomalies, the PL, CSb, and BPL models are used in the testing of different fit models and normalization of separate spectra, but are excluded from any subsequent analysis.

The CS model (third panel in Figure \ref{fig:J2318_DV2_norm}) resulted in the best $\chi_\nu^2$ and reasonably represented the continuum at all wavelengths. 
Although the $\alpha_\lambda=-4.0$ value is unphysical in this model, the other parameters are reasonable and the continuum is best normalized for the analysis of its characteristics.
The SMC model (second panel in Figure \ref{fig:J2318_DV2_norm}) is the third best fit; its different reddening law may offer a contrasting normalization of the Ly$\alpha$ forest region.
Thus, the SMC and CS models are used to measure the broad absorption properties of the \ion{C}{4} and \ion{Si}{4} BALs.

For an additional estimate of systematic uncertainties, we used the MJD 57328 and 59188 spectra `morphed' to match the MJD 60251 spectrum using the fit discussed in \S \ref{sec:specmatch}.
We normalized these matched spectra with the same SMC and CS continuum fits used for MJD 60251+60291.

\clearpage
\subsection{Quasar Luminosity and Spectral Energy Distribution}\label{sec:sed}
\begin{figure}\centering\includegraphics[width=18cm]{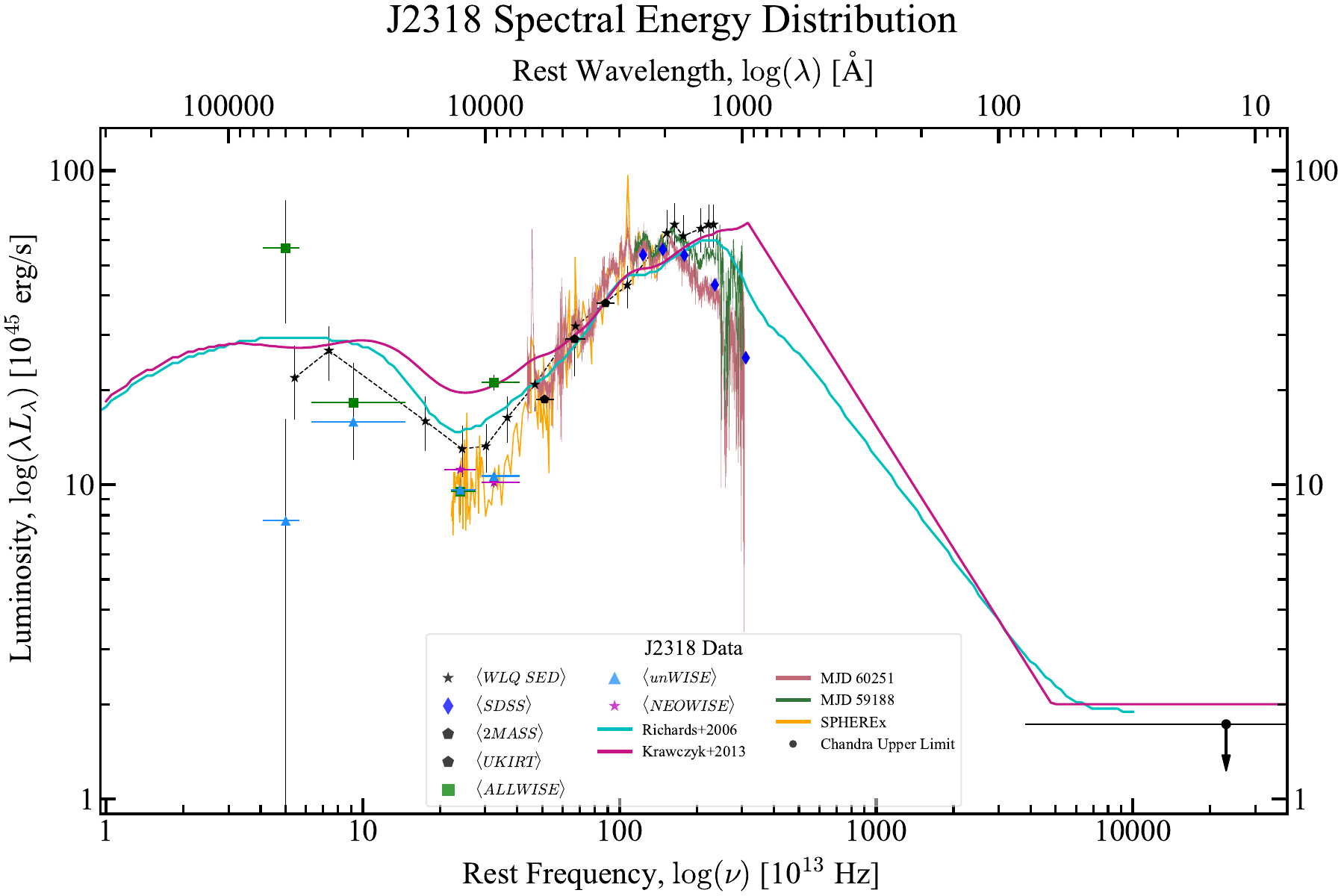}\caption{J2318 Average Photometric, Spectroscopic, and SED Luminosity versus Frequency.
    J2318's log-scaled rest-frame luminosity for the mean WISE photometry data points (colored squares, stars, and triangles), mean SDSS photometry (blue diamonds), 2MASS and UKIRT photometry (black pentagons), Chandra X-ray luminosity (black circle), SEDs and UV/optical+NIR spectrum (colored curves) as a function of rest-frame frequency. 
    All photometric data points lie at their effective rest wavelength, with colored horizontal lines indicating the band's wavelength interval.
    The downward pointing arrow indicates that the Chandra X-ray luminosity is an upper limit.
    The MJD 60291 NIR spectrum is matched with the optical spectrum of MJD 60251 (red curve), while only an optical spectrum is available for MJD 59188 (green curve), and the NIR spectrum from SPHEREx (orange curve) is smoothed for $\lambda_{r}\geq12,000$ \A{} and scaled by 0.7 at all wavelengths.
    The comparison SEDs are from \citet{gtr06} (cyan), \citet{2013ApJS..206....4K} (violet), and \citet{2011ApJ...743..163L} (black), and have been scaled to lie at the $\langle UKIRT-J\rangle$ band's luminosity for visual comparison.
}\label{fig:Photo+SED_Luminosity}\end{figure}

For comparison with other quasars' spectral energy distributions \citep[e.g.,][]{2015MNRAS.446.3427C,2016MNRAS.460..212C} we convert the spectra and the photometric data points to luminosity values using
\begin{equation}
    \lambda_r L_\lambda = 4\pi d_L^2 ~\lambda_o f_\lambda ~,
\end{equation}
where $\lambda_r L_\lambda$ is the rest-frame wavelength times rest-frame luminosity density, $d_L$ is the luminosity distance, $\lambda_o f_\lambda$ is the observed wavelength times observed flux density, and $\lambda_o=\lambda_r(1+z)$.
With our adopted cosmological parameters, the luminosity distance to an object at $z=2.6781$ is $d_L = 2.27\times10^{10}~{\rm pc} = 7.02\times10^{28}~{\rm cm}$.
Figure \ref{fig:Photo+SED_Luminosity} displays the luminosity of the spectra, dereddened from Galactic extinction, for MJD 60251+60291 and MJD 59188 as a function of frequency.

When determining the luminosity of the photometric data points, the magnitudes are first converted to AB magnitudes, \citep[$m_{\rm AB}$;][]{og83}, before they  can be represented as flux densities.
The SDSS photometry \citep{fuk96} is already provided in the AB magnitude system, while the ALLWISE, NEOWISE \citep{NEOWISE,NEOWISEreact}, and unWISE photometry is converted to AB magnitudes by adding a correction factor to their Vega magnitudes.
The correction factors for bands 1, 2, 3, and 4 are 2.699, 3.339, 5.174, and 6.620, respectively.
Furthermore, the 2MASS, UKIRT, and SDSS photometry are corrected for Galactic extinction, while the WISE photometry is not corrected for Galactic extinction, which is a $<1.5$\% effect at W1 and $<1$\% at longer wavelengths \citep{2013MNRAS.430.2188Y}
We compute the weighted average of each band's magnitude, $\langle m_{\rm AB}\rangle$, and use it to calculate the average flux density, $\langle f_\lambda \rangle$, per unit wavelength using the following formula,
\begin{equation}
    f_\lambda = \langle f_\lambda \rangle = \frac{c}{\lambda^2}10^{-\left(\langle m_{\rm AB}\rangle + 48.6\right)/2.5} ~,
\end{equation}
where $c$ is the speed of light and $\lambda$ is taken as the effective observed wavelength of each bandpass.
Table \ref{tab:Luminosity} lists the calculated average AB magnitude, flux density, and luminosity values for each bandpass in the observed frame, while a visual representation is provided in Figure \ref{fig:Photo+SED_Luminosity}.

\citet{2011ApJ...743..163L} created composite WLQ spectral energy distributions (SEDs) from their sample of 17 SDSS WLQs, and binned them into six rest-frame UV, two near-IR, and seven near-to-mid-IR flux bins. 
The fluxes of these SED bins were unnormalized by multiplying them by the mean flux density of the quasar sample at rest-wavelength \A{1445}, before shifting them to the observed-frame at $z=2.6781$ and converting them into luminosities.
A linear interpolation was done between the WLQ SED bins bounding the UKIRT-J band to determine the luminosity at that wavelength.
This value was then scaled to match J2318's UKIRT-J band luminosity, and all other WLQ SED bins were scaled by that same factor.
\citet{gtr06} created composite SEDs of 259 quasars using SDSS and Spitzer photometry when studying the mid-infrared and optical properties of broad-line quasars.
\citet{2013ApJS..206....4K} provide a composite SED for 119,652 luminous broad-line quasars.
These composite SEDs have been included in Figure \ref{fig:Photo+SED_Luminosity} and have been scaled so that they intersect with J2318's UKIRT-J band luminosity.

The later spectrum obtained by SPHEREx is brighter than the earlier observations made by GNIRS and SDSS.
The brightness could be explained by the quasar’s variability or SPHEREx’s lower resolution and larger aperture allowing the nearby star’s light to contribute to the brighter spectrum.
To account for the brighter spectrum, and to match to the UKIRT J-band luminosity, we scaled the luminosity at all wavelengths by 0.7 (Figure \ref{fig:Photo+SED_Luminosity}).
Once scaled the SPHEREx spectrum is in good agreement with the W1 and W2 photometry, and conforms well with the optical+NIR spectrum.
Thus confirming our choice of recalibrating the GNIRS spectrum to match the UHS observed J-K color before merging it with the SDSS spectrum.

Figure \ref{fig:Photo+SED_Luminosity} shows that J2318 is less luminous in the mid-infrared (rest-frame $7-70$~$\mu{\rm m}$) relative to the near-infrared than are the typical quasar SEDs of \citet{gtr06}, \citet{2013ApJS..206....4K} and \citet{2011ApJ...743..163L}

Figure \ref{fig:Photo+SED_Luminosity} also shows that in the rest-frame optical, J2318 is bluer than the typical quasar SEDs of \citet{gtr06}, \citet{2013ApJS..206....4K} and \citet{2011ApJ...743..163L}, even after adjusting the GNIRS spectrum to match the UHS $J-K$ color (\S~\ref{sec:gemini}).
The best-fit power-law slope (without reddening) for the GNIRS spectrum is $\alpha_\lambda=-2.095$, equivalent to $\alpha_\nu=0.095$ ($F_\nu\propto\nu^{\alpha_\nu}$).
For comparison, \citet{2007ApJ...668..682D} measured $\alpha_\nu$ between \A{2200} and \A{4000} for a large sample of SDSS quasars and found an average of $-0.37$ and a maximum of $\simeq 0.5$; a value of $0.095$ places J2318 among the bluest $\sim$15\% of quasars.
J2318 is nonetheless not quite as blue at optical wavelengths as predicted for a standard thin disk ($\alpha_\lambda=-2.33$, $\alpha_\nu=0.33$).

The apparent peak in the SED near rest-frame 2000$-$3000~\A{} or (1-1.5)$\times$10$^{15}$~Hz occurs at a factor of $\sim$2 longer wavelength than the `big blue bump' seen in typical quasars or WLQs.
This offset may be due to dust reddening shifting the peak of a normal big blue bump to longer wavelengths. 
However, the fact that our best-fit models of the optical+NIR spectrum including reddening are physically implausible (\S\ref{sec:contfit}, \S\ref{sec:norm}) casts some doubt on that possibility.
This offset may instead (or also) reflect a relatively high black hole mass (e.g., \citealt{Lai+23}) or mass loss rate from disk winds \citep{2012MNRAS.426..656S}, though the effects of black hole spin and deviations from idealized thin disks are also important (e.g., \citealt{2016MNRAS.460..212C}).

\citet{2018MNRAS.480.5184N} studied 32 WLQs with \ion{C}{4} EW $\lesssim$ \A{15} to investigate the nature of X-ray weakness. 
They split their sample into an `extreme subsample' where \ion{C}{4} EW $<$ \A{7.0} and a `bridge subsample' where \ion{C}{4} EW $=7.0-$\A{15.5}. 
Comparison with their Figure 6 reveals that J2318 appears to be like their `bridge' subsample, in which the luminosity drops off in the optical and UV.
\citet{2015ApJ...805..122L} conducted an X-ray multiwavelength study of 32 WLQs and 18 quasar analogs of the extreme WLQ, PHL 1811. 
The 32 WLQs were split into a subsample of 14 X-ray weak quasars and a subsample of 17 X-ray normal quasars. 
Their Figure 8 compares the two subsamples to the SEDs of \citet{2011ApJ...743..163L} and \citet{gtr06}, and the ultraviolet luminosity of the X-ray weak subsample (their Figure 18a) lies below the WLQ SEDs of \citet{2011ApJ...743..163L}, similar to that of J2318, suggesting that J2318 is likely to be X-ray-weak. 

The current X-ray upper limit for J2318, however, does not distinguish between it being an X-ray normal or an X-ray weak WLQ \citep{2022MNRAS.511.5251N}.

\begin{table*}\centering\begin{tabular}{cccccc}
    \hline
    Survey & Filter (abbr.) & $\lambda_{o}$ & $\langle m_{\rm AB}\rangle$
         & $\langle f_\lambda\rangle$
         & $\langle\lambda L_\lambda\rangle$ \\
      & & ($\mu$m) & 
         & ($10^{-13}~{\rm erg}~{\rm s}^{-1}~{\rm cm}^{-2}~{\mu\rm m}^{-1}$)
         & ($10^{45}~{\rm erg}~{\rm s}^{-1}$) \\
    \hline
    unWISE & W1 (uW1) & $3.4$ & $18.17\pm0.03$ & $0.51\pm0.01$ & $10.66\pm0.25$ \\
    unWISE & W2 (uW2) & $4.6$ & $17.95\pm0.04$ & $0.34\pm0.01$ & $9.64\pm0.36$ \\
    unWISE & W3 (uW3) & $12.0$ & $16.37\pm0.13$ & $0.21\pm0.03$ & $15.89\pm1.90$ \\
    unWISE & W4 (uW4) & $22.0$ & $16.5\pm1.2$ & $0.06\pm0.06$ & $7.69\pm8.50$ \\
    \hline
    ALLWISE & W1 (AW1) & $3.4$ & $17.43\pm0.06$ & $1.01\pm0.06$ & $21.14\pm1.19$ \\
    ALLWISE & W2 (AW2) & $4.6$ & $17.97\pm0.17$ & $0.33\pm0.05$ & $9.51\pm1.51$ \\
    ALLWISE & W3 (AW3) & $12.0$ & $16.22\pm0.37$ & $0.25\pm0.08$ & $18.24\pm6.25$ \\
    ALLWISE & W4 (AW4) & $22.0$ & $14.33\pm0.46$ & $0.42\pm0.18$ & $56.62\pm24.02$ \\
    \hline
    NEOWISE & W1 (NW1) & $3.4$ & $18.22\pm0.04$ & $0.48\pm0.02$ & $10.17\pm0.36$ \\
    NEOWISE & W2 (NW2) & $4.6$ & $17.79\pm0.06$ & $0.39\pm0.02$ & $11.18\pm0.60$ \\
    \hline
    SDSS & $u$ (SDSSu) & $0.3551$ & $19.684\pm0.037$ & $11.55\pm0.40$ & $25.37\pm0.87$ \\
    SDSS & $g$ (SDSSg) & $0.4686$ & $18.803\pm0.009$ & $14.93\pm0.12$ & $43.28\pm0.36$ \\
    SDSS & $r$ (SDSSr) & $0.6165$ & $18.269\pm0.007$ & $14.11\pm0.09$ & $53.81\pm0.35$ \\
    SDSS & $i$ (SDSSi) & $0.7481$ & $18.014\pm0.006$ & $12.12\pm0.07$ & $56.09\pm0.30$ \\
    SDSS & $z$ (SDSSz) & $0.8931$ & $17.864\pm0.018$ & $9.76\pm0.16$ & $53.92\pm0.89$ \\
    \hline
    2MASS & J (2MASS-J) & $1.25$ & $17.99\pm0.22$ & $4.42\pm0.90$ & $34.17\pm6.92$ \\
    2MASS & H (2MASS-H) & $1.65$ & $17.87\pm0.26$ & $2.85\pm0.68$ & $29.10\pm6.97$ \\
    2MASS & Ks (2MASS-Ks) & $2.16$ & $18.07\pm0.42$ & $1.38\pm0.53$ & $18.46\pm7.14$ \\
    \hline
    UKIRT & J (UKIRT-J) & $1.25$ & $17.88\pm0.01$ & $4.89\pm0.06$ & $37.82\pm0.49$ \\
    UKIRT & K (UKIRT-K) & $2.16$ & $18.06\pm0.03$ & $1.40\pm0.04$ & $18.70\pm0.50$ \\
    \hline
\end{tabular}\caption{Weighted-average AB magnitude, observed-frame flux density, and observed-frame luminosity for the bandpasses of the WISE photometry data points, the SDSS, the 2MASS, and the UHS. uW1 through uW4 represent unWISE coadd forced photometry (DR13 version), bands AW1 through AW4 correspond to observations before MJD 55545, and bands NW1 and NW2 were observed after MJD 56645. The SDSS observations were conducted between MJD 53265 and 60251. 2MASS observations are from MJD 51087, while the UHS observations were between MJD 56211 and 58013.}
\label{tab:Luminosity}\end{table*}

\clearpage
\subsection{Black hole mass estimate} \label{sec:mbh}
To estimate the mass of the black hole (BH) we follow two approaches: \citet[][their Eq.3; hereafter SL12]{Shen+Liu2012} and \citet[][their Eq.5; hereafter A11]{Assef+2011}.
We provide a generalized formula for the two methods,
\begin{equation}
    M_{BH} = A \left(\frac{L_{5100}}{10^{44}~{\rm erg~s}^{-1}}\right)^b \left(\frac{\rm FWHM_{H\alpha}}{v_{\rm norm}}\right)^c~,
\end{equation}
where $A$, $b$, $c$, and $v_{\rm norm}$ are parameters specific to each method (see Table \ref{tab:BHmass_parameters}), $L_{5100}$ is the continuum luminosity at $\A{5100}$, and ${\rm FWHM_{H\alpha}}$ is the full width at half maximum (FWHM) of the ${\rm H\alpha}$ emission feature.
Each estimate of BH mass has a statistical error propagated from the uncertainty in the FWHM$_{H\alpha}$ measurement and the uncertainty in $L_{5100}$,
\begin{equation}
    \delta M_{\rm BH,vir} = \sqrt{\left(b\frac{\delta L_{5100}}{L_{5100}}\right)^2 + \left(c\frac{\delta{\rm FWHM}}{\rm FWHM}\right)^2}~M_{\rm BH,vir} ~.
\end{equation}

\begin{table*}
	\centering
	\begin{tabular}{ccccc}
		\hline
		Model    & $A$            & $b$ & $c$  & $v_{\rm norm}$ (\kms{})   \\
		\hline
        $M_{SL12}$ & $10^{1.39}\simeq24.55$ & $0.555$ & $1.873$ & $1$  \\
		$M_{A11}$  & $8.9856\times10^6$ & $0.520$ & $2.060$ & $10^3$  \\
		\hline
	\end{tabular}
	\caption{Parameter values for black hole mass estimates. The model subscripts $SL12$ and $A11$ refer to the work done by \citet{Shen+Liu2012} and \citet{Assef+2011}, respectively.}
	\label{tab:BHmass_parameters}
\end{table*}

The NIR spectrum covering the $\A{5100}$ region is quite noisy, so we extract the $\A{5100}$ flux values of the fitted models that adequately represent the continuum at this wavelength, convert them to luminosity values, and calculate the average.
The SMC reddened power law, the CS reddened power law, and PQF model match the continuum best around $\A{5100}$ and yield $L_{5100}$ values of $2.52\times10^{46}~{\rm erg~s}^{-1}$, $2.49\times10^{46}~{\rm erg~s}^{-1}$, and $2.30\times10^{46}~{\rm erg~s}^{-1}$, respectively, with a systematic rms uncertainty of $\pm0.25\times10^{46}~{\rm erg~s}^{-1}$.
For comparison, we compute the average luminosity of the unreddened versions of the SMC and CS models along with the reddened PQF model.
The results for the reddened and unreddened luminosity averages of MJD 60251 are $\langle L_{5100}\rangle = \left(2.51\pm0.01\right)\times10^{46}~{\rm erg~s}^{-1}$ and $\langle L_{5100}\rangle = \left(3.82\pm0.01\right)\times10^{46}~{\rm erg~s}^{-1}$, respectively.

One of PyQSOFit's features is the ability to determine various properties of emission or absorption features, by fitting these features with Gaussian distributions.
In our case, we are interested in the H$\alpha$ emission feature, which we assume can be fit by two broad Gaussians and one narrow Gaussian, all centered at $\A{6564.61}$.
The PQF measurement of $\rm FWHM_{H\alpha}$ is determined to be $\A{3084 \pm 91}$, and the weighed average is $\A{3059 \pm 76}$ when including the previous fit discussed in \S~\ref{sec:redshift} of \kms{3000\pm140}.
We adopt the mean of the two measurements as our value of FWHM$_{\rm H\alpha}$.

Table \ref{tab:Ledd_result} lists our results for MJD 60251, where we calculate the BH measurements of $M_{SL12} = (17.47\pm0.94)\times10^8M_\odot$ and $M_{A11} = (15.66\pm0.86)\times10^8M_\odot$, with intrinsic scatter of $\pm0.12$ dex and $\pm0.14$ dex, respectively. 
We adopt their weighted average of $\langle M_{\rm BH}\rangle = (16.52\pm0.65)\times10^{8}M_\odot$ as our $M_{\rm BH}$ value.
When SMC and CS reddening is removed, $M_{\rm BH}$ is $+14.84\%$ larger, thus providing an upper estimate for the mass.

\subsubsection{Eddington Luminosity} \label{sec:Edd}
The Eddington luminosity, $L_{\rm Edd}$, of an accreting BH can be calculated given its mass and the mean number of protons and neutrons for every electron in the ionized matter, $\mu$, that is falling onto its event horizon,
\begin{equation}\label{eq:Eddington}
    L_{\rm Edd} = \frac{4\pi c G \mu m_{\rm p}M_{\rm BH}}{\sigma_T} ~,
\end{equation}
where $G$ is the gravitational constant, $\sigma_T$ is the Thompson cross section and $m_{\rm p}$ is the proton mass.
The standard uncertainty of $L_{\rm Edd}$ is calculated via a root mean square of the uncertainties associated with $G$, $m_p$, $M_{\rm BH}$, and $\sigma_T$.
We assume that the infalling gas has a metallicity similar to that of our sun, making $\mu=1.17$ \citep{net13}.
Table \ref{tab:Ledd_result} lists $L_{\rm Edd}$ results for MJD 60251, calculated using the weighted average of $M_{\rm BH}$.

The Eddington ratio, $\lambda_{\rm Edd}$, can be estimated from $L_{\rm Edd}$ and the bolometric luminosity, $L_{\rm bol}$, of the quasar:
\begin{equation}\label{eq:Edd_ratio}
    \lambda_{\rm Edd} = \frac{L_{\rm bol}}{L_{\rm Edd}} \approx \frac{\langle L_{5100}\rangle}{L_{\rm Edd}}{\rm BC}.
\end{equation}
Here, $L_{\rm bol}$ can be determined by applying a correction factor, BC, to the average optical luminosity, $\langle L_{5100}\rangle$.
We use the bolometric correction estimate of ${\rm BC}=4.33\pm1.29$ from \citet{2013ApJS..206....4K} to calculate $L_{\rm bol}=(1.09\pm 0.32)\times 10^{47}$ erg s$^{-1}$. 
Table \ref{tab:Ledd_result} lists the Eddington ratios for MJD 60251, after applying the bolometric correction to $\langle L_{5100}\rangle$.
Using the weighted average of $M_{\rm BH}$ yields $L_{\rm Edd} = (2.430\pm0.096)\times10^{47}$ erg s$^{-1}$ 
which corresponds to an Eddington ratio of $\lambda_{\rm Edd}=0.447\pm0.134$.
Here, the resultant systematic uncertainty of $\delta\lambda_{Edd}=\pm0.04$ from the rms uncertainty of $L_{5100}$ is dwarfed by the uncertainty of the bolometric correction.
When SMC and CS reddening is removed, $L_{\rm Edd}$ and $\lambda_{\rm Edd}$ are $+15\%$ and $+23\%$ larger and thus provide upper estimates of these values.

\begin{table*}
\centering
\begin{tabular}{cccccccc}
    \hline
    Continuum & $\langle\rm FWHM_{H\alpha}\rangle$ & $\langle L_{5100}\rangle$ & $M_{\rm SL12}$ & $M_{\rm A11}$ & $\langle M_{\rm BH}\rangle$ & $L_{\rm Edd}$ & $\lambda_{\rm Edd}$ \\
    Model & (\kms{}) & ($10^{46}$ erg s$^{-1}$) & ($10^8M_\odot$) & ($10^8M_\odot$) & ($10^8M_\odot$) & ($10^{46}$ erg s$^{-1}$) & \\
    \hline
    red & $3059\pm76$ & $2.51\pm0.01$ & $17.47\pm0.94$ & $15.66\pm0.86$ & $16.52\pm0.65$ & $24.30\pm0.96$ & $0.45\pm0.13$ \\
    \hline
    non-red & $3059\pm76$ & $3.82\pm0.01$ & $20.63\pm1.05$ & $18.29\pm0.93$ & $19.40\pm0.72$ & $28.54\pm1.06$ & $0.58\pm0.17$ \\
    \hline
\end{tabular}
\caption{Eddington Luminosity calculations for the unsmoothed MJD 60251+60291 spectra.
The first column is the version of the SMC and CS continuum fitting models used to determine $\langle L_{5100} \rangle$, where `red' and `non-red' refer to the inclusion and exclusion of the reddening factor in each model.
The second column is the weighted-average $\rm FWHM_{H\alpha}$, determined from the NIR spectrum and optical+NIR spectrum measurements.
The third column is the weighted-average luminosity at \A{5100} determined from the SMC, CS, and PQF fitting models.
The fourth, fifth, and sixth columns are the BH mass estimates using our two methods and their weighted average.
The seventh and eighth columns are the calculated Eddington Luminosity and ratio.
}
\label{tab:Ledd_result}
\end{table*}

\subsection{Identification of Absorption Troughs} \label{sec:troughs}
\begin{figure}\centering\includegraphics[width=18cm]{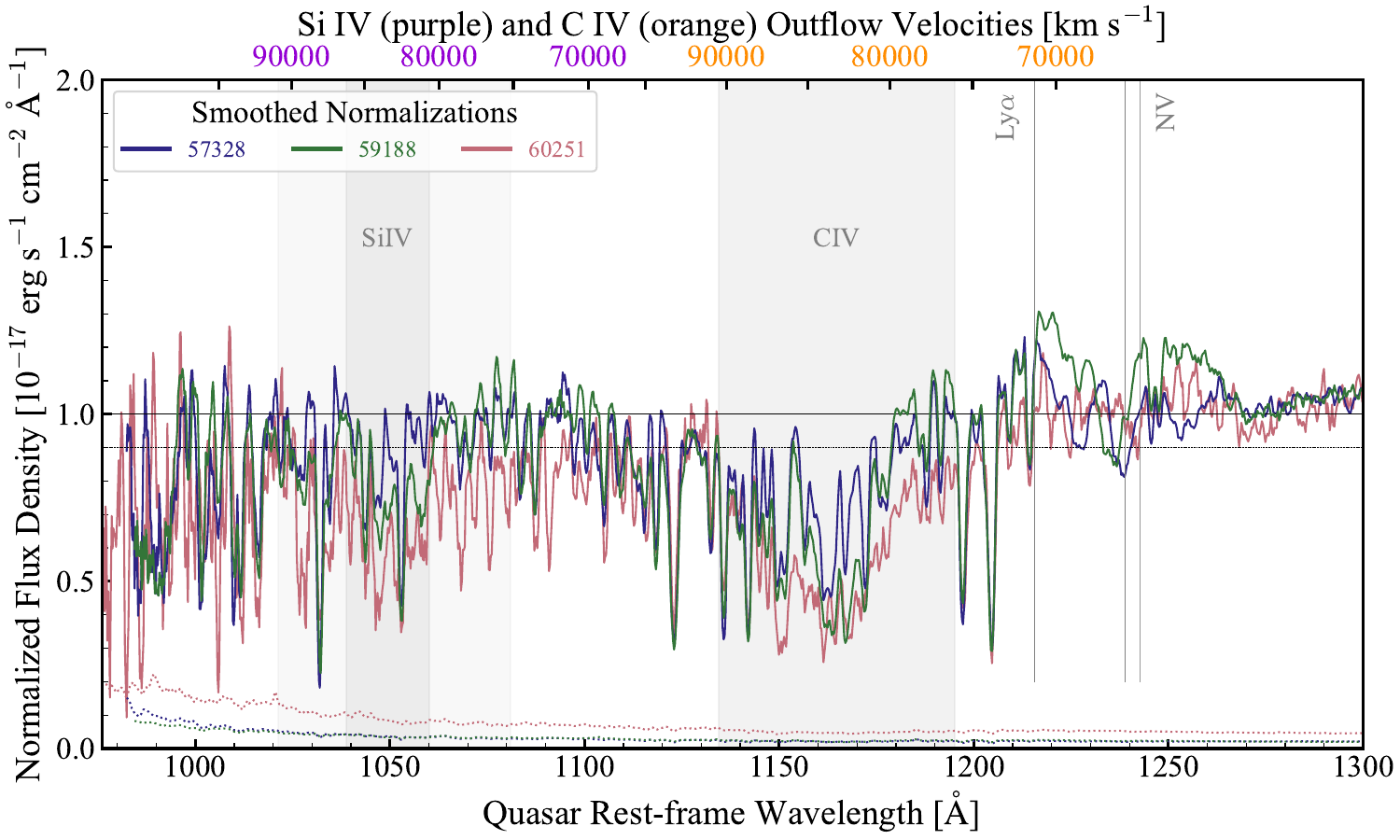}\\\includegraphics[width=18cm]{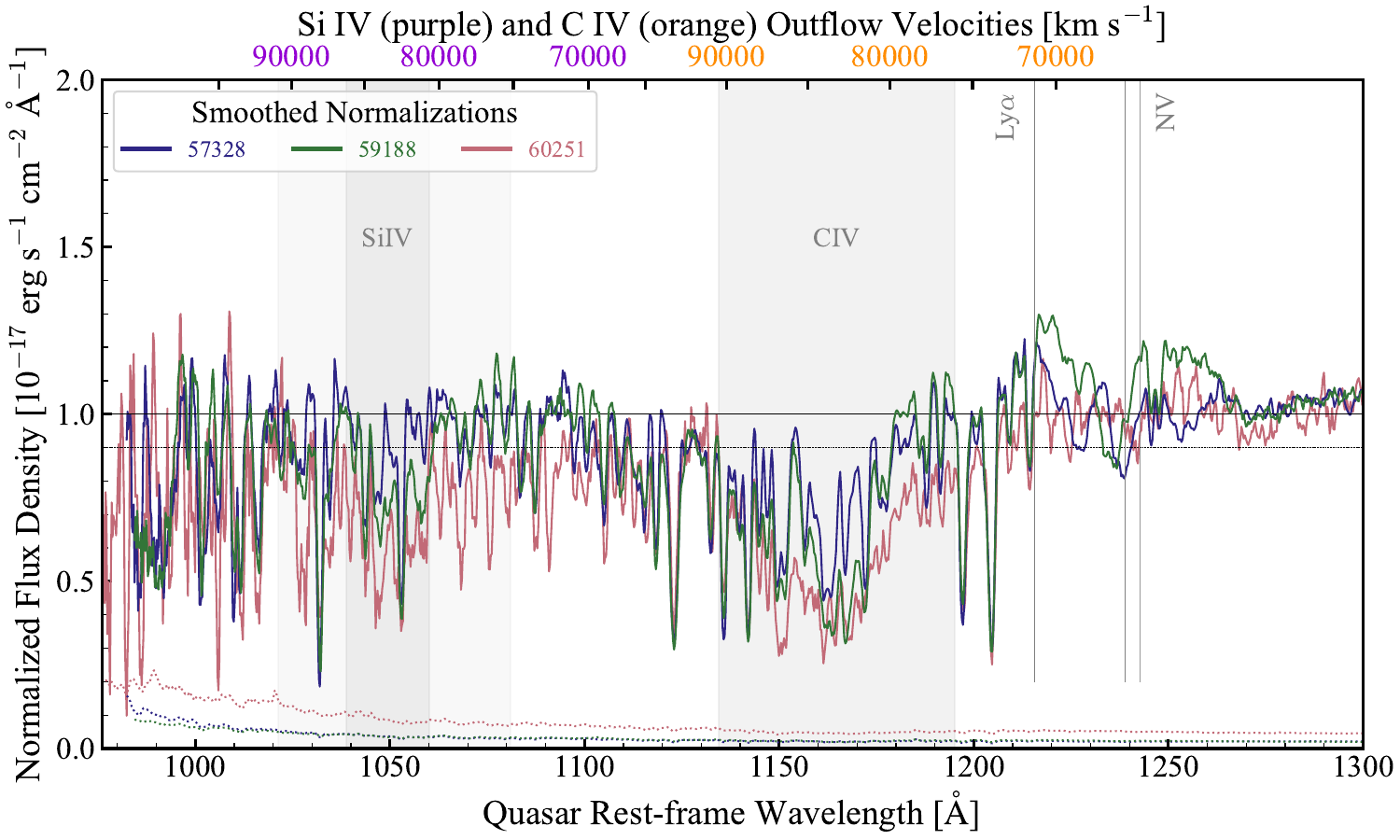}\caption{J2318 EHVO Zoom-In.
		J2318's \ion{C}{4} and \ion{Si}{4} BAL regions for the SMC (top panel) and CS (bottom panel) normalized spectra of MJD 57328 (blue), MJD 59188 (green) and MJD 60251 (red), with errors (dotted lines) provided in their respective colors. 
        All spectra have been smoothed by a 5-pixel boxcar. 
        The emission wavelengths of Ly$\alpha$ and \ion{N}{5} at the quasar redshift are marked by the vertical lines. 
        The velocity ranges used for direct integration of the \ion{Si}{4} and \ion{C}{4} BAL troughs are shown with the gray background. 
        The full velocity range spanned by \ion{C}{4} is shown for \ion{Si}{4} by the light gray background.
        Outflow velocities for the short-wavelength members of the \ion{C}{4} (orange) and \ion{Si}{4} (purple) doublets are marked along the upper axis every \kms{5000} from $70,000$ \kms{} to $95,000$ \kms{}.
        The flux of the normalized continuum is at 1.0 and flux levels that are less than 0.9 are typically where BAL troughs are defined. 
	}\label{fig:J2318_BAL_Zoom1}\end{figure}

Figure \ref{fig:J2318_BAL_Zoom1} shows the Ly$\alpha$ forest region of J2318, with BAL troughs indicated.

We identify the two troughs in the Ly$\alpha$ forest as troughs of \ion{Si}{4} and \ion{C}{4} in the same outflow from the quasar. 
The trough wavelengths matching \ion{C}{4} and \ion{Si}{4} at the same redshift support our identification. 
Those ions are commonly seen together in quasar outflows, with \ion{Si}{4} weaker (as in this case) or at most as strong as \ion{C}{4}; the only unusual factor in this case is the speed of the outflow.
Also, the fact that both troughs strengthened from 2015 to 2023 is most easily explained if they both arise in the same outflow.

Other possible identifications can be excluded.
As discussed in \cite{2018MNRAS.476..943H}, for the troughs to be identified as two independent Ly$\alpha$ troughs without accompanying metal-line absorption would require very low ionization parameter and column density values not seen among other BAL outflows and would make it difficult to explain the observed variability. 
Moreover, this identification would require two unprecedented  outflows along the same line of sight.

An identification of the stronger trough (\ion{C}{4} at $\lambda_{\rm cent}\simeq1165$ \AA) with \ion{N}{5} at $v=-19,000$ \kms{} would place \ion{P}{5} at the wavelength of the weaker trough (\ion{Si}{4} at $\lambda_{\rm cent}\simeq1050$ \AA). 
However, \ion{P}{5} is found in the same ionization conditions as \ion{C}{4} and is about a thousand times less abundant \citep{borgaravPV}. 
Therefore, the absence of \ion{C}{4} at $v=-19,000$ \kms{} rules out this possibility. 
Even the very high ionization outflow noted by \cite{tea98} has just as strong \ion{C}{4} absorption as \ion{N}{5}.

We searched for other typical ions in these outflows \citep[see][]{2011MNRAS.411..247R,2020ApJS..247...37A,2020ApJS..247...38X} and did not confidently find any. 
\ion{N}{5} and \ion{O}{6} would appear outside of our wavelength range, as would even higher-ionization lines such as are seen in X-ray UFOs with similar velocities. 
\ion{Mg}{2} and \ion{Al}{3}, if present, would lie on top of the \ion{Fe}{2} complex emission and the \ion{C}{4} emission line, respectively, and we do not detect them unambiguously: when we compare the wavelength regions of these absorbing ions we can confidently say that they are not showing as strong absorption or large variability as the \ion{C}{4} absorption does.

Finally, strong Ly$\alpha$ forest features outside the two troughs are consistent through the two observations, indicating that there is not a problem with the normalization of the entire Ly$\alpha$ forest region of the spectra that might give rise to spurious absorption troughs.

There is some apparent variability in the Ly$\alpha$ and \ion{N}{5} emission-line region over the three epochs of observation.
It is unclear if the cause is variable lower-velocity absorption, variable emission, or both. We defer study of these potential changes until further SDSS-V spectra of J2318 are analyzed.

\subsection{Absorption Trough Measurements Through Direct Integration} \label{sec:direct}
We first measured absorption rest-frame equivalent widths through direct integration (Table \ref{tab:direct}), although this approach will be contaminated by Ly$\alpha$ forest absorption, which we attempt to correct for in an average sense in \S\ref{sec:abs}.
Our reported EWs are from spectra normalized by the CS fits, as those fits have the smallest $\chi^2_\nu$ values.
To estimate systematic errors at each epoch, we used the rms scatter among those measurements, measurements from the spectra normalized by the SMC fits, and (for MJD 57328 and 59188) measurements from both the CS and SMC normalizations of those MJDs' spectra morphed to match the shape of the MJD 60251 spectrum (\S~\ref{sec:specmatch}).

We defined the edges of the \ion{C}{4} trough as the wavelengths in the MJD 60251 epoch at which the CS-normalized flux in the 5-pixel-boxcar-smoothed spectrum crossed 0.9 in the \ion{C}{4} outflow region, excluding 5.5 \AA\ at the long-wavelength end of this region due to a strong narrow Ly$\alpha$ absorber.
This trough extends from 1195.16\,\AA\ to 1134.56\,\AA, or $-76,370$ \kms{} (seen in \ion{C}{4} $\lambda$1550.774) to $-90,300$ \kms{} (seen in \ion{C}{4} $\lambda$1548.202), both uncertain by $\pm 40$ \kms{}.
Thus, the outflow had a velocity width of at least $13,930$ \kms{} at MJD 60251.
(Note that in the rest frame of \ion{C}{4} $\lambda$1548.202, used for plotting purposes, this trough extends from $-75,900$ \kms{} to $-90,300$ \kms{}.)
These velocity limits exclude a possible higher-outflow-velocity component reaching $v\simeq-96,000$ \kms{}; see the next section.

For the \ion{Si}{4} trough, we adopted two approaches.
First, we measured the EW in the same velocity range as seen in \ion{C}{4} on MJD 60251.
Second, we measured the EW between the \ion{Si}{4} trough edges: the wavelengths at which the the CS-normalized flux in the 5-pixel-boxcar-smoothed spectrum crossed 0.9 in the \ion{Si}{4} outflow region.
Given the wavelengths of both members of the \ion{Si}{4} doublet, that corresponds to 1038.8--1060.1~\AA, respectively.

For each trough at each epoch, we summed the difference between unity and the normalized flux in the unsmoothed spectrum times the pixel width (\kms{69}) between the appropriate wavelengths, then corrected for Ly$\alpha$ forest absorption.
The average Ly$\alpha$ forest absorption in J2318, measured between 1096--1134~\AA\ in the quasar rest frame, is $18.5\% \pm 3.8\%$ of our adopted continuum.
This number and its uncertainty are the average and rms scatter of all SMC- and CS-fit-normalized spectra at all epochs in that wavelength range.\footnote{We estimate the true Ly$\alpha$ forest absorption in the trough regions by dividing each overall Gaussian fit from \S \ref{sec:abs} into its normalized spectrum. 
In these divided spectra we measure the average Ly$\alpha$ forest absorption in the same velocity ranges used in Table \ref{tab:direct}. 
We find $15.3\%\pm2.9\%$ for the \ion{C}{4} region, $12.7\%\pm7.2\%$ for the matching \ion{Si}{4} region, and $16.3\%\pm9.1\%$ for the narrower \ion{Si}{4} region. 
Thus, our adopted correction of $18.5\%\pm3.8\%$ is conservative in the sense that it may underestimate the absorption (although it may also underestimate the systematic uncertainty for \ion{Si}{4}).}
Therefore, our \ion{C}{4} EW measurements are estimated to include \kms{2885\pm 593} of Ly$\alpha$ forest absorption, and our \ion{Si}{4} measurements include \kms{3166\pm650} or \kms{1149\pm236} for our first and second approaches, respectively.
We also recorded the average trough depth $d_{BAL}={\rm EW}_{\rm corrected}/(v_{start}-v_{end})$ and measured the centroid velocity $v_{cent}$ for each trough, defined as the weighted-average velocity within the trough limits, with the weight given by the depth below the normalized continuum \citep{jarthesis}.

Our direct-integration results are given in Table \ref{tab:direct}.
The corrected direct-integration \ion{C}{4} EWs, including first the statistical and second the systematic error estimates, are
$660\pm50\pm690$ \kms{}, $1680\pm50\pm870$ \kms{}, and $3730\pm120\pm620$ \kms{} for MJD 57328, 59188, and 60251, respectively.
Measuring the \ion{Si}{4}
trough within the \ion{C}{4} trough velocity limits has significantly larger uncertainties. 
The \ion{Si}{4} EWs measured using the \ion{Si}{4} trough velocity limits: $-580\pm60\pm330$ \kms{}, $230\pm60\pm490$ \kms{}, and $1100\pm140\pm240$ \kms{} for MJD 57328, 59188, and 60251, respectively.
The \ion{Si}{4} trough was not formally detected on MJD 57328; the absorption in the \ion{Si}{4} trough wavelength range in that epoch is consistent with Ly$\alpha$ forest absorption only.
There is a decrease in $v_{cent}$ for \ion{C}{4} in the MJD 60251 epoch, consistent with the observed trough depth increasing more at the lower-velocity end of the trough than at the higher-velocity end.

\begin{table}[ht]
\centering
\small
\begin{tabular}{cccccrrrrr}
\hline
\text{\parbox[t]{0.95cm}{\centering MJD}}
& \text{Trough} 
& \text{\parbox[t]{2.10cm}{\centering Adopted $v_{start}$ \\ (km s$^{-1}$)}} 
& \text{\parbox[t]{1.95cm}{\centering Adopted $v_{end}$ \\ (km s$^{-1}$)}} 
& \text{\parbox[t]{1.30cm}{\centering $v_{cent}$ \\ (km s$^{-1}$)}} 
& \text{\parbox[t]{1.35cm}{\centering EW \\ (km s$^{-1}$)}}
& \text{\parbox[t]{1.65cm}{\centering EW$_{\rm corrected}$ \\ (km s$^{-1}$)}}
& \text{\parbox[t]{1.55cm}{\centering $\sigma_{\rm EW}$, stat \\ (km s$^{-1}$)}} 
& \text{\parbox[t]{1.45cm}{\centering $\sigma_{\rm EW}$, sys \\ (km s$^{-1}$)}} 
& \text{$d_{BAL}$} \\ \hline
57328 & \ion{C}{4}  & $-$76,370 & $-$90,300 & $-$84,900 & 3550 &     660 &  50 & 690  & 0.048 \\ 
59188 & \ion{C}{4}  & $-$76,370 & $-$90,300 & $-$85,200 & 4570 &    1680 &  50 & 870  & 0.121 \\ 
60251 & \ion{C}{4}  & $-$76,370 & $-$90,300 & $-$84,000 & 6610 &    3730 & 120 & 620  & 0.267 \\ \hline 
57328 & \ion{Si}{4} & $-$76,370 & $-$90,300 &   ...    &  980 & $-$2180 & 100 & 990  & ... \\ 
59188 & \ion{Si}{4} & $-$76,370 & $-$90,300 &   ...    & 2230 &  $-$940 &  90 & 1400 & ... \\ 
60251 & \ion{Si}{4} & $-$76,370 & $-$90,300 &   ...    & 4640 &    1470 & 230 & 650  & 0.106 \\ \hline 
57328 & \ion{Si}{4} & $-$81,850 & $-$85,670 & $-$83,600 &  570 &  $-$580 &  60 & 330  & ... \\ 
59188 & \ion{Si}{4} & $-$81,850 & $-$85,670 & $-$83,100 & 1380 &     230 &  60 & 490  & 0.060 \\ 
60251 & \ion{Si}{4} & $-$81,850 & $-$85,670 & $-$83,400 & 2250 &    1100 & 140 & 240  & 0.289 \\ \hline 
\end{tabular}
\caption{Results of direct integration absorption measurements. The first three entries for each ion use the MJD 60251 \ion{C}{4} trough velocity limits. 
The second three entries for \ion{Si}{4} use the MJD 60251 \ion{Si}{4} trough velocity limits. 
The adopted $v_{start}$ is in the rest frame of the short-wavelength member of the relevant doublet, the adopted $v_{end}$ is in the rest frame of the long-wavelength member, and $v_{cent}$ is in the rest-frame of the oscillator-strength-weighted-average wavelength.
For \ion{C}{4}, the statistical uncertainties on $v_{cent}$ are between \kms{50} and \kms{100}, and the systematic uncertainties are about \kms{150}. For \ion{Si}{4}, the systematic uncertainties are similar but the statistical uncertainties are about twice as large.}
\label{tab:direct}
\end{table}

\subsection{Absorption Trough Measurements through Gaussian Fitting} \label{sec:abs}
We next performed Gaussian fitting of the \ion{C}{4} and \ion{Si}{4} doublets  to measure the broad absorption at each epoch. 
We fit both the SMC- and CS-normalized spectra.
To exclude Ly$\alpha$ forest lines from the fitting, a $\sigma$-clipping function developed by Joseph Choi is used (for more information, see Appendix A of \citealt{J1646}).
This function follows an iterative process of (1) smoothing the spectra using a fixed Gaussian kernel width, (2) flagging and removing points that are at a specified $\sigma$ below the smoothed spectra, and (3) repeating the process after interpolating between non-flagged points. 
The settings for $\sigma$-clipping are kept consistent across all observations and spectral versions, but whenever visual inspection suggested the omission of some additional regions, they are excluded manually and uniquely for each spectrum. 
We minimized the number of Gaussian doublets used, only adding a new doublet when the fit was improved significantly. 
We performed an $F$-test using the same method described in \S\ref{sec:redshift}. 
We found that two, three, and two doublets were the most ideal fits for \ion{C}{4} absorption in observations MJD 57328, 59188, and 60251, respectively: these fittings resulted in $F_\chi$ in the range 18 - 30, higher than the critical values $\sim$5.7 (indicating a probability $<0.01\%$ that the doublets improved the fit by chance); an additional doublet resulted in $F_\chi$ values lower than the critical values.

\begin{figure}
    \includegraphics[width=0.49\linewidth]{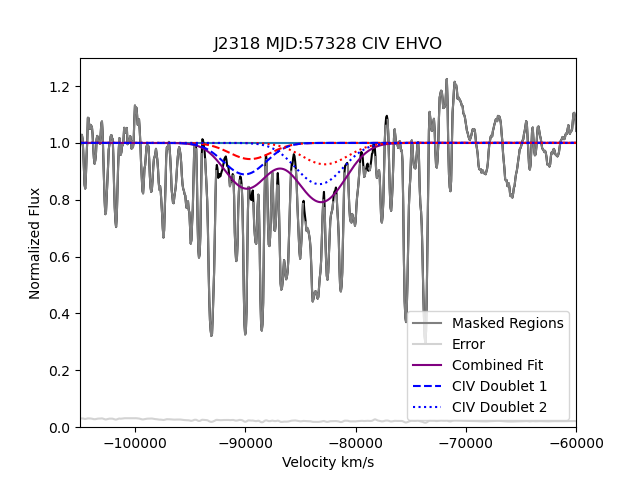}\includegraphics[width=0.49\linewidth]{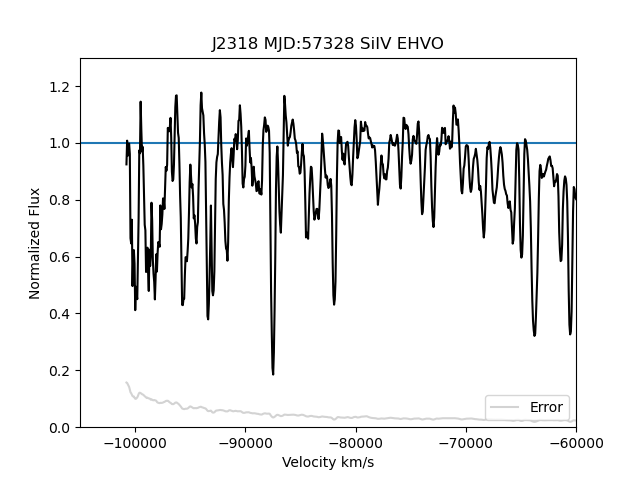}
    \centering
    \includegraphics[width=0.49\linewidth]{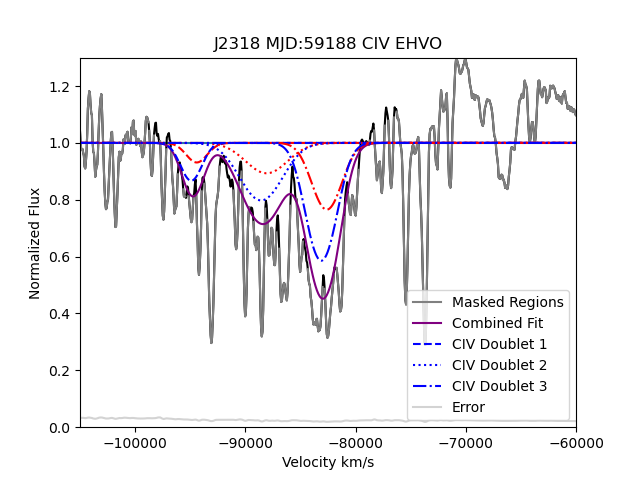}
    \includegraphics[width=0.49\linewidth]{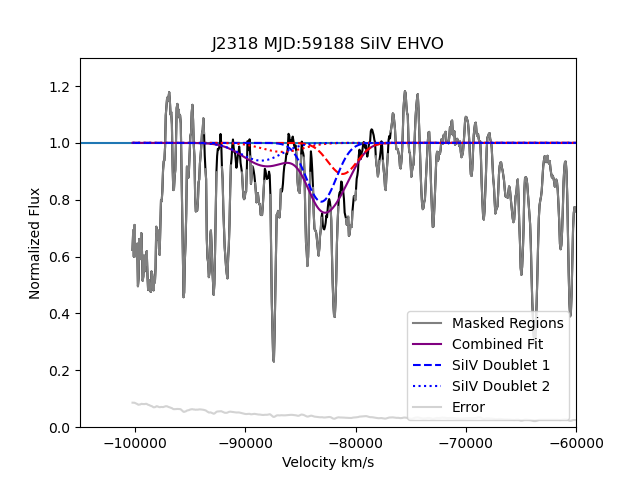}
    \includegraphics[width=0.49\linewidth]{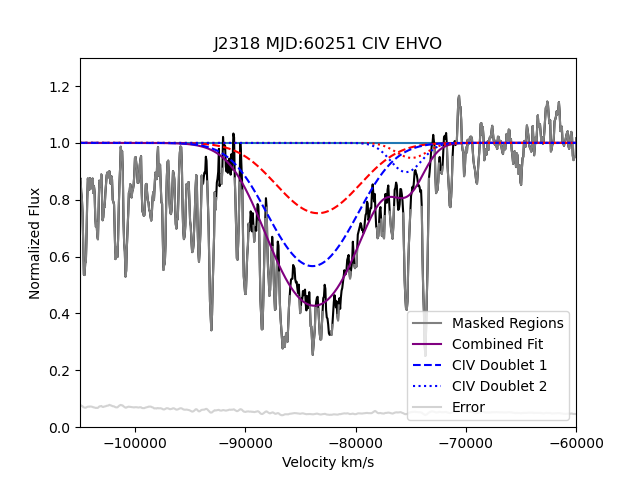}
    \includegraphics[width=0.49\linewidth]{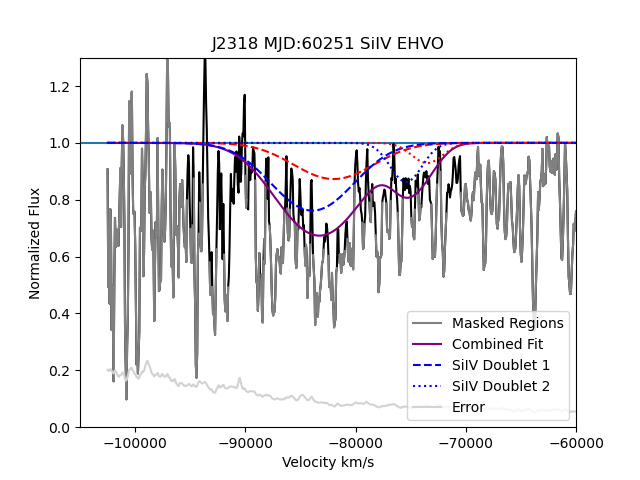}
    \caption{Fits to the absorption trough in CS-normalized, smoothed spectra, with \ion{C}{4} in the left column and \ion{Si}{4} in the right column. MJDs run from top to bottom. The horizontal axis is the outflow velocity for the short-wavelength member of each doublet transition.
    Fits are performed using doublets: two Gaussian curves represented by blue and red dashed/dotted lines. 
    }
    \label{fig:placeholder0}
\end{figure}

Figure \ref{fig:placeholder0} shows the fits to the CS-normalized spectra and Table \ref{tab:abs_measurements} shows the integrated absorption measurements performed, including Balnicity Index ($BI_{EHVO}$), maximum velocity ($v_{max}$), minimum velocity ($v_{min}$), and EW.
We measured the Balnicity Index \citep[BI;][]{wea91} modified for EHVO troughs ($\rm BI_{\rm EHVO}$) by setting the integral limits to account for absorption between $70,000$ \kms{} and $100,000$ \kms{}, but keeping the parameter $C$ to account for absorption larger than \kms{2000} (see \citealt{EHVO} for more details).
Velocity limits $v_{min,0.9}$ and $v_{max,0.9}$ were measured at the start and end of the absorption troughs where the absorption is below 90$\%$ the normalized flux. 
We also calculated the velocity limits at 99$\%$ of the normalized flux level ($v_{min,99}$ and $v_{max,99}$). 
EW values represent the integrated absorption using the full range of velocities at either the 90\% or 99\% threshold. 
Errors were measured by shifting the spectrum up and down by 50$\%$ of the error array values, refitting the absorption, and comparing to original-fit measurements.
In Table \ref{tab:abs_measurements} the top errors come from shifting the spectrum downward, and the bottom errors come from shifting upward.
When shifting the spectra upwards and using the original BI we used a value of \kms{450} instead of the regular \kms{2000} to avoid absorption being undetected; some derived errors are quite large due to a whole absorption trough not being detected anymore. 
Two exceptions to this methodology include: (1) at MJD 57328 the \ion{Si}{4} absorption is undetected in the fit (see Figure \ref{fig:placeholder0}), so only upper limit measurements are provided for $v_{min,99}$, $v_{max,99}$, and EW at the 99$\%$ threshold, and (2) at MJD 60251 the \ion{Si}{4} absorption bottom errors reflect that the absorption is undetected when the spectrum is shifted upwards: the errors are the measured values for BI and EW and are NaN for the velocities.

\begin{table*}[]
    \centering
    \small
    \renewcommand{\arraystretch}{1.5}
    
\begin{tabular}{ccccccccc}
\hline
\text{\parbox[t]{1cm}{\centering MJD}} & \text{\parbox[t]{1cm}{\centering Trough}}  & \text{\parbox[t]{1.3cm}{\centering $BI_{EHVO}$ \\ (km s$^{-1}$)}} & \text{\parbox[t]{1.3cm}{\centering $v_{max, 0.9}$ \\ (km s$^{-1}$)}} & \text{\parbox[t]{1.3cm}{\centering $v_{min, 0.9}$ \\ (km s$^{-1}$)}} & \text{\parbox[t]{1.3cm}{\centering $EW, 0.9$ \\ (km s$^{-1}$)}} \    & \text{\parbox[t]{1.3cm}{\centering $v_{max, 0.99}$ \\ (km s$^{-1}$)}} & \text{\parbox[t]{1.3cm}{\centering $v_{min, 0.99}$ \\ (km s$^{-1}$)}} & \text{\parbox[t]{1.3cm}{\centering $EW, 0.99$ \\ (km s$^{-1}$)}} \\\hline

57328 & CIV & \ensuremath{160^{+160}_{-140}} & \ensuremath{-91,000^{-800}_{+6000}} & \ensuremath{-81,200^{+500}_{-500}} & \ensuremath{330^{+190}_{-140}} & \ensuremath{-93,600^{-1200}_{+1200}} & \ensuremath{-79,000^{+700}_{-700}} & \ensuremath{1300^{+400}_{-500}} \\
 & SiIV & - & - & - & - & $>-84,814$ & $<-81,884$ & $<29$ \\
59188 & CIV & \ensuremath{1800^{+400}_{-400}} & \ensuremath{-95,400^{-400}_{+600}} & \ensuremath{-80,840^{+130}_{-130}} & \ensuremath{2300^{+400}_{-400}} & \ensuremath{-97,400^{-500}_{+600}} & \ensuremath{-79,620^{+190}_{-190}} & \ensuremath{3600^{+500}_{-500}} \\
 & SiIV & \ensuremath{200^{+200}_{-200}} & \ensuremath{-84,700^{-500}_{+400}} & \ensuremath{-82,300^{+400}_{-600}} & \ensuremath{300^{+200}_{-200}} & \ensuremath{-90,000^{-1400}_{+4000}} & \ensuremath{-80,350^{+190}_{-190}} & \ensuremath{800^{+500}_{-300}} \\
60251 & CIV & \ensuremath{3000^{+1000}_{-900}} & \ensuremath{-89,300^{-900}_{+800}} & \ensuremath{-75,000^{+1000}_{-800}} & \ensuremath{4000^{+1000}_{-900}} & \ensuremath{-92,100^{-1100}_{+1000}} & \ensuremath{-73,000^{+1400}_{-1000}} & \ensuremath{4700^{+1100}_{-1000}} \\
 & SiIV & \ensuremath{200^{+1400}_{-200}} & \ensuremath{-86,000^{-1800}_{+NaN}} & \ensuremath{-82,000^{+7000}_{-NaN}} & \ensuremath{300^{+1400}_{-300}} & \ensuremath{-90,200^{-900}_{+NaN}} & \ensuremath{-73,600^{+700}_{-NaN}} & \ensuremath{1300^{+1700}_{-1300}} \\
\hline \end{tabular}
    \caption{Results of Gaussian fitting \ion{C}{4} and \ion{Si}{4} absorption measurements on unsmoothed CS-normalized spectra. Balnicity index (BI), maximum velocity ($v_{max}$), minimum velocity ($v_{min}$), and equivalent width (EW) are given for all three observations of J2318 measured at 90\% and 99\% of the normalized flux. $v_{max}$ is calculated using the long wavelength reference while $v_{min}$ uses the short wavelength reference. See Section \S\ref{sec:abs} for error constraining method and exceptions.
    }
    \label{tab:abs_measurements}
\end{table*}

\section{Discussion} \label{sec:dis}
We have detected \ion{C}{4} and \ion{Si}{4} absorption with large EWs at outflow speeds up to $v\sim-0.3c$ in the $z=2.6781$ quasar J2318. 
This is the fastest \ion{Si}{4} outflow ever detected, and the second-fastest \ion{C}{4} outflow after PDS 456.
An important open question is whether the extreme properties of J2318 arise primarily from orientation effects or from a distinct physical state of the accretion flow. 
While orientation alone cannot be excluded, the combination of weak UV emission lines, pronounced spectral and photometric variability, and the presence of an extremely high-velocity outflow suggests that J2318 may represent a peculiar accretion state rather than a purely geometrical viewing effect.

It is difficult for current simulations to reproduce how outflows accelerate such relatively low-ionization gas to highly supersonic speeds 
\citep[e.g.,][and references therein]{2020MNRAS.499.2173B} or entrain and accelerate it to the wind velocity \citep[e.g.][]{2020MNRAS.491.4325Z}.
Formation of low-ionization gas in situ in outflows via cooling and shielding is one explanation for the observations (e.g., \citealt{2022ApJ...931..134Waters}; \S 4.3.1 of \citealt{2024MNRAS.528.6496H}). 
Another is that the gas avoids overionization during acceleration by being highly clumped and dense (e.g., \citealt{2025MNRAS.541.2393M}; see also \S~\ref{sec:SS433} below), although how the clumps would maintain their integrity is unclear.

Independent of how the gas is accelerated, in \S \ref{sec:mass} we estimate the mass outflow rate in the observed high-velocity outflow and its potential feedback on the surrounding host galaxy.
Then, in \S \ref{sec:variability} we explore likely drivers if the outflow and compare J2318 to other quasars with broad absorption in \S \ref{sec:compare}.

\subsection{Constraints on absorbing column density, mass outflow rate, kinetic luminosity, and distance} \label{sec:mass}
From the observed UV troughs of \ion{C}{4} and \ion{Si}{4}, we cannot precisely determine the mass in the outflow. 
However, we can place a firm lower limit on the outflow mass by conservatively assuming unsaturated absorption and 100\% covering of the continuum source by the outflow; doing so yields the minimum optical depth in the outflow. 
We note that the absorption may instead be highly saturated, in which case the actual outflow mass will be much higher than our lower limit (e.g., \citealt{borgaravPV,2018ApJ...857...60A}).

A given transition's column density in cm$^{-2}$ is
$N=\frac{3.7679\times 10^{14}}{\lambda f} \langle\tau\rangle \Delta V$ where $\Delta V$ is the width of the absorption trough, $\langle\tau\rangle$ is the average optical depth in the trough, $\lambda$ is the wavelength in \AA\ of the absorption line, and $f$ is the dimensionless oscillator strength of the line.
We treat doublet transitions as a single transition using the sum of the oscillator strengths and the oscillator-strength-weighted-average wavelength.
For \ion{C}{4}, $\lambda=1549.06$ and $f=0.286$.
For \ion{Si}{4}, $\lambda=1396.75$ and $f=0.768$.
We adopt $\Delta V=13,930$ \kms{} as measured via direct integration on MJD 60251
(\S \ref{sec:direct}).
Assuming unsaturated absorption, our MJD 60521 average trough depth measurements $d_{BAL}$ become $\langle\tau\rangle=0.311$ for \ion{C}{4} and $\langle\tau\rangle=0.112$ for \ion{Si}{4}.

Our estimate of the hydrogen column density of the outflow, $N_H$, is given by $N_H = {\rm max}(N_{\rm CIV}/f_{\rm CIV}A_{C},N_{\rm SiIV}/f_{\rm SiIV}A_{Si})$ where $f_{\rm CIV}$ is the fraction of C in ionization stage IV in the outflow, $A_C$ is the fractional abundance of C relative to H, and similarly for Si.
We adopt solar abundances of $A_C=10^{-3.57}$ and $A_{Si}=10^{-4.49}$ \citep{2009ARA&A..47..481A}.
To ensure a lower limit on $N_H$ we adopt $f_{\rm CIV}=f_{\rm SiIV}=0.5$, the largest values seen in the calculations of \citet{ham97}.\footnote{As noted by \citet{2020A&A...633A..55L}, special relativitistic deboosting reduces the ionizing luminosity seen by outflowing gas and therefore its ionization level, thus increasing the column density $N_H$ needed to produce an observed level of ionized absorption. 
From their Figure 3, we estimate that this could increase the minimum $N_H$ by a factor of two in J2318.}

This approach produces $N_{\rm CIV}=3.68\times 10^{15}$ cm$^{-2}$ and $N_{\rm SiIV}=5.48\times 10^{14}$ cm$^{-2}$, corresponding to $N_H\geq2.73\times 10^{19}$ cm$^{-2}$ and $N_H\geq3.39\times 10^{19}$ cm$^{-2}$, respectively. 
The latter is our lower limit on $N_H$.

If we approximate the outflow as a shell of solid angle $4\pi\Omega$ and uniform hydrogen density $n_H$ at radius $R$ with thickness $\Delta R \ll R$ and volume filling factor $f_{\rm vol}$, then $M = f_{\rm vol} n_H \mu m_p 4\pi\Omega R^2 \Delta R$ where $\mu=1.4$ is the mean atomic mass per hydrogen nucleus. 
We assume $f_{\rm vol}=1$ because we are ignorant of the internal structure of the outflow, but see the discussion in \S~\ref{sec:Xray} and \S~\ref{sec:feedback}. 
Since $N_H = n_H \Delta R$, we can write $M = N_H \mu m_p 4\pi\Omega R^2$.
To convert this approach to an average outflow mass-loss rate, we assume the outflow has traveled at constant observed velocity $v$ for time $t$ to reach a distance $R=vt$ and therefore that $\langle\dot{M}\rangle = M/t = N_H \mu m_p 4\pi\Omega R v$.
We adopt $\Omega=0.2$ for consistency with previous work (e.g., \citealt{dunnarav10}).
However, in the similar radio-quiet EHVO quasar PDS 456 ($z=0.184$), analysis of the P-Cygni profile of the Fe K region indicates that the X-ray-absorbing outflow covers the entire source \citep{2025Natur.641.1132XRISM.PDS456}.
We write $R=R_{pc}/{\rm (1~pc})$ where $R_{pc}$ is the outflow distance in units of parsecs, so that our results can be easily scaled for different distances.
We adopt $v=84,000$ \kms{} (Table \ref{tab:direct}).

For the above values, we have $\langle\dot{M}\rangle = (2.42 \times 10^{-21}) (\Omega/0.2) N_H R_{pc}$ M$_\odot$/yr for $N_H$ in units of cm$^{-2}$.
With our lower limit $N_H=3.39\times 10^{19}$ cm$^{-2}$, the mass-loss rate will be $\dot{M} > 0.82 ~M_\odot~ {\rm yr}^{-1}$ for $R_{pc}=10$ (an outflow distance of 10~pc). 
The corresponding kinetic luminosity lower limit is $L_K=\frac{1}{2}\dot{M}v^2>1.82\times 10^{45}$ erg s$^{-1}$, as compared to an estimated $L_{\rm Edd}=2.43\times 10^{47}$ erg s$^{-1}$ for J2318.
The ratio of the two luminosities $L_K/L_{\rm Edd}$ is $>$0.75\%, above the 0.5\% threshold for significant feedback found by \cite{he10}.  
We conclude that the outflow in J2318 will have significant feedback effects on the host galaxy. We discuss this topic further in \S \ref{sec:feedback}.

The study of EHVOs is relatively new, with at least 40 quasars exhibiting UV/optical EHVOs \citep{EHVO} travelling at speeds $0.1c<v<0.2c$, and 97 more found in SDSS DR16 (R. Candelaria-Stoner in prep.). \
We do not have any direct measurements for J2318 beyond the UV, making our calculated values, by necessity, uncertain. 
For the following distance measurements we contrast J2318 to J1646, which has an UV/optical EHVO that was extensively analyzed by \citealt{J1646}.

To estimate constraints on the distance of the outflow from the central SMBH, we adopt two methods.
First, to determine the minimum distance $R$, we use the definition of the ionization parameter, $U = Q / 4\pi c R^2 n_H$, where the number of photoionizing photons per second emitted from the central engine, $Q$, is estimated through analysis of the quasar’s bolometric luminosity, while the density of the gas and the ionization parameter $U$ are determined from spectral analysis, following \citet{J1646}.
We scale the bolometric luminosity of J2318 to that of J1646 to estimate a value of $\log Q=56.99$ (photon s$^{-1}$), and assume $\log U=-0.7$ \citep{J1646}.
Previously, the hydrogen density was $\log n_H = 8.43$ for our assumed $\log U=-2$ and $\log R = 10$ pc, so with our new value of $U$ we assume a larger hydrogen density of $\log n_H = 9.73$.
Using the new values yields a minimum outflow distance of $R\gtrsim0.5$ pc.

For the maximum distance we assume that an energy-driven BAL outflow, that may be radiatively driven, cannot transport energy greater than the quasar’s radiative power.
With this assumption, we follow the approach outlined in \citet{2024ApJ...970....9B}, where we equate the kinetic luminosity to the bolometric luminosity and isolate for the radius, $R_{pc} = 2L_{\rm bol} / 4\pi\mu m_pv^3\Omega N_H$. 
Since we assume that all of the bolometric luminosity is converted into kinetic luminosity, for the maximum distance estimate we adopt a covering fraction of $\Omega = 1$.
Inputting all our values yields a maximum outflow distance of $R\lesssim120$ pc.
Thus, we constrain the outflow's distance to lie between $0.5\lesssim R\lesssim120$ pc.

\subsection{Outflow Variability Drivers} \label{sec:variability}
The spectra of J2318 reveal an outflow seen in two ions, \ion{C}{4} and \ion{Si}{4}, traveling along our line of sight at $v\lesssim0.3c$.
Both troughs are measured to have similar velocity ranges and relative redshifts to their host quasar.
\ion{C}{4} absorption in BAL quasars is common, and due to its larger abundance and higher ionization levels can be the only observed ion within the absorbing gas \citep[e.g.,][]{2014ApJ...791...88F}. 
However, the lower abundances and ionization levels of \ion{Si}{4} prevent it from being solely observed; when found, it is seen at similar velocities of the  
\ion{C}{4} absorber \citep[e.g.,][]{ham97,2014ApJ...791...88F}.
Thus, the \ion{Si}{4} absorption must arise from the same gas producing the \ion{C}{4} outflow.

The \ion{C}{4} trough is easily distinguishable in all three spectra, whereas the \ion{Si}{4} trough is not distinguishable from the noise of the Ly$\alpha$ forest in MJD 57328, but is distinct in MJD 59188 and MJD 60251 (Figure \ref{fig:J2318_BAL_Zoom1}).
As seen in Figures \ref{fig:J2318_rawSDSS} and \ref{fig:J2318_Mag}, the continuum flux increases (most notably for rest-wavelengths of $\lesssim$\A{1900}) from MJD 57328 to 59188 at which time the \ion{Si}{4} trough appears in the spectrum and the already present \ion{C}{4} trough becomes stronger.
Between MJD 59188 and 60251 the continuum flux decreases to lower than its first epoch (though the ZTF photometry shows less of a flux decrease than the SDSS spectrophotometry) and both troughs have $\rm\Delta EW> 0$.
In the same time span, both troughs undergo an increase in their width to include lower-velocities, while the higher-velocity portion of their troughs deepens (Figure \ref{fig:placeholder0}).
This coordinated variability across the troughs, along with a changing continuum flux density, can be attributed to changes in the ionizing continuum flux \citep[e.g.,][]{2012ApJ...757..114F,2013ApJ...777..168F,2015ApJ...806..111G,wywf,DeCicco+2018,2022SciA....8.3291H}.
When the ionizing flux that can create or destroy \ion{C}{4} increases, it is expected that the \ion{C}{4} BAL troughs will get weaker.
However, since the ionizing flux doesn't vary identically to the rest-frame $1000$-\A{2000} continuum, the \ion{C}{4} BAL troughs are not always seen to get weaker when the continuum gets stronger within this rest-wavelength interval (e.g., Figure 7 of \citealt{wywf}).
Ionization simulations are beyond the scope of this paper due to the limitations of our data. 
These simulations are more useful for other objects that have more ions available for absorption studies. 
As stated earlier (Section \ref{sec:troughs}), higher ionization transitions appear in the far-/extreme-UV, which requires new-generation X-ray observations for J2318.

\subsubsection{Transverse Motion}\label{sec:transverse_motion}

Variability in BAL troughs could also be driven by transverse motion across our line of sight of outflows that block all or some portion of the continuum emitting region \citep[e.g.,][]{Vivek+2012,2013MNRAS.429.1872C,Yi+2022}.
The transverse velocity of an outflow across our line of sight can be determined by dividing the distance travelled by the time elapsed, $\Delta t_{rest}$. 
To calculate the distance we need to determine the relative size of the continuum emitting region at \A{1100} and the outflow.
Following the method of \cite{J0230}, we find that a diameter of $D_c=10.9$ light-days includes 95\% of the emission at that wavelength.
If we assume that the variability in a BAL is due entirely to an optically thick cloud traversing our line of sight, then the change in covering fraction, $\Delta C$, is equal to the change in the depth of the trough, $\Delta d_{bal}$, during the time between observations.

We consider the time elapsed between successive observations, and the first and last epoch since it shows the most dramatic change. 
This is especially true for \ion{Si}{4} that shows no absorption in the first epoch. 
Here, we assume that the absorbing cloud showed up between the first two observations and fully established itself by the third epoch. 
The minimum and maximum distances and velocities of the outflow are estimated by assuming the cloud has occulted more of the disk or it has traversed the entire disk. 
Minimum distance and velocity are calculated via $d_{min}=\sqrt{\Delta C D_c}$ and $v_{min}=c\sqrt{\Delta C}D_c/\Delta t_{rest}$, respectively, and their maximum counterparts are calculated as $d_{max}=\Delta C D_c$ and $v_{max}=c\Delta C D_c/\Delta t_{rest}$, respectively. 
Table \ref{tab:transverse_motion} lists our results for the \ion{C}{4} and \ion{Si}{4} BALs.

\begin{table*}
\centering
\begin{tabular}{ccccccc}
    \hline
    Ion & Epochs & $\Delta t_{rest}$ & $\Delta d_{bal}$ & ${v_{max}}^{(a,b)}$ & ${v_{min}}^{(a)}$ & ${v_{min}}^{(b)}$ \\
     & & (days) & & (\kms{}) & (\kms{}) & (\kms{}) \\
    \hline
    \ion{C}{4} & 1 to 2 & 506 & 0.073 & 6480 & 1750 & 473 \\
    \ion{C}{4} & 2 to 3 & 289 & 0.146 & 11,300 & 4330 & 1650 \\
    \ion{C}{4} & 1 to 3 & 795 & 0.219 & 4120 & 1930 & 902 \\
    \hline
    \ion{Si}{4} & 1 to 2 & 506 & 0.060 & 6480 & 1590 & 389 \\
    \ion{Si}{4} & 2 to 3 & 289 & 0.229 & 11,300 & 5420 & 2600 \\
    \ion{Si}{4} & 1 to 3 & 795 & 0.289 & 4120 & 2220 & 1190 \\
    \hline
\end{tabular}
\caption{Calculated transverse velocities for an outflow that appears (a) smaller than the emitting region or (b) larger than the emitting region. This assumes that the change in trough depth, $d_{bal}$, over the elapsed rest-frame time, $\Delta t_{rest}$, is due to a circular cloud occulting the continuum region with a diameter of $D_c$.}
\label{tab:transverse_motion}
\end{table*}

If we assume that the absorber is much larger than the continuum region the distance travelled by the outflow is $d_{min}=\Delta C D_c$, and the minimum velocity is lowered due to the cloud requiring slower speeds to achieve the same change in trough depth (Table \ref{tab:transverse_motion}).

The absorption troughs deepen across our spectroscopic observations, so if a moving cloud is responsible it is occulting more of the disk and has not moved out of our line of sight. 
If the flux is inhomogeneous and akin to that of a SS73 accretion disk, where the greater concentration of flux is near the centre, then changes in the trough depth could help determine the cloud’s relative size. 
A larger cloud would increase its covering fraction over time resulting in the deepening of the BALs, until it moves out of our line of sight thereby weakening the troughs. 
A smaller cloud would occult more of the flux as it approaches the disk’s center, strengthening the troughs, and cover less of the flux as it moves away from the disk’s centre, weakening the troughs.

For J2318 both scenarios are plausible. 
A larger cloud would explain the observed BAL variability, whereas a smaller cloud would be partially through its journey since there is no evidence of the troughs weakening. 
Newer observations, coupled with their respective velocity and trough depth changes, would help in understanding the relative size of the cloud. 
For similar velocities, if the troughs were to weaken we could conclude that the cloud was smaller and moving away from the disk’s center; if the troughs were to further strengthen then we could conclude that the cloud was bigger. 
If the troughs were to further strengthen and then weaken the smaller cloud would be traversing the centre of the disk before travelling towards the disk’s edge.

Since \ion{Si}{4} is always seen with \ion{C}{4}, a possible explanation is that a flowtube of infinite extent \citep[e.g.,][]{J0230} travels across our line of sight, with a spatially dependent column density profile, due to changes in the cloud’s extent along the line of sight, or density, or combination of the two.
Here, the inferred column density in the part of the flow tube along our line of sight increases with time, resulting in more absorption in \ion{C}{4} and new absorption in \ion{Si}{4}.
This flowtube would be similar to that of a relatively larger cloud: the troughs would strengthen until the flowtube fully established itself after which time the depth of the troughs would remain constant. 
With only three spectroscopic observations, our data set is too small to adequately simulate this scenario. 
A simulation would be needed to determine what flowtube velocities, widths, and distances from the BH are required to match the depth of the BALs for each successive observation. 
However, since the BALs observed in J2318 are still deepening, we can say that a flowtube would be in the beginning stages of establishing itself.

The variability seen in the BAL troughs could also be explained by a combination of ionization changes and tangential motion of the gas \citep[e.g.,][]{J0230,2017arXiv170503019M,2019ApJ...872...21H}.
With future observations and subsequent analysis of BAL variability, we hope to narrow down what the most likely driver is.

\subsection{Comparison to similar quasars} \label{sec:compare}
It is instructive to compare J2318 and its outflow to previously studied BAL and EVHO quasars and their outflows.

\citet{2014ApJ...791...88F} have studied the properties of BAL quasar outflows in which absorption is seen in only \ion{C}{4} (denoted \ion{C}{4}$_{\rm 00}$), in both \ion{C}{4} and \ion{Si}{4} (\ion{C}{4}$_{\rm S0}$), and in those transitions plus \ion{Al}{3} $\lambda\lambda1854,1862$ (\ion{C}{4}$_{\rm SA}$). 
As mentioned in Section 3.8, J2318 has no obvious \ion{Al}{3} absorption trough.
In comparison to the \ion{C}{4}$_{\rm S0}$ subsample's average BAL-trough value properties of \citet{2014ApJ...791...88F}, J2318 has a $\sim$50\% larger EW and trough width, but a 43\% smaller average depth (their Table 8). 
J2318 is also an outlier in terms of its two-epoch $\rm \Delta EW/\langle EW\rangle$ (their Figures 10 and 11).
For \ion{C}{4}, measured in (rest-frame) \AA\ for comparison, J2318 had $\rm \Delta EW/\langle EW\rangle =14.8/17.0 =0.87$ between MJD 57328 and 59188, and $\rm \Delta EW/\langle EW\rangle =29.6/39.2 =0.76$ between MJD 59188 and 60251; for \ion{Si}{4}, J2318 had $\rm \Delta EW/\langle EW\rangle =12.7/9.7 =1.31$ between MJD 59188 and 60251.
These $\langle {\rm EW}\rangle$ values are larger than any values in \citet{2014ApJ...791...88F}, and the latter $\rm \Delta EW/\langle EW\rangle$ value is larger than any for $\langle EW\rangle > 20$ \A{}.
Nonetheless, J2318 falls on the trend line seen in their Figure 14b relating variations in \ion{C}{4} and \ion{Si}{4}, with $\rm \Delta EW/\langle EW\rangle=0.76$ for \ion{C}{4} and 1.31 for \ion{Si}{4}. 

The comparison between $\rm\Delta EW_{C~IV}$ and $\rm\Delta EW_{Si~IV}$ can only be made between MJD 59188 and 60251.
The point of (12.7, 29.6) is above the trend line seen in their Figure 14a, however, the scatter of their $\rm\Delta EW_{C~IV}$ also tends to fall above the trend line for $\rm\Delta EW_{Si~IV}>2$.
Using their Bayesian linear-regression model (their equation 2), we calculate expected values of $\rm\Delta EW_{C~IV}=15.29\pm1.06$ for $\rm\Delta EW_{Si~IV}=12.7$ and $\rm\Delta EW_{Si~IV}=24.62\pm1.69$ for $\rm\Delta EW_{C~IV}=29.6$. 
Our measured values are $\sim50\%$ larger and are not in agreement, within errors, of the model's results nor the intrinsic scatter of $\sigma_{\rm IS}=3.4$.
For MJD 57328 to 59188, we do a similar approach to find an expected value of $\rm\Delta EW_{Si~IV}=12.29\pm0.86$ for $\rm\Delta EW_{C~IV}=14.8$.

We can place J2318 in the \ion{C}{4} emission-line blueshift/EW space studied by \cite{EHVORHR} using our measurements from \S~\ref{sec:wlq}. J2318 is an outlier at low emission-line EW (log$_{10}$(\ion{C}{4}~EW=0.7) relative to both BAL and EHVO quasars. Its \ion{C}{4} emission-line blueshift (\kms{3760\pm 150}) is large for BAL quasars but is unremarkable for EHVO quasars. 

In terms of outflow comparisons, first is the SDSS sample of \cite{jarthesis}, which includes outflows at velocities reaching $-60,000$ \kms{} (see their Figures 4 through 7).
Despite J2318's larger centroid velocity of $-84,000$ \kms{}, its average corrected trough depth of $d_{\rm BAL}=0.267$ on MJD 60251 rivals the highest $d_{BAL}$ values in that sample at $v<-30,000$ \kms{}. 
Its maximum trough width of at least $13,930$ \kms{} (and possibly as high as $15,500$ \kms{}) and maximum BI of \kms{3000} are both among the top $\simeq$3\% in that sample\footnote{Our BI$_{EHVO}$ measurements required \kms{2000} of contiguous absorption and are thus more conservative than the BI$^{\rm *}$ measurements of \cite{jarthesis} which only required \kms{1000}.}, 
and the highest at $v<-30,000$ \kms{}.
The highest-velocity trough in that sample is in the weak-lined quasar J0230, with $v_{cent}=-56,200$ \kms{} \citep{J0230}. That trough is less remarkable than that of J2318, having width \kms{7400}, $d_{BAL}=0.194$, and maximum EW = \kms{1450\pm 120}.

Second, from the information given in \cite{2018MNRAS.476..943H}, the \ion{C}{4} trough in PDS 456 has $v_{cent}=-90,000$ \kms{}, width $11,250$ \kms{}, $d_{BAL}=0.213$, and EW = \kms{2400\pm 130} (no Ly$\alpha$ forest correction being needed due to its low redshift). These quantities are all similar to those in J2318.
For PDS 456, \ion{Si}{4} cannot be detected because its expected wavelength is the same as that of Galactic Ly$\alpha$.

Finally, J2318 also has a maximum \ion{C}{4} EW$_{\rm corrected}$ and $BI_{EHVO}$ which are equaled or exceeded by only one quasar in the EHVO sample of \cite{EHVO}. 
That quasar is SDSS J164653.72+243942.2 (J1646), which has an EHVO $16,100$ \kms{} wide at $v_{cent}\simeq-43,000$ \kms{} with EW ranging from \kms{3900} to \kms{6800} and $BI_{EHVO}$ from \kms{3700} to \kms{6400} \citep{J1646}.

Thus J2318 joins J1646 and PDS 456 in hosting exceptional UV-absorbing outflows.
The outflow in J2318 is weaker and somewhat narrower than that in J1646 but at almost double its $v_{cent}$, and stronger and deeper than that in PDS 456 but at a slightly lower $v_{cent}$.

These three quasars, J2318, J1646, and PDS 456, have high luminosities, black hole masses, and Eddington ratios, spanning a range of only $\sim$4 in each quantity.
PDS 456 has an estimated black hole mass of $5\times 10^8$ $M_\odot$ and an estimated Eddington ratio of 1.7 \citep{2025Natur.641.1132XRISM.PDS456}, as compared to our estimates of $2\times 10^9$ $M_\odot$ and 0.4$-$0.6 for J2318 and to $3\times 10^9$ $M_\odot$ and 0.4 for J1646 \citep{J1646}.
Again, we caution that such high Eddington ratios may be lower limits and the corresponding black hole masses upper limits; see \S~\ref{sec:Edd}.

\subsection{Potential implications from X-ray studies of PDS 456} \label{sec:Xray}
PDS 456 has been observed recently with the XRISM satellite, revealing five discrete velocity components seen in absorption in hard X-rays, and a sixth seen in soft X-rays, in an outflow spanning $v=-0.22c$ to $v=-0.33c$ \citep{2025Natur.641.1132XRISM.PDS456,2025PASJ..tmp...73XuPDS456XRISM}.
The X-ray observations can be explained by a clumpy and stratified wind. 
The UV detection of \ion{C}{4} in PDS 456 at $v=-(0.30\pm0.03)c$ could be part of this outflow if the column density in at least some clumps is large enough to enable such relatively low-ionization ions to exist in them.

In the model of \citet{2025Natur.641.1132XRISM.PDS456}, the clumps are of comparable size to the X-ray emitting region ($R\simeq 8~GM_{BH}/c^2$).
Because the size of the UV emitting region is larger than that ($R\simeq 100~GM_{BH}/c^2$), the coverage of the UV emitting region will depend on the volume filling factor of the clumps, estimated to be $f_{\rm vol}=0.1-0.3$ \citep{2025Natur.641.1132XRISM.PDS456}. 
That filling factor range corresponds to a projected areal covering factor range of $f_{\rm vol}^{2/3}=0.22-0.45$. 
The maximum depth of \ion{C}{4} absorption in PDS 456 is $\simeq$30\%, consistent with that range if most clumps contain both X-ray and UV-absorbing gas. 
Alternatively, the physical structure of the UV-absorbing outflow might yield high covering of the UV-emitting region but low covering of the X-ray-emitting region, complicating the extrapolation of results from one wavelength range to the other.

In any case, it would be useful to conduct X-ray studies of J2318 and J1646 (which has no existing X-ray observations) to determine how similar their overall outflows are to that of PDS 456. Such observations will require next-generation X-ray facilities such as NewAthena \citep{2025NewAthena}.

\subsection{Comparison to the Galactic microquasar SS 433} \label{sec:SS433}
The Galactic microquasar SS~433 \citep{1984ARA&A..22..507M} exhibits emission from excited neutral hydrogen and helium travelling outward from an accretion disk at $0.24c < v < 0.28c$ \citep{2004ApJ...616L.159B,2005ApJ...622L.129B} in two oppositely directed, precessing jets.
Line emission from \ion{H}{1} in the jets is seen close enough to the central source from which the jets are launched \citep{2017A&A...602L..11G} that the neutral gas must have survived rapid acceleration in the outflow. 
The outflowing neutral gas in SS~433 must be highly clumped to be dense enough for recombination to counteract ionization: \citet{1980ApJ...238..722B} find volume filling factor $f\lesssim 3\times 10^{-5}$ from general considerations (their Eq.~14) and \citet{1984ApJ...283..295P} find $f\lesssim 10^{-7}$ from the density contrast $n/\bar{n}\gtrsim 10^{7}$ required for a \ion{He}{2}-driven line-locking model (their \S~IV), and $f\lesssim 10^{-8}$ for an \ion{H}{1}-driven line-locking model.
Intriguingly, these values are comparable to the EHVO dense gas filling factor $f\lesssim10^{-6}$ inferred by \citet{2013MNRAS.435..133H}.

The line-locking model for the outward jet velocities of $0.24c < v < 0.28c$ in SS~433 is potentially relevant to J2318 (outward velocities $0.25c<v<0.32c$) and PDS 456 (outward velocities $0.27c<v<0.33c$ in the UV and $0.22c<v<0.33c$ in the X-ray). 
The line-locking model was proposed for SS~433 based on the fact that the Doppler shift between Ly$\alpha$ and the Lyman limit for hydrogenic ions corresponds to $v=0.28c$. 
If gas is accelerated by scattering and absorption of Ly$\alpha$ photons in such transitions, and the accelerating continuum is sharply reduced at the relevant Lyman limit, the acceleration of the gas will terminate at $0.26c<v<0.28c$ depending on the strength of Lyman series absorption \citep{1986ApJS...60..393S}. 
As \citet{1984ARA&A..22..507M} states, it is also possible that this line-locking mechanism serves only to stabilize the velocity once it is reached via some other process. 
Nonetheless, the similar velocities of the absorbers in SS 433, PDS 456, and J2318 may have their origin in this line-locking  mechanism, which is related to but distinct from lower-velocity line-doublet line-locking mechanisms known to occur in quasar outflows \citep[e.g.,][]{2014MNRAS.445..359B}.

PDS 456 and J2318 have no spectral coverage down to the Lyman limit ($\lambda_{obs}$ of 1080\,\AA\ and 3354\,\AA, respectively). But composite spectra of samples of BAL quasars show that they are not uniformly absorbed below the Lyman limit \citep{2015MNRAS.449.1593B}. 
If this line-locking mechanism is at work in J2318 and PDS 456, the somewhat larger outward velocities reached in them could correspond to a gradual reduction in flux down to $\lambda_{rest}\simeq 863-873$~\AA\ rather than a sharp cutoff at the Lyman limit. 
Conversely, the presence of flux in these objects below those rest-frame wavelengths would argue against \ion{H}{1} Ly$\alpha$/Lyman-limit line-locking as an explanation for their observed UV outflow speeds. 
The relevant $\lambda_{obs}$ of 1019~\AA\ for PDS 456 and 3207~\AA\ for J2318 mean such observations are challenging but possible.
Alternatively, SDSS-V, DESI \citep{DESIDR1} and other such spectroscopic surveys eventually may provide large enough samples of EHVOs to determine if there is a significant overabundance of outflows at outward velocities $v\sim0.3c$, which would potentially be a signature of Ly$\alpha$/Lyman-limit line-locking at work.

\subsection{Feedback} \label{sec:feedback}
In \S \ref{sec:mass}, we used conservative assumptions to show that at minimum the J2318 UV-absorbing outflow is about a factor of two below the threshold to have significant feedback, as quantified by the ratio of kinetic to bolometric luminosities.
The only parameter that could decrease our assumed feedback level is the volume filling factor of the outflow. 
We assumed $f_{\rm vol}=1$ but if the outflow in J2318 is accompanied by an X-ray-absorbing outflow like that in PDS 456 and if the X-ray and UV-absorbing gas is cospatial, a value $f_{\rm vol}=0.1-0.3$ would be more appropriate.
Otherwise, the feedback will be larger if the UV absorption is saturated, it could be up to five times higher if the outflow fully covers the source, it will increase linearly with distances larger than 10 pc, and it could be one or two orders of magnitude larger if the absorbing column is dominated by higher ionization stages than \ion{C}{4} and \ion{Si}{4}, which is often the case in lower-velocity BAL outflows \citep{borgaravPV,2020MNRAS.499.1522M}. 

If rapid trough variability is observed in future, it might usefully constrain the distance and density of the UV-absorbing outflow (e.g., \citealt{wywf}, \citealt{2019ApJ...872...21H}). 
If J2318 is found to have a normal X-ray luminosity, as some weak-lined quasars do, that would constrain the column density of higher-ionization absorption. 
Studies of EHVOs at similar redshifts with lower-velocity outflows ($v=-30,000$ \kms{} to $-60,000$ \kms{}) will have better prospects for observing transitions useful for constraining UV-absorbing outflow properties, such as \ion{S}{4} $\lambda$1062 and \ion{C}{3} $\lambda$1175 \citep{borgaravPV}.

For example, we can compare the \ion{C}{4} absorption in PDS 456 to the absorbing column inferred from its X-ray absorption.
Using the same methods as above with $\tau=0.24$ and $R=10$~pc, we find $N_{\rm CIV}=3.44\times 10^{15}$ cm$^{-2}$, $N_H>1.3 \times 10^{19}$ cm$^{-2}$, $\langle\dot{M}\rangle > 0.3~M_\odot~ {\rm yr}^{-1}$ and $L_{kin}=7.5\times 10^{44}$ erg s$^{-1}$ = 0.6\% of $L_{\rm Edd}$.
The distance assumed above is $10^3$ times larger than the distance $R=0.01$ pc inferred for the X-ray absorbing gas in PDS 456 by \citet{2025Natur.641.1132XRISM.PDS456}.
In contrast, the total absorbing column in the X-ray UFO in PDS 456 is $\simeq 4\times 10^{23}$ cm$^{-2}$ \citep{2025PASJ..tmp...73XuPDS456XRISM}, the inferred $\langle\dot{M}\rangle \simeq 60-170~M_\odot~ {\rm yr}^{-1}$, and the inferred $L_{kin}\simeq 10^{47}$ erg s$^{-1}$ $\simeq$ $L_{\rm Edd}$ \citep{2025Natur.641.1132XRISM.PDS456}. 
Furthermore, PDS 456 has a molecular outflow reaching $-$\kms{1000} detected in CO which dominates the feedback in terms of mass outflow rate ($\sim 290 M_\odot {\rm ~yr^{-1}}$) but not kinetic luminosity
\citep{2019A&A...628A.118B}.

The above shows the stark contrast between the lower limit on feedback calculable in EHVOs from UV absorption only and the comprehensive view of feedback obtainable from studying outflowing gas in all ionization phases.

\section{Conclusions} \label{sec:con}
We have reported the discovery and analysis of an extremely high-velocity \ion{C}{4} and \ion{Si}{4} outflow in the weak-lined quasar SDSS J231854.31+243954.2 (J2318). 
With a centroid velocity of $-$84,000 km\,s$^{-1}$, this is the highest-velocity quasar outflow first discovered in the ultraviolet, the fastest \ion{Si}{4} outflow known, and the second-fastest \ion{C}{4} outflow known.

(1) The spectra reveal a \ion{C}{4} and \ion{Si}{4} BAL trough outflowing from J2318 at velocities of $\sim84,700$ \kms{} and $\sim83,400$ \kms{}, respectively. 
The CIV trough is easily distinguishable in all three spectra, whereas the \ion{Si}{4} trough is only observed in MJD 59188 and 60251, see Figure \ref{fig:J2318_BAL_Zoom1}.
Both of these absorption features are assumed to be part of the same outflow, due to their similar velocities and coordinated variability, which could likely be driven by the transverse motion of the outflows crossing our line of sight that block all, or some portion, of the continuum emitting region, changes in the ionization parameter of the outflow, or a combination of the two, see \S \ref{sec:variability}.
Transverse velocity calculations (see \S \ref{sec:transverse_motion}) of the \ion{C}{4} and \ion{Si}{4} BALs across all epochs yield  $473<v~({\rm km~s}^{-1})<11300$ and $389<v~({\rm km~s}^{-1})<11300$, respectively, while ionization parameters changes are beyond the scope of this paper. 
Thus, future observations and subsequent BAL variability analysis are necessary to determine what the most likely driver is.

(2) Photometry of J2318 highlights that the quasar’s brightness increased from SDSS imaging epochs into the PTF and ZTF observations, before a faintening and reddening between MJD 59000 and 59750, followed by a small recovery in brightness through the second SDSS-V observation. 
See Figure \ref{fig:J2318_Mag}.

(3) Various continuum models were used to fit the spectrum; the CS and SMC models have the lowest, and third lowest $\chi^2_\nu$ values, respectively. 
The $\alpha_\lambda$ best-fit parameter of the CS model was nonphysical, but the overall model adequately normalized the spectrum for trough and continuum measurements. 
See Figure \ref{fig:J2318_DV2_SpecFit}.

(4) The CS, SMC, and PQF continuum model’s reasonable fitting of the longer wavelengths in the spectra, along with PyQSOFit’s built-in emission line fitting, enabled a measurement of the flux at \A{5100} and the FWHM of the H$\alpha$ feature. 
The measurements for MJD 60251 were used to estimate a virial black hole mass of $M_{\rm BH}=1.65\pm0.06\times10^9~M_\odot$, an Eddington luminosity of $(2.43\pm0.10)\times10^{47}$ erg s$^{-1}$, and an Eddington ratio of $0.45\pm0.13$.
See Table \ref{tab:Ledd_result}.

(5) We compared the luminosity of J2138 to SEDs of larger quasar samples, to find that J2138 is more luminous and bluer than typical SDSS quasar SEDs.
See Figure \ref{fig:Photo+SED_Luminosity}.

(6) Using the CS normalized spectra, the EW and BI were measured for the BAL troughs via direct integration and Gaussian fitting.
The EW of the EHVO increased over the three epochs, MJD 57328 to 60251, spanning $\sim2.2$ rest-frame years total.
Direct integration reveals that \ion{C}{4} increased from an EW of 660 to \kms{3740}, and \ion{Si}{4} increased from consistent with zero to \kms{1100}.
While Gaussian fitting reveals that \ion{C}{4} increased from and EW of 510 to \kms{4000}, and \ion{Si}{4} increased from consistent with zero to \kms{250}.
Maximum velocity ranges of ($-76,360$ to $-90,300$) \kms{} and ($-81,850$ to $-85,670$) \kms{} were measured for \ion{C}{4} and \ion{Si}{4}, respectively, where their initial centroid velocities of $\sim-84,900$ \kms{} and $\sim-83,600$ \kms{} have little variation across epochs.
See Table \ref{tab:direct} and Table \ref{tab:abs_measurements} for the results of direct integration and Gaussian fitting, respectively.

(7) Under the assumption that $100\%$ of the continuum source is covered by the outflow, the lower limit of the outflow's column density is $N_H\geq3.39\times10^{19}~{\rm cm}^{-2}$.
This result leads to a mass-loss rate of $\dot{M}>0.82~M_\odot~{\rm yr}^{-1}$ for an outflow distance of 10 pc and a kinetic luminosity of $L_K > 1.82\times10^{45}~{\rm erg~s}^{-1}$.
The outflow in J2318 is expected to have significant feedback on its host galaxy since its kinetic luminosity to Eddington luminosity ratio is 0.75\%, above the 0.5\% threshold for significant feedback found by \citet{he10}.
With assumptions about the outflow's physical and energetic properties, the radius from the central SMBH was determined to be $0.5\lesssim R \lesssim120$ pc.
See \S\ref{sec:mass}.

\begin{acknowledgments}\label{sec:Acknowledgements}
Funding for the Sloan Digital Sky Survey V has been provided by the Alfred P. Sloan Foundation, the Heising-Simons Foundation, the National Science Foundation, and the Participating Institutions. 
SDSS acknowledges support and resources from the Center for High-Performance Computing at the University of Utah. 
SDSS telescopes are located at Apache Point Observatory, funded by the Astrophysical Research Consortium and operated by New Mexico State University, and at Las Campanas Observatory, operated by the Carnegie Institution for Science. 
The SDSS web site is \url{www.sdss.org}.

SDSS is managed by the Astrophysical Research Consortium for the Participating Institutions of the SDSS Collaboration, including the Carnegie Institution for Science, Chilean National Time Allocation Committee (CNTAC) ratified researchers, Caltech, the Gotham Participation Group, Harvard University, Heidelberg University, The Flatiron Institute, The Johns Hopkins University, L'Ecole polytechnique f\'{e}d\'{e}rale de Lausanne (EPFL), Leibniz-Institut f\"{u}r Astrophysik Potsdam (AIP), Max-Planck-Institut f\"{u}r Astronomie (MPIA Heidelberg), Max-Planck-Institut f\"{u}r Extraterrestrische Physik (MPE), Nanjing University, National Astronomical Observatories of China (NAOC), New Mexico State University, The Ohio State University, Pennsylvania State University, Smithsonian Astrophysical Observatory, Space Telescope Science Institute (STScI), the Stellar Astrophysics Participation Group, Universidad Nacional Aut\'{o}noma de M\'{e}xico, University of Arizona, University of Colorado Boulder, University of Illinois at Urbana-Champaign, University of Toronto, University of Utah, University of Virginia, Yale University, and Yunnan University.

Funding for the Sloan Digital Sky Survey IV has been provided by the Alfred P. Sloan Foundation, the U.S. Department of Energy Office of Science, and the Participating Institutions. 

This work was enabled by observations made from the Gemini North telescope, located within the Maunakea Science Reserve and adjacent to the summit of Maunakea. 
We are grateful for the privilege of observing the Universe from a place that is unique in both its astronomical quality and its cultural significance.
The international Gemini Observatory is a program of NSF's NOIRLab, which is managed by the Association of Universities for Research in Astronomy (AURA) under a cooperative agreement with the National Science Foundation on behalf of the Gemini Observatory partnership: the National Science Foundation (United States), National Research Council (Canada), Agencia Nacional de Investigaci\'{o}n y Desarrollo (Chile), Ministerio de Ciencia, Tecnolog\'{\i}a e Innovaci\'{o}n (Argentina), Minist\'{e}rio da Ci\^{e}ncia, Tecnologia, Inova\c{c}\~{o}es e Comunica\c{c}\~{o}es (Brazil), and Korea Astronomy and Space Science Institute (Republic of Korea).

This research has made use of the NASA/IPAC Infrared Science Archive, which is funded by the National Aeronautics and Space Administration and operated by the California Institute of Technology.
This publication makes use of data products from the Wide-field Infrared Survey Explorer, which is a joint project of the University of California, Los Angeles, and the Jet Propulsion Laboratory/California Institute of Technology, and NEOWISE, which is a project of the Jet Propulsion Laboratory/California Institute of Technology. 
WISE and NEOWISE are funded by the National Aeronautics and Space Administration.
This research has made use of the UKIRT Hemisphere Survey, a partnership between the UK STFC, The University of Hawaii, The University of Arizona, Lockheed Martin and NASA.
This publication makes use of data products from the Two Micron All Sky Survey, which is a joint project of the University of Massachusetts and the Infrared Processing and Analysis Center/California Institute of Technology, funded by the National Aeronautics and Space Administration and the National Science Foundation.

PH acknowledges support from the Natural Sciences and Engineering Research Council of Canada (NSERC), funding reference number 2023-05068, and from the Research at York program to support MV and ZZ.
ALR acknowledges support from a Leverhulme Early Career Fellowship.
CAN acknowledges the support from projects CONAHCyT CBF2023-2024-1418, PAPIIT IA104325 and IN119123.
RJA was supported by FONDECYT grant number 1231718 and by the ANID BASAL project FB210003.
\end{acknowledgments}

\begin{contribution} 
    Author Marianna Veltri noted an unusual absorption trough in the MJD 60251 SDSS-V spectrum of J2318 during visual inspection of new SDSS-V quasar spectra in mid-November 2023.
    Author Patrick B. Hall started this project, proposed for and reduced the GNIRS data, and analyzed the \ion{C}{4} and \ion{Si}{4} absorption troughs through direct integration.
    Author Lucas M. Seaton took over the project and normalized the spectra with fitted models, determined J2318's SED and compared it to other quasars, and estimated the BH mass and Eddington luminosity.
    Authors Liliana Flores and Paola Rodr\'iguez Hidalgo performed the absorption trough measurements through Gaussian fitting.
    All aforementioned authors contributed to the writing and editing of this paper and all listed authors provided valuable insight and discussion.
\end{contribution}

\facilities{Sloan, Gemini:Gillett, IRSA, WISE, NEOWISE, PS1, CXO, UKIRT, SPHEREx}

\software{
Astropy \citep{2013A&A...558A..33A,2018AJ....156..123A},
IRAF \citep{IRAFV218}
}

\bibliographystyle{aasjournalv7}
\bibliography{pathall}

\end{document}